\documentclass[12pt]{article}
\usepackage[utf8]{inputenc}
\usepackage[DIV=13]{typearea}\usepackage{ragged2e}
\usepackage{calligra,amsmath,amsfonts,bbm,mathrsfs,amssymb}
\usepackage{slashed,cancel}
\usepackage{cite}
\usepackage{hyperref}
\hypersetup{colorlinks=true,urlcolor=magenta,anchorcolor=blue,citecolor=blue,filecolor=blue,linkcolor=magenta,menucolor=blue,linktocpage=true,pdfproducer=medialab,pdfauthor={Jorge Alda, Marta Fuentes Zamoro, Luca Merlo, Xavier Ponce Diaz and Stefano Rigolin},pdftitle={Comprehensive ALP Searches in Meson Decays}}
\usepackage[english]{babel}
\usepackage{indentfirst}
\usepackage{graphicx}
\usepackage{float}
\usepackage{enumerate}
\usepackage{caption}
\usepackage{subcaption}
\usepackage{multirow}
\usepackage{xcolor}
\usepackage{booktabs}
\usepackage{physics}
\usepackage{url}
\usepackage{soul}
\usepackage{verbatim}
\usepackage{multicol}
\usepackage{bbold}
\usepackage{fontawesome}
\usepackage{tablefootnote}
\usepackage[normalem]{ulem}
\newcommand{\email}[1]{\href{mailto:#1}{\tt #1}}

\numberwithin{equation}{section}

\newcommand{\blue}[1]{\color{blue} #1 \color{black}}
\newcommand{\magenta}[1]{\color{magenta} #1 \color{black}}
\newcommand{\be}{\begin{equation}}
\newcommand{\ee}{\end{equation}}
\newcommand{\ba} {\begin{equation}\begin{aligned}}
\newcommand{\ea} {\end{aligned}\end{equation}}
\newcommand{\bea}{\begin{eqnarray}}
\newcommand{\eea}{\end{eqnarray}}

\newcommand{\cO}{\mathcal{O}}

\newcommand{\hc}{\text{h.c.}}

\newcommand{\nn}{\nonumber}
\renewcommand{\vev}[1]{\langle #1\rangle}
\newcommand{\ov}[1]{\overline{#1}}

\newcommand{\unity}{\mathbbm{1}}

\def\Tr{{\mathrm{Tr}}}

\newcommand\subsetsim{\mathrel{
  \ooalign{\raise0.2ex\hbox{$\subset$}\cr\hidewidth\raise-0.8ex\hbox{\scalebox{0.9}{$\sim$}}\hidewidth\cr}}}

\newcommand{\TeV}{\ \text{TeV}}
\newcommand{\GeV}{\ \text{GeV}}

\newcommand{\alpaca}{{\fontfamily{cmss}\selectfont ALPaca }}
\usepackage{color}
\usepackage{colortbl}
\definecolor{rossoc}{cmyk}{0,1,1,0.2}
\definecolor{dgreen}{rgb}{0.0, 0.5, 0.0}

\newcommand{\X}{\mathcal{X}}

\definecolor{orangered}{rgb}{1.0, 0.27058823529411763, 0.0}
\definecolor{darkred}{rgb}{0.5450980392156862, 0.0, 0.0}
\definecolor{goldenrod}{rgb}{0.8549019607843137, 0.6470588235294118, 0.12549019607843137}
\definecolor{darkgreen}{rgb}{0.0, 0.39215686274509803, 0.0}
\definecolor{darkblue}{rgb}{0.0, 0.0, 0.5450980392156862}
\definecolor{mediumturquoise}{rgb}{0.2823529411764706, 0.8196078431372549, 0.8}
\definecolor{orchid}{rgb}{0.8549019607843137, 0.4392156862745098, 0.8392156862745098}
\definecolor{lightslategrey}{rgb}{0.4666666666666667, 0.5333333333333333, 0.6}
\definecolor{yellowgreen}{rgb}{0.603922, 0.835294, 0}

\begin{document} 
\renewcommand*{\thefootnote}{\fnsymbol{footnote}}

\begin{titlepage}

\vspace*{-1cm}
\flushleft{\magenta{IFT-UAM/CSIC-25-55}} 
\\[1cm]
\vskip 1cm

\begin{center}
\blue{\bf \Large Comprehensive ALP Searches in Meson Decays}
\centering
\vskip .3cm
\end{center}
\vskip 0.5  cm
\begin{center}
{\large\bf Jorge Alda}$^{a,b}$~\footnote{\email{jorge.alda@pd.infn.it}},
{\large\bf Marta Fuentes Zamoro}$^{c}$~\footnote{\email{marta.zamoro@uam.es}},
{\large\bf Luca Merlo}$^{c}$~\footnote{\email{luca.merlo@uam.es}},\vskip 0.5cm
{\large\bf Xavier Ponce D\'iaz}$^{d}$~\footnote{\email{xavier.poncediaz@unibas.ch}},
and 
{\large\bf Stefano Rigolin}$^{a}$~\footnote{\email{stefano.rigolin@pd.infn.it}},
\vskip .7cm
{\footnotesize
$^a$~Dipartamento di Fisica e Astronomia ``G.~Galilei" and Istituto Nazionale di Fisica Nucleare,\\
Sezione di Padova, Universit\`a degli Studi di Padova, I-35131 Padova, Italy{\par\centering \vskip 0.25 cm\par}
$^b$~Centro de Astropart\'iculas y F\'isica de Altas Energ\'ias (CAPA)\\ Pedro Cerbuna 12, E-50009 Zaragoza, Spain
{\par\centering \vskip 0.25 cm\par}
$^c$~Departamento de F\'isica Te\'orica and Instituto de F\'isica Te\'orica UAM/CSIC,\\
Universidad Aut\'onoma de Madrid, Cantoblanco, 28049, Madrid, Spain \\
{\par\centering \vskip 0.25 cm\par}
$^d$~Department of Physics, University of Basel,  Klingelbergstrasse 82, \\ CH-4056 Basel, 
Switzerland
}

\end{center}
\vskip 2cm
\begin{abstract}
\justify

We present a comprehensive study of axion-like particles (ALPs) in meson decays, combining effective field theory and 
ultraviolet models within the open-source tool \href{https://github.com/alp-aca/alp-aca}{\alpaca  \faicon{github}}. 
The analysis accounts for running and matching effects across energy scales, including non-perturbative QCD corrections 
via chiral perturbation theory. We discuss several benchmark models, both flavour-universal and non-universal, using 
the most up-to-date theoretical computations for ALP decays and branching ratios. Experimental signatures such as prompt, 
displaced, and invisible decays are included. A dedicated analysis of the Belle~II anomaly in the decay $B^+\to K^+\nu\ov{\nu}$ is performed. Our results 
highlight the power of flavour observables in constraining ALPs and provide a versatile foundation for future searches.

\end{abstract}
\end{titlepage}
\setcounter{footnote}{0}

\pdfbookmark[1]{Table of Contents}{tableofcontents}
\tableofcontents
\renewcommand*{\thefootnote}{\arabic{footnote}}

\bigskip

\section{Introduction}
The appearance of global $\mathrm{U}(1)$ symmetries is  common in a large variety of Beyond Standard Model (BSM) 
constructions. In general, they must be broken in order to ensure consistency with the Standard Model (SM) at 
low energies, that only allows invariance under $B-L$ at the quantum level. When these global symmetries undergo 
a spontaneous breaking, new light degrees of freedom, i.e. Goldstone bosons, appear in the low energy spectrum.
One of these frameworks is the well-known axion solution to the Strong CP problem~\cite{Peccei:1977hh,Weinberg:1977ma,
Wilczek:1977pj}, which provides a dynamical explanation for the absence of CP violation (CPV) in the pure QCD 
Lagrangian, as indicated by the neutron electric dipole moment measurements~\cite{Abel:2020pzs}. 

In addition to the QCD-axion proposal, part of the community intensively investigated, both theoretically 
and experimentally, the so-called Axion-Like Particle (ALP) alternative: a CP-odd scalar field whose 
Lagrangian is invariant under a shift symmetry only broken by gauge anomalous terms and by a small 
soft-breaking mass $m_a$.

There are many examples of ALPs in the literature: appearing in string theories \cite{Witten:1984dg,Choi:2006qj,
Svrcek:2006yi,Arvanitaki:2009fg,Cicoli:2012sz}, in composite Higgs models~\cite{Merlo:2017sun,Brivio:2017sdm,
Alonso-Gonzalez:2018vpc,Alonso-Gonzalez:2020wst}, or even supersymmetric contexts~\cite{Bellazzini:2017neg}; 
playing the role of a Dark Matter candidate~\cite{Gelmini:1984pe,Berezinsky:1993fm,Lattanzi:2007ux,
Bazzocchi:2008fh,Lattanzi:2013uza,Queiroz:2014yna}, or having an impact on cosmological observables~
\cite{Ferreira:2018vjj,DEramo:2018vss,Escudero:2019gvw,Arias-Aragon:2020qtn,Arias-Aragon:2020qip,
Arias-Aragon:2020shv,Ferreira:2020bpb,Escudero:2021rfi,Araki:2021xdk,DEramo:2021psx,DEramo:2021lgb,
DEramo:2022nvb}; associated to flavour model dynamics~\cite{Davidson:1981zd,Wilczek:1982rv,Ema:2016ops,
Calibbi:2016hwq,Arias-Aragon:2017eww,Arias-Aragon:2022ats,DiLuzio:2023ndz,Greljo:2024evt} or neutrino 
mass generation~\cite{Chikashige:1980qk,Chikashige:1980ui,Gelmini:1980re,deGiorgi:2023tvn}. Alongside the 
model building activity, the effective description of ALPs has been a main goal for part of the community. 
Several studies appeared to define the ALP Effective Field Theory (EFT)~\cite{Choi:1986zw,Salvio:2013iaa,
Brivio:2017ije,Alonso-Alvarez:2018irt,Gavela:2019wzg,Chala:2020wvs,Bonilla:2021ufe,Arias-Aragon:2022byr,
Arias-Aragon:2022iwl} and study its possible signals both at colliders~\cite{Jaeckel:2012yz,Mimasu:2014nea,
Jaeckel:2015jla,Alves:2016koo,Knapen:2016moh,Brivio:2017ije,Bauer:2017nlg,Mariotti:2017vtv,Bauer:2017ris,
Baldenegro:2018hng,Craig:2018kne,Bauer:2018uxu,Gavela:2019cmq,Haghighat:2020nuh,Wang:2021uyb,deGiorgi:2022oks,
Bonilla:2022pxu,Ghebretinsaea:2022djg,Vileta:2022jou,Marcos:2024yfm,Arias-Aragon:2024gpm} and low-energy 
facilities~\cite{Izaguirre:2016dfi,Merlo:2019anv,Aloni:2019ruo,Bauer:2019gfk,Bauer:2020jbp,Bauer:2021mvw,Carmona:2021seb,
Guerrera:2021yss,Gallo:2021ame,Bonilla:2022qgm,Bonilla:2022vtn,deGiorgi:2022vup,Guerrera:2022ykl,
Bonilla:2023dtf,Arias-Aragon:2023ehh,DiLuzio:2024jip,deGiorgi:2024str,Alda:2024cxn,Alda:2024xxa,
Arias-Aragon:2024qji,Arias-Aragon:2024gdz}.

ALPs have traditionally not been considered a solution to the strong CP problem, but recent axion model building~\cite{DiLuzio:2016sbl,DiLuzio:2017pfr,Gaillard:2018xgk,Hook:2019qoh,DiLuzio:2020wdo,DiLuzio:2020oah,DiLuzio:2021pxd,
DiLuzio:2021gos,Gavela:2023tzu,Cox:2023dou,deGiorgi:2024elx} showed that the inverse proportionality between $m_a$ and its characteristic scale $f_a$ can be relaxed, opening up the QCD-axion parameter space. In particular, this implies that ALPs may be considered as effective descriptions of axions that actually solve the Strong CP problem, but whose mass falls outside the traditional QCD band. This is the context on which we develop our study.

Over the past decade, the search for promising ALP signals has been highly intensive, involving a wide range of experimental setups. Simultaneously, experimental collaborations have conducted dedicated data analyses to identify potential evidence of New Physics (NP) consistent with ALPs. As a result, they have successfully constrained large regions of the ALP parameter space. The field has reached a maturity point to provide the community with a tool that facilitates the combination among the different pieces of information on ALP physics. Our project presents an open-access program, \alpaca — ALP Automatic Computing Algorithm — designed to explore the ALP parameter space by computing a wide range of observables. The program~\cite{ALPACA} accepts as input either user-defined couplings for the EFT at a chosen energy scale, $\Lambda$, or ultraviolet (UV) completions defined at the same scale, such as those described in Section~\ref{sec:models}. \alpaca then evolves the Lagrangian parameters down to the  energy scale $\mu$, and matches them to the appropriate EFT. Once the relevant couplings are determined at an energy scale suitable for testing flavour observables, the program can generate exclusion plots for the parameters of interest, such as specific Wilson coefficients, $m_a$, or $f_a$, using the available experimental datasets. The user has full control over which experimental constraints to include. 
In addition, the program contains a routine that computes the decay length of the ALP in each considered experiment, allowing one to distinguish between visible and invisible ALPs, something that has sometimes been overlooked in past analyses.

On the other hand, this manuscript outlines the formal background of the program and presents a phenomenological analysis using the most up-to-date data from flavour facilities. This is not only the latest available analysis of ALP couplings based on flavour observables, but also serves as a laboratory playground to showcase the potential of the \alpaca program. A word of caution is in order: searches for long-lived particles in beam-dump experiments are not considered here, as they require Monte Carlo simulations to compute ALP production across all relevant channels. Furthermore, we do not include astrophysical bounds or searches, as these may rely on specific model-dependent assumptions. Both types of observables could be implemented in future versions of \alpaca\!\!; however, doing so lies beyond the scope of the present study.
 
More into detail, Section~\ref{sec:lagrangian} introduces the ALP effective Lagrangian, discussing the possible redundancies among parameters and the different bases commonly used in the literature. We stress out the relevance of identifying the scale at which the EFT is defined, making the difference between exact and broken electroweak (EW) phase and below $\sim2\GeV$, where chiral perturbation theory is the correct tool. A key aspect to take into consideration is the quantum corrections of running the theory from one given scale down to the energy relevant for the considered observables.

In Section~\ref{sec:models}, we move from the effective description to specific benchmark models. Indeed, \alpaca can be fed with an effective ALP description or with a renormalizable Lagrangian. The models we focus on are {\it \`a la} Dine-Fischler-Sredinicki-Zhitnitsky (DFSZ)~\cite{Zhitnitsky:1980tq,Dine:1981rt} and Kim-Shifman-Vainshtein-Zakharov (KSVZ)~\cite{Kim:1979if,Shifman:1979if} that are based on the original formulations of the invisible Axion. Additionally, we also consider a top-philic ALP construction that is deeply studied especially at colliders~\cite{Esser:2023fdo,Rygaard:2023dlx,Blasi:2023hvb}, and the flaxion or axiflavon model~\cite{Ema:2016ops,Calibbi:2016hwq} as a setup that presents flavour-violating couplings at tree level. We also considered a few ALP-coupling textures that are often of interest in the community and show the impact of the quantum corrections. This section does not pretend to include all the ALP models presented in the literature, but to provide a showcase of classes of models with specific properties in common.  

The rest of the sections deal with the ALP observables. Section~\ref{sec:decays} discusses the various ALP decays: depending on the ALP mass, different channels may be open and we compute all the possible decays for an ALP with a mass in the window $[0.01,\,10]$~GeV, comparing our results with the literature. In Section~\ref{sec:signatures}, we delve into the crucial yet often overlooked aspect of many analyses in the literature: determining, for each experimental facility considered, whether an ALP is long-lived or gives rise to prompt or displaced vertices. 

Finally, in Section~\ref{sec:pheno}, we provide a phenomenological analysis on the UV models discussed in Section~\ref{sec:models} as well as specific 2-parameter effective descriptions. We also present a dedicated study of the Belle II anomaly, discussing the conditions an ALP needs to satisfy to explain the excess in the $B^+\to K^+ \nu\ov{\nu}$ decay. The code used in our phenomenological analysis can be found in the repository \href{https://github.com/alp-aca/examples/tree/main/comprehensive_meson_searches}{\alpaca\!\!/examples  \faicon{github}}.

\section{The ALP Lagrangian}
\label{sec:lagrangian}

The Axion-Like Particle (ALP) Lagrangian is usually defined as the $D\le 5$ effective Lagrangian invariant 
under the constant shift symmetry of the ALP field, $a \to a + \epsilon$, only broken by gauge anomalous 
terms and by a light soft-breaking mass $m_a \ll f_a$ uncorrelated to the symmetry breaking scale $f_a$. 
Therefore, the QCD-axion relation $m_a f_a \approx m_\pi f_\pi$ is not assumed for a generic ALP. We 
furthermore assume that the ALP, $a$, is odd under CP and that the only source of CP violation arises 
from the SM Yukawa sector\footnote{The CP-violating ALP case is considered for example in \cite{DiLuzio:2020oah,
DiLuzio:2023cuk}.}. The corresponding effective Lagrangian at the UV scale $\Lambda=4\pi f_a$ can be written as
\begin{align}
\label{eq:UVLag}
    \mathcal{L}_{\textrm{ALP}} = &\frac{1}{2}(\partial_\mu a)^2 - \frac{1}{2}m_a^2 a^2 
    +\frac{a}{f_a} \left(\frac{g'^2 c_{B}}{16\pi^2} B_{\mu\nu}\tilde{B}^{\mu\nu} 
    +\frac{g^2 c_{W}}{16\pi^2} W^i_{\mu\nu}\tilde{W}^{i,\mu\nu} 
    +\frac{g_s^2 c_{G}}{16\pi^2}G^a_{\mu\nu}\tilde{G}^{a,\mu\nu}\right)+ \nn \\ 
  & \hskip -1cm +\frac{\partial^\mu a}{f_a} \left(\bar{q}_L\, c_{q} \gamma_\mu q_L  +\bar{u}_R \,c_{u} \gamma_\mu u_R 
    +\bar{d}_R\,c_{d}\gamma_\mu d_R +\bar{\ell}_L\,c_{\ell}\gamma_\mu\ell_L +\bar{e}_R\,c_{e}\gamma_\mu e_R \right)  + \mathcal{O}\left(\frac{a^2}{f^2_a}\right) .
\end{align}
Here $g_s$, $g$ and $g'$ denote the strong, weak and hypercharge coupling respectively; $G^a_{\mu\nu}$, $W^i_{\mu\nu}$ 
and $B_{\mu\nu}$ are the $\mathrm{SU}(3)_C \times \mathrm{SU}(2)_L \times \mathrm{U}(1)_Y$ field strengths and $\tilde{X}^{\mu\nu}\equiv 
\frac{1}{2} \epsilon^{\mu\nu\rho\sigma} {X}_{\rho\sigma}$, with $\epsilon^{0123}=1$, their associated dual fields. The left-handed (LH) and right-handed (RH) fermions $f=\{q_L,u_R,d_R,\ell_L,e_R\}$ are triplets in flavour space, 
while the corresponding ALP couplings, $c_f=\{c_{q},c_{u},c_{d},c_{\ell},c_{e}\}$, are general $3\times 3$ hermitian 
matrices. No $\nu_R$ multiplet has been introduced in $\mathcal{L}_{\textrm{ALP}}$, following the massless neutrinos 
SM \textit{ansatz}. Notice also that no ALP--Higgs operator has been included in the Lagrangian of Eq.~\eqref{eq:UVLag}, being 
redundant when ALP--fermion operators are included \cite{Georgi:1986df}. 

It is important to remark that not all the 49 real parameters appearing in the Lagrangian of Eq.~\eqref{eq:UVLag} 
are independent, due to the accidental global symmetries\footnote{The global $\mathrm{U}(1)_Y$ symmetry is 
associated to the redefinition of the ALP--Higgs coupling already not introduced in the Lagrangian of 
Eq.~\eqref{eq:UVLag}, see for example Refs.~\cite{Bauer:2020jbp,Bauer:2021mvw,Bonilla:2021ufe}.} of the SM. 
For example, by imposing total baryon and lepton number conservation, as we will customarily do throughout the whole 
paper, two real parameters can be reabsorbed by means of the following field-dependent transformations: 
\begin{equation}
\begin{split}\label{eq:BLtransformation}
   & q \to e^{i \alpha \frac{a}{f_a}} q \,   \qquad\qquad\qquad\qquad  \textrm{for } q= q_L,\, u_R,\, d_R , \\ 
   & \ell \to e^{i \beta \frac{a}{f_a}} \ell\, \,\, \qquad\qquad\qquad\qquad  \textrm{for } \ell = \ell_L,\, e_R\, .
\end{split}
\end{equation}
These transformations are anomalous under $\mathrm{SU}(2)_L\times \mathrm{U}(1)_Y$ but not under $\mathrm{SU}(3)_C\times \mathrm{U}(1)_{\textrm{QED}}$, 
leading to the following shifts in the couplings of Eq.~\eqref{eq:UVLag}:
\begin{align}
    \label{eq:fieldredefinitions1}
    & c_q \to c_q - \alpha  \qquad\qquad\qquad\qquad\textrm{for } q= q_L,\, u_R,\, d_R , \\
    \label{eq:fieldredefinitions2}
    & c_\ell \to c_\ell -\beta \qquad\qquad\qquad\qquad \,\textrm{for } \ell= \ell_L,\, e_R , \\
    \label{eq:fieldredefinitions3}
    & c_{W/B} \to c_{W/B} - (3\,\alpha + \beta)\, n_F/2\,,
\end{align}
with $n_F$ the number of flavours. These redefinitions can be used in flavour-universal ALP--fermion scenarios 
to remove the LH ALP--fermion couplings and define everything in terms of the RH ALP--fermion ones 
(or equivalently in terms of the ALP--fermion axial coefficients).

In the literature one often refers to the Lagrangian in Eq.~\eqref{eq:UVLag} as the ``derivative'' or 
``chirality-preserving'' basis Lagrangian. By mean of the following field-dependent redefinition: 
\begin{equation}
   f \to e^{i c_f \frac{a}{f_a}} \,f \,\qquad\qquad  \textrm{for } f=q_L,\,u_R,\,d_R,\,\ell_L,\,e_R,
\label{eq:tochiral}   
\end{equation}
one can, instead, write the ALP Lagrangian in the so called ``chirality-flipping'' basis, with the derivative 
couplings introduced in Eq.~\eqref{eq:UVLag} substituted for the ALP--fermion Yukawa-type interactions: 
\begin{eqnarray}
 \label{eq:Lag-chi-flipping}
\mathcal{L}_{\textrm{ALP}}&=& \frac{1}{2}(\partial_\mu a)^2 - \frac{1}{2}m_a^2 a^2  +
     \frac{a}{f_a} \left(\frac{g'^2 \tilde{c}_{B}}{16\pi^2} B_{\mu\nu}\tilde{B}^{\mu\nu} 
    + \frac{g^2 \tilde{c}_{W}}{16\pi^2} W^i_{\mu\nu}\tilde{W}^{i,\mu\nu}
    + \frac{g_s^2 \tilde{c}_{G}}{16\pi^2} G^a_{\mu\nu}\tilde{G}^{a,\mu\nu}\right)+ \nn \\
&&+ i\frac{a}{f_a} \left(\bar{q}_L H \tilde{c}_{d} d_R + \bar{q}_L \tilde{H} \tilde{c}_{u} u_R 
    + \bar{\ell}_L H \tilde{c}_{e} e_R \,+\mathrm{h.c.} \right) + \mathcal{O}\left(\frac{a^2}{f^2_a}\right) .
    \end{eqnarray}
Here $H$ is the Higgs doublet and $\tilde{H}\equiv i\tau_2 H^*$, while the ALP--fermion coefficients, $\tilde{c}_{f}$, are now complex $3\times 3$ matrices defined as: 
\begin{equation}
\tilde{c}_u = Y_u \,c_u - c_q Y_u \,,\qquad  
\tilde{c}_d = Y_d \,c_d - c_q Y_d \,, \qquad 
\tilde{c}_e = Y_e \,c_e - c_\ell Y_e \,,
\label{eq:chiralbasis}
\end{equation}
with $Y_i$ denoting the SM Yukawa matrices at the UV scale $\Lambda$. Notice that the two Lagrangians in 
Eq.~\eqref{eq:UVLag} and Eq.~\eqref{eq:Lag-chi-flipping} differ by $\mathcal{O}(a^2/f^2_a)$ terms. In the 
``chirality-flipping'' basis the shifted ALP--gauge anomalous coupling are related to the ones introduced in 
Eq.~\eqref{eq:UVLag} by:
\begin{align}
\tilde{c}_G &= c_G -\frac{1}{2} \Tr\Big(2\,c_{q}-c_{u}-c_{d}\Big)\,, \\
\tilde{c}_W &= c_W-\frac{1}{2} \Tr\Big(3\,c_{q}+c_{\ell}\Big)\,, \\
\tilde{c}_B &=c_B -\Tr \Big(\frac{1}{6}c_{q}+\frac{1}{2}c_{\ell}-\frac{4}{3}c_{u}-\frac{1}{3}c_{d}-c_{e}\Big)\,. 
\label{eq:gaugechiralbasis}
\end{align}
It turns out that these specific combinations of gauge and fermion ALP couplings will be very useful in the next 
sections as they enter in the running group equations (RGE) of the ALP--gauge Wilson coefficients.

By means of $\mathrm{SU}(2)_L \times \mathrm{U}(1)_Y$ invariant (field-independent) rotations of the SM fermion multiplets it is always 
possible to choose a basis in which the up-quark and lepton Yukawa matrices are diagonal, denoted by $\widehat{Y}$, 
while the down-quark one is proportional to the CKM matrix, at the UV scale $\Lambda$, times a diagonal matrix:
\begin{equation}
Y_u \rightarrow  \widehat{Y}_u (\Lambda)\,, \qquad\qquad
Y_d  \rightarrow  V_\text{CKM}(\Lambda) \, \widehat{Y}_d (\Lambda)\,,
\qquad\qquad
Y_e \rightarrow  \widehat{Y}_e(\Lambda) \,.
\label{eq:yukbasis}
\end{equation}
In the following discussion we will assume the ALP--fermion couplings $c_f$ in Eq.~\eqref{eq:UVLag} (and 
consequently the couplings $\tilde{c}_f$ Eqs.~\eqref{eq:Lag-chi-flipping}-\eqref{eq:gaugechiralbasis}) as
defined in this particular fermion basis at the UV scale $\Lambda$.

\subsection{The ALP Lagrangian: from the UV to the EW scale}\label{sec:UVtoEW}

\begin{figure}[t]
    \centering
    \includegraphics[width=0.85\linewidth]{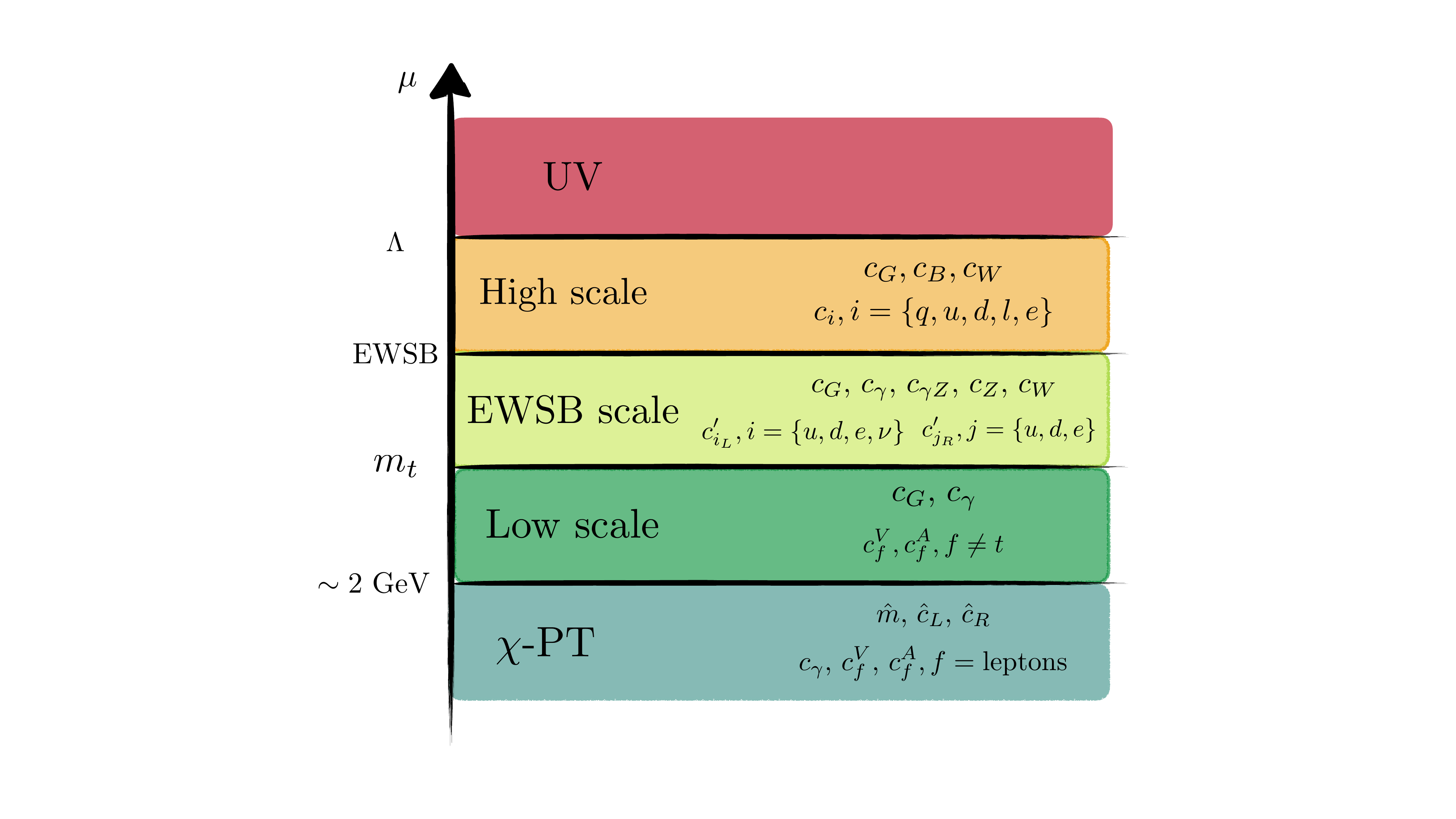}
\caption{\em Summary diagram of all scales and couplings implemented in \alpaca\!\!: A selection of UV models is presented in Section~\ref{sec:models}. The couplings at the High scale are defined in Section~\ref{sec:lagrangian}, at the EWSB scale in Section~\ref{sec:UVtoEW}, at the Low scale in Section~\ref{sec:belowEW}, and finally in $\chi$PT in Section~\ref{ALPchiPT} and App.~\ref{appendix: ALP_chiPT_details}.}
    \label{fig:scale-diagram}
\end{figure}

The renormalisation group (RG) evolution of the ALP couplings from the UV scale down to a generic scale $\mu$ is illustrated in Fig.~\ref{fig:scale-diagram}. The figure delineates the validity range of each effective regime and highlights the characteristic Wilson coefficients associated with the corresponding ALP Lagrangians. The running is obtained by solving the corresponding differential equations
\begin{equation}
    \frac{d}{d\log\mu} c_i(\mu) = \frac{1}{16\pi^2}\gamma_{ij}(\mu) c_j(\mu)\,.
\end{equation}
Here $c_i$ and $c_j$ are any pair of ALP couplings in Eq.~\eqref{eq:UVLag} (or generally on any other basis, after 
the corresponding change of variables) and $\gamma$ is the anomalous dimension matrix. The $\mu$-dependence of the 
anomalous dimension matrices enters through the SM Yukawa and gauge couplings running. In general, this system of 
differential equations can be efficiently solved by numerical integration. While the SM-like running of the SM 
parameters is often included in the numerical solution of RGE, the ALP contribution to the SM couplings running is 
typically neglected as suppressed by powers of $m_a/f_a$. The Renormalisation Group (RG) evolution from the UV to EW 
scale can significantly modify the ALP--fermion couplings, especially if $\Lambda \gg v_\mathrm{EW}$, while the ALP--gauge 
boson couplings $c_V$ are scale invariant at least up to the two-loop order. However, notice that their tilde 
siblings, $\tilde{c}_V$, may run. The RG equations for the ALP--SM field couplings in the ``derivative'' and 
``chirality-flipping'' basis have been derived in \cite{MartinCamalich:2020dfe,Bauer:2020jbp,Bonilla:2021ufe,DasBakshi:2023lca,Bresciani:2024shu} where 
all the details are reported. For the numerical evaluation of this paper, the ``derivative'' basis has been 
used and the corresponding RGE have been summarised in App.~\ref{app:Running}. 

As a final comment,  notice that a simplifying assumption, common to many proposed UV models, is that the Wilson 
coefficients $c_f$ are diagonal and generation universal, at the high energy scale $\Lambda$. However, as appears 
evident in the $\tilde{c}_f$ basis of Eq.~\eqref{eq:chiralbasis} and Eq.~\eqref{eq:yukbasis}, non-universal and 
non-diagonal ALP fermion couplings are inevitably generated through the running and have to be necessarily 
accounted for at low energy.

At the scale $\mu\approx v_\mathrm{EW}$ the EWSB mechanism takes place and appropriate bi-unitary transformations 
have to be introduced in order to rotate onto the physics mass basis. The Yukawa matrices, at the scale $\mu$, can be diagonalised through 
\begin{eqnarray}
\label{eq:Diagonalization}
\widehat{Y}_f(\mu) = L^\dagger_f Y_f(\mu) R_f \qquad \rightarrow \qquad \left\{ 
\begin{array}{lcr} f_L &\rightarrow & L_f f_L \\ f_R &\rightarrow& R_f f_R \\  \end{array}\right. \qquad 
\qquad \textrm{for } f=u,\,d,\,,e.\end{eqnarray}
with $L_{f},R_{f}$ the unitary matrices that diagonalise the corresponding fermion mass matrices.
In the physical basis, therefore, the ALP--fermion Wilson coefficients get accordingly rotated: 
\begin{align}
    \label{eq:rotatedcoefficients}
    \mathcal{L}_{\textrm{ALP}}(\mu) \supset & \frac{\partial_\mu a}{f_a} \Big(
    \bar{u}_L c'_{u_L}(\mu) \gamma^\mu u_L + 
    \bar{u}_R c'_{u_R}(\mu) \gamma^\mu u_R + 
    \bar{d}_L c'_{d_L}(\mu) \gamma^\mu d_L + \bar{d}_R c'_{d_R}(\mu) \gamma^\mu d_R \nn \\  
 & \hspace{0.5cm} + 
    \bar{\nu}_L c'_{\nu}(\mu)\gamma^\mu \nu_L + \bar{e}_L c'_{e_L}(\mu)\gamma^\mu e_L + \bar{e}_R c'_{e_R}(\mu) \gamma^\mu e_R  \Big) \,,
\end{align}
where the ALP--fermion couplings at the EW scale, in the fermion mass basis read:
\begin{equation}
\begin{aligned}
c'_{u_L}(\mu) & =  L_{u}^\dagger c_{q}(\mu) L_{u} \,,\qquad 
&c'_{u_R}(\mu) = R_{u}^\dagger c_{u}(\mu) R_{u}  \,, \\ 
c'_{d_L}(\mu) & =   L_{d}^\dagger c_{q}(\mu) L_{d}  \,, \qquad 
&c'_{d_R}(\mu) =  R_{d}^\dagger c_{d}(\mu) R_{d}  \,, \\
c'_{\nu_L}(\mu) & =  c_{\ell}(\mu) \,, \\
c'_{e_L}(\mu) & =   L_{e}^\dagger c_{\ell}(\mu) L_{e}  \,,\qquad  
&c'_{e_R}(\mu) =  R_{e}^\dagger c_{e}(\mu) R_{e} \,.
\end{aligned}
\label{PrimedCouplingsPhysicalBasis}
\end{equation}
The ALP--fermion Lagrangian at the EW scale, in the ``chirality flipping'' basis, reads instead:
\begin{eqnarray}
\mathcal{L}_{\textrm{ALP}}(\mu) & \supset & i \frac{a}{f_a} \left(\frac{v+h}{\sqrt{2}}\right)
\Big(\bar{u}_L \tilde{c}'_{u}(\mu) u_R + \bar{d}_L \tilde{c}'_{u}(\mu) d_R + \bar{e}_L \tilde{c}'_{e}(\mu) 
e_R \,+ \,\mathrm{h.c.} \Big) \,,
\end{eqnarray}
with
\begin{equation}
\begin{split}
\tilde{c}'_u (\mu)&=\widehat{Y}_\mathrm{U}(\mu)\Big(R^\dagger_u c_\mathrm{U}(\mu) R_u\Big) - 
                    \Big(L^\dagger_u c_q(\mu) L_u\Big)\widehat{Y}_\mathrm{U}(\mu) \\ 
\tilde{c}'_d (\mu)&=\widehat{Y}_d(\mu)\Big(R^\dagger_d c_d(\mu) R_d\Big) -
                    \Big(L^\dagger_d c_q(\mu) L_d\Big)\widehat{Y}_d(\mu) \\
\tilde{c}'_e (\mu)&=\widehat{Y}_e(\mu)\Big(R^\dagger_e c_e(\mu) R_e\Big) - 
                    \Big(L^\dagger_e c_\ell(\mu) L_e\Big)\widehat{Y}_e(\mu) \,.
\end{split}
\label{eq:chiralbasisrun}
\end{equation}
The anomalous couplings to the physical bosons are given by
\begin{equation}
\begin{split}
    \mathcal{L}_\mathrm{ALP} \supset& \frac{a}{f_a} \left(\frac{\alpha_s}{4\pi}c_G G_{\mu\nu}^a \tilde{G}^{a,\mu\nu} 
    + \frac{\alpha_\mathrm{em}}{4\pi}c_\gamma F_{\mu\nu}\tilde{F}^{\mu\nu} \right. \\ 
    &+ \left. \frac{\alpha_\mathrm{em}}{2\pi s_w c_w}c_{\gamma Z} F_{\mu\nu}\tilde{Z}^{\mu\nu}+ 
    \frac{\alpha_\mathrm{em}}{4\pi s_w^2 c_w^2}c_Z Z_{\mu\nu}\tilde{Z}^{\mu\nu} 
    + \frac{\alpha_\mathrm{em}}{2\pi s_w^2}c_W W^+_{\mu\nu}\tilde{W}^{-\mu\nu}\right)\,,
\end{split}
\end{equation}
in terms of the Weinberg angle $\theta_w$ ($s_w = \sin \theta_w$, $c_w = \cos\theta_w$) and the strong and 
electro-magnetic coupling constants $\alpha_s=g^2_s/(4\pi)$ and $\alpha_\mathrm{em}= e^2/(4\pi)$ with $e=g s_w=g' c_w$, 
and with the relations
\begin{equation}
\label{eq:cVphysical}
    c_\gamma = c_W + c_B\,,\qquad\qquad c_{\gamma Z} = c_w^2 c_W-s_w^2 c_B\,,\qquad\qquad c_Z = c_w^4 c_W + s_w^4 c_B\,.
\end{equation}
In the broken EW phase it is sometimes customary to define the dimensionful couplings with the physical bosons as
\begin{equation}
    \mathcal{L}_{\textrm{ALP}} \supset a \!\left(\frac{g_{agg}}{4}G_{\mu\nu}^a\tilde{G}^{a,\mu\nu}\!\!+\! 
    \frac{g_{a\gamma\gamma}}{4} F_{\mu\nu}\tilde{F}^{\mu\nu}\!\!+\!\frac{g_{a\gamma Z}}{4} Z_{\mu\nu}\tilde{F}^{\mu\nu}\!\!+\!
    \frac{g_{aZZ}}{4} Z_{\mu\nu}\tilde{Z}^{\mu\nu}\!\!+\!\frac{g_{aWW}}{2}W^+_{\mu\nu} \tilde{W}^{-\mu\nu}\right) \, ,
    \label{eq:LagafterEW}
\end{equation}
with the corresponding coefficients defined as
\begin{align}
    g_{agg} = &\frac{\alpha_s}{\pi f_a} c_{G}\, , \qquad 
    g_{a\gamma\gamma} = \frac{\alpha_{\textrm{em}}}{\pi f_a} c_\gamma\, , \qquad 
    g_{\gamma Z} = 2\frac{\alpha_{\textrm{em}}}{\pi s_w c_w f_a}c_{\gamma Z} \,,\nn \\  
 &  g_{aZZ} = \frac{\alpha_{\textrm{em}}}{\pi s_w^2 c_w^2 f_a} c_Z\, , \qquad 
    g_{aWW} = \frac{\alpha_{\textrm{em}}}{\pi s_w^2 f_a} c_{W}\, . 
\end{align}
When the ``chirality-flipping" basis is used the corresponding $\tilde{c}_V$ siblings have to be used in 
Eq.~\eqref{eq:cVphysical}.

\subsection{The ALP Lagrangian: below the EW scale}\label{sec:belowEW}

Tight constraints on light ALPs, $m_a\lesssim 10\,$GeV, can be obtained from studying Low Energy (LE) observables, 
i.e. at $\sqrt{s} \ll v_\mathrm{EW}$. At these energies the heavy particles, i.e. weak gauge bosons, the Higgs scalar and 
the top quark, can be integrated out at the EW scale and the high energy (HE) Lagrangian of Eq.~\eqref{eq:UVLag} 
matched to the LE one where only the light d.o.f. are propagating:
\be
\label{EWALP}
\begin{split}
\mathcal{L}^{\textrm{LE}}_{\textrm{ALP}} (\mu \le v_\mathrm{EW})  =&   (\partial_\mu a)^2 - \frac{1}{2}m_a^2 a^2 
+\frac{\alpha_\mathrm{em}}{4\pi}\frac{c_\gamma}{f_a} a F\widetilde{F}
+\frac{\alpha_s}{4\pi}\frac{c_G}{f_a} a G\widetilde{G}+ \\ 
&+ \frac{\partial_\mu a}{2f_a}\sum_{f,f'}\bar{f}\gamma_\mu\left(c^{V}_{ff'}+c^{A}_{ff'}\gamma_5\right) f'\,,
\end{split}
\ee
where $c_f^{V,A}$ are $3\times3$ hermitian matrices in flavour space for charged leptons and down-type quarks, 
but $2\times2$ for the up-type ones, having integrated out the top quark.

The LE Wilson coefficients in Eq.~\eqref{EWALP} differ from the analogous HE ones by the inclusion of the 
matching contribution, which arise at the considered loop level when the heavy d.o.f. are integrated out:
\begin{equation}
c^{\textrm{LE}}_i (\mu \le v_\mathrm{EW}) = c^{\textrm{HE}}_i(\mu) + \Delta c_i (v_\mathrm{EW}) 
\,, \qquad (i = \mbox{light~ d.o.f.}).
\end{equation}
The one-loop matching coefficients $\Delta c_i (v_\mathrm{EW})$ and the LE Wilson coefficients RGE have been derived in 
Ref.~\cite{Bauer:2021mvw} and collected for completeness in Apps.~\ref{app:matching} and \ref{app:runningbelow}. 
In particular, notice that additional off-diagonal contributions arise from integrating out the $W^\pm$ boson and the top quark. In 
the LE Lagrangian, moreover, it is phenomenologically advantageous to parameterise the vector and axial combinations 
of the fermionic couplings, $ c^{V/A}_f= c^{L}_{f} \pm  c^{R}_{f}$, rather than the left-handed and right-handed 
combinations.

\subsection{The ALP Lagrangian in the non-perturbative region}
\label{ALPchiPT}

At the scale $\mu \approx \Lambda_\mathrm{QCD}$ the perturbative QCD approach breaks down and chiral perturbation theory ($\chi$PT) has to be introduced to derive ALP interactions with light hadrons. The derivation of the ALP interactions via the $\chi$PT description has been discussed by several authors in the recent years, see Refs.~\cite{Aloni:2018vki,Bauer:2021wjo,Bai:2024lpq,Ovchynnikov:2025gpx,Bai:2025fvl,Balkin:2025enj}. The starting point is to consider the hadronic part of the 
Lagrangian of Eq.~\eqref{EWALP} at the scale $\mu \approx \Lambda_\mathrm{QCD}$, with only the three lightest quark flavours included\footnote{The proper procedure involves accounting for threshold effects associated with the $b$ and $c$ quarks. These effects are discussed in Ref.~\cite{Bauer:2020jbp}. The estimated impact of the integration of a quark is approximately $0.03\%$. In our analysis, we neglect these threshold contributions.}, i.e. $q=(u, d, s)^T$.

The first step for matching the fundamental Lagrangian with the effective Chiral one is to remove the anomalous ALP coupling 
to gluons via the chiral rotation  
\begin{equation}
    q \to \exp\!\!\left(-i \frac{a}{f_a} c_G \boldsymbol{\kappa} \gamma_5\right)q\,,
    \label{eq:rotchirallag}
\end{equation}
where $\boldsymbol{\kappa}$ is an arbitrary matrix with unit trace. The resulting Lagrangian then reads:
\begin{align}
    \label{eq:lagr_prechiral_eff}
    \mathcal{L}^\mathrm{LE}_\mathrm{ALP} \supset & \, \bar{q}(i \slashed{D} - \boldsymbol{\hat{m}}) q + \frac{1}{2} (\partial_\mu a)^2 - 
    \frac{1}{2}m_a^2 a^2  +
    \frac{\alpha_\mathrm{em}}{4\pi f_a}(c_\gamma +\mathcal{C}_\gamma^\chi) a F_{\mu\nu} \tilde{F}^{\mu\nu} \nonumber \\
    & +\frac{\partial_\mu a}{f_a} \bar{q} (\boldsymbol{\hat{c}_R} \gamma^\mu P_R + \boldsymbol{\hat{c}_L} \gamma^\mu P_L)q\,,
\end{align}
where $\boldsymbol{m}=\mathrm{diag}(m_u, m_d, m_s)$ and the ALP-dependent transformations of the quark masses and couplings are given by
\begin{eqnarray}
\boldsymbol{\hat{m}} &=& \exp\!\!\left(-i \frac{a}{f_a} c_G \boldsymbol{\kappa}\right) \boldsymbol{m} 
    \exp\!\!\left(-i \frac{a}{f_a} c_G \boldsymbol{\kappa}\right)=\boldsymbol{m}-2i\frac{a}{f_a}c_G 
    \boldsymbol{\kappa}\boldsymbol{m}+\mathcal{O}(a/f_a)\,, \\
\boldsymbol{\hat{c}_R} &=& \exp\!\!\left(i \frac{a}{f_a} c_G \boldsymbol{\kappa} \right)(\boldsymbol{c_R} + 
   c_G \boldsymbol{\kappa}) \exp\!\!\left(-i \frac{a}{f_a} c_G \boldsymbol{\kappa} \right) = \boldsymbol{c_R} + 
   c_G \boldsymbol{\kappa} + \mathcal{O}(a/f_a)\,, \\
\boldsymbol{\hat{c}_L} &=& \exp\!\!\left(-i \frac{a}{f_a} c_G \boldsymbol{\kappa}\right)(\boldsymbol{c_L} - 
   c_G \boldsymbol{\kappa}) \exp\!\!\left(i \frac{a}{f_a} c_G \boldsymbol{\kappa} \right) = \boldsymbol{c_L} - 
   c_G \boldsymbol{\kappa} + \mathcal{O}(a/f_a)\,. 
\label{eq:effCRL}
\end{eqnarray}
The tree-level shift to the ALP--photon coupling is, instead, defined as 
\be
\mathcal{C}_\gamma^\chi = - c_G N_c \langle\boldsymbol{\kappa}\boldsymbol{Q}^2\rangle 
\label{eq:effCgamma}\,,
\ee
where $\boldsymbol{Q} = \mathrm{diag}(2/3, -1/3, -1/3)$ is the light quarks charge matrix, and we are using the notation 
$\langle \cdots\rangle = \frac{1}{2}\Tr(\cdots)$ to denote the trace over quark flavours. The coefficient in 
Eq.~\eqref{eq:effCgamma} has been calculated in Ref.~\cite{GrillidiCortona:2015jxo} up to the NLO corrections giving 
$\mathcal{C}_\gamma^\chi = -(1.92\pm0.04)\,c_G$, value that is going to be used in the numerical evaluations. 

At the QCD scale $\Lambda_\mathrm{QCD}$ the fundamental Lagrangian of Eq.~\eqref{eq:lagr_prechiral_eff} is mapped into the Effective 
$\chi$PT Lagrangian, written in terms of the light pseudoscalar fields $\boldsymbol{\Phi}$. Up to $\mathcal{O}(p^2)$ one has:
\begin{equation}
\begin{split}
    \mathcal{L}_{\mathrm{ALP}}^{\chi\mathrm{PT}} =& \frac{F_0^2}{2} \langle\partial_\mu \boldsymbol{U} (\partial^\mu \boldsymbol{U})^\dagger\rangle - i \frac{F_0^2}{2 f_a} (\partial_\mu a)\langle\boldsymbol{\hat{c}_A}(\boldsymbol{U}\partial^\mu \boldsymbol{U}^\dagger-\boldsymbol{U}^\dagger \partial^\mu \boldsymbol{U})\rangle  \\
    &+ F_0^2 B_0\langle\boldsymbol{m}(\boldsymbol{U}+  \boldsymbol{U}^\dagger)\rangle+2i\frac{F_0^2 B_0}{f_a}c_G  a\langle\boldsymbol{\kappa}\boldsymbol{m}(\boldsymbol{U}-\boldsymbol{U}^\dagger)\rangle -\frac{1}{2}m_{\eta_0}^2 \eta_0^2 \\
    &+ \frac{1}{2}\partial_\mu a\, \partial^\mu a - \frac{1}{2} m_a^2 a^2 + \frac{\alpha_\mathrm{em}}{4\pi f_a}(c_\gamma + \mathcal{C}_\gamma^\chi) a F_{\mu\nu} \tilde{F}^{\mu\nu}\,.
\end{split}
\label{ChiPTFinalyExpression}
\end{equation}
where 
\begin{equation}
   \boldsymbol{U} \equiv \exp(i\boldsymbol{\Phi}/F_0)
\end{equation}
is the non-linear chiral field, $F_0$ is the pion decay constant $F_0=93\,\mathrm{MeV}$, $B_0 \approx m_\pi^2/(m_u+m_d)$ comes from the 
chiral condensate and $m_{\eta_0}$ is the ``anomalous'' mass term 
that has to be introduced to account for the $\eta_0$ mass.

Going back to the expression in Eq.~\eqref{ChiPTFinalyExpression}, the second term in the first line induces a kinetic mixing between the ALP and the pseudoscalar mesons, while the terms in the second line describe a mass mixing. Therefore, the physical field must be obtained by appropriate redefinitions. The corresponding mixing pattern is detailed in App.~\ref{sec:alp_meson_mix}. Recently, Refs.~\cite{Bai:2024lpq,Bai:2025fvl} and \cite{Ovchynnikov:2025gpx} have proposed two different procedures to ensure that physical results are free from the spurious $\boldsymbol{\kappa}$ dependence, and have checked the agreement between them. For simplicity, we have adopted in \alpaca the procedure of Ref.~\cite{Ovchynnikov:2025gpx}, as outlined in App.~\ref{app:kappa_dependence}

In the phenomenological analysis, when relevant, also the ALP coupling to vector, scalar and tensor mesons are included. The procedure is summarised in Apps.~\ref{sec:chiral_VMD} and~\ref{sec:alp_scalar_tensor}. As a complete description of these terms is beyond the present work, we refer to \cite{Aloni:2018vki,Ovchynnikov:2025gpx} for a more detailed description.

\section{Benchmark Models}
\label{sec:models}
While the ALP-EFT enables an unbiased exploration of the whole allowed parameter space, the alternative approach of considering specific UV models offers several advantages: i) the scale $f_a$ is more precisely constrained for a given ALP model; ii) all the Wilson coefficients of the EFT description at a given expansion order are determined in terms of a significantly lower number of fundamental parameters, possibly reducing or eliminating unconstrained regions in the parameter space due to the presence of flat directions; iii) UV models typically highlight the experimental strategies to better constrain specific frameworks. 

In this section, we will introduce a set of complementary examples of ALP UV models with the aim of illustrating the different phenomenological impact, by choosing specific ALP--SM particle interactions. We will start, in Section~\ref{sec:universal}, by considering two complementary benchmark models, respectively dubbed as DFSZ-like and KSVZ-like models, with the common feature of introducing flavour-universal ALP--fermions couplings, at the lowest order. In this class of frameworks, non-universality arises through higher order effects, like running and/or matching conditions. Then, in Section~\ref{sec:nonuniversal}, we will introduce well-known examples of flavour non-universal ALP--fermions couplings frameworks, like the top-philic and flaxion scenarios. In this 
latter case non-universality arises at the lowest order in the Lagrangian parameters and then propagates to all the different observables. These few examples, in our opinion, can be useful to highlight possible strategies for disentangling the UV structure from available experimental data. 

Notice that, however, in analysing these models we do not provide a rationale for the origin of the ALP mass. 
Understanding the mechanism that explicitly, and softly, breaks the ALP shift symmetry is an open problem in 
ALP physics~\cite{Frigerio:2011in,deGiorgi:2023tvn,deGiorgi:2024str} and it is beyond the scope of this paper 
to account for it. Therefore, the ALP mass $m_a$ and the ALP characteristic scale $f_a$ will be considered 
free and uncorrelated parameters.

\subsection{Flavour-Universal Benchmark Models}
\label{sec:universal}

Two complementary flavour-universal classes of models can be envisaged, depending on whether in the fundamental UV Lagrangian 
the tree-level ALP couplings to SM fermions are present or not. One can construct a set of models with 
the desired properties by modifying or extending the original invisible axion models, namely the DFSZ
~\cite{Zhitnitsky:1980tq,Dine:1981rt} and KSVZ~\cite{Kim:1979if,Shifman:1979if} realisations. 

\subsubsection*{DFSZ-like ALP Models}

The DFSZ axion models extend the SM particle content by introducing a second Higgs doublet and a complex scalar 
SM-singlet field, $\phi$. The SM gauge symmetry content is enlarged to include the Peccei-Quinn (PQ) global Abelian 
group $\mathrm{U}(1)_\text{PQ}$, anomalous under QCD. This extension decouples the axion scale $f_a$ from the EW scale, 
a scenario that is not accessible to experimental investigation~\cite{Davier:1986ps}. To explore a broader range 
of DFSZ-like benchmark models, we further enlarge the spectrum by considering a third Higgs doublet and use the 
following EW-invariant Lagrangian: 
\begin{align}
    \label{eq:generic_DFSZ}
    -\mathcal{L}^{\textrm{DFSZ-like}} \supset &
    \, \bar{q}_L Y_u \, \tilde{H}_u u_R  + \, \bar{q}_L Y_d  \, H_d d_R + \, \bar{\ell}_L Y_e\, H_e e_R + \nonumber \\ 
    &+ \lambda_{H\phi} H^\dagger_i H_j \phi^m + 
    \lambda'_{H\phi} H_i^\dagger H_k \phi^n + \mathrm{h.c.} \,.
\end{align}
In Eq.~\eqref{eq:generic_DFSZ} we introduced a different Higgs field for each type of Yukawa interaction: $H_u$, $H_d$, 
and $H_e$. Additionally, we included a renormalizable coupling term $\lambda_{H\phi}$ ($\lambda'_{H\phi}$) between the 
PQ scalar $\phi$ and two different Higgs doublet fields, $H_i$ and $H_{j(k)}$, where $i,j,k=\{u,d,e\}$ ($i \neq j \neq k$) 
and $m \, (n) = (\pm 1,\pm 2)$, with the negative values corresponding to $(\phi^\dagger)^{|m|}$. Recall also that no $\nu_R$ component is assumed in the SM Lagrangian \textit{ansatz}.
Notice that in Eq.~\eqref{eq:generic_DFSZ} we only considered the case with at most two scalar $H-\phi$ interactions. 
This condition guarantees that only one ALP is generated when the original $\mathrm{U}(1)^4$ global symmetry is broken. The 
explicit choice of $i,j,k$ will depend on the concrete UV realisation, that is the specific charge assignments under 
the PQ-symmetry.

Adopting the exponential parameterisation,
\be
H_{i,j,k}\supset \dfrac{v_{{i,j,k}}}{\sqrt{2}}\,e^{i\X_{H_{{i,j,k}}} a_{{i,j,k}} /v_{{i,j,k}}} \begin{pmatrix}
    0 \\ 1
\end{pmatrix} \, ,\qquad\qquad
\phi\supset \dfrac{v_\phi}{\sqrt{2}}\,e^{i\X_{\phi} a_\phi /v_\phi} \, ,
\label{ExpNotVevs}
\ee
we define with $a_{i,j,k}$ and $\X_{H_{i,j,k}}$ the pseudoscalar components and the PQ charges of the corresponding Higgs 
doublets. Then, the axion emerges as the linear combination of all the pseudoscalar fields~\cite{DiLuzio:2020wdo}, 
\begin{equation}
    a = \frac{1}{f_a}\sum_{l} \X_{H_l} v_l a_l \, ,
    \quad \textrm{with} \quad f_a^2 =\sum_l \X_{H_l}^2 v_l^2 \, ,
\end{equation}
with the sum running over $l=\left\{\phi,\,u, \,d,\, e\right\}$. The scalars PQ charges can be determined by imposing 
the invariance of the $H^\dagger_i H_j \phi^m$ and $H^\dagger_i H_k \phi^n$ terms and the absence of the ALP--$Z$ boson  
kinetic mixing, see for example Ref.~\cite{Bjorkeroth:2019jtx}. By assuming for simplicity $\X_\phi\equiv 1$ one obtains:

\begin{equation}
    \X_{H_i} = \frac{m v^2_j+n v_k^2}{v^2_\textrm{EW}},\quad \X_{H_j}= \frac{n v_k^2-m(v_i^2+v_k^2)}{v^2_\textrm{EW}}, 
    \quad \X_{H_k}= \frac{m v^2_j-n (v_i^2+v_j^2)}{v^2_\textrm{EW}} \, .
    \label{eq:Higgs_charges}
\end{equation}
with $v^2_\textrm{EW}\equiv v_i^2+v_j^2+v_k^2$. These equations apply to the generic formulation of the Lagrangian in 
Eq.~\eqref{eq:generic_DFSZ}, regardless of the number of physical Higgs fields introduced and independent of which 
Higgs field is charged under the $\mathrm{U}(1)_\textrm{PQ}$ symmetry. 

Choosing the basis in which $c_q=0=c_\ell$, which is always possible by means of a $\mathrm{U}(1)_B$ and $\mathrm{U}(1)_L$ field 
redefinition, see Eqs.~\eqref{eq:fieldredefinitions1}--\eqref{eq:fieldredefinitions3}, we obtain the generic 
expression for the coupling at $\mu=\Lambda = f_a$ of the ALP Lagrangian in Eq.~\eqref{eq:UVLag}: 
\be 
\label{eq:ALPfermions_DFSZgeneric}
\begin{aligned}
    c_{u}=& -(\delta_{iu} \X_{H_i}+\delta_{ju} \X_{H_j} +\delta_{ku}\X_{H_k} )\,\unity_3\, ,\\ 
    c_{d}=&(\delta_{id} \X_{H_i}+\delta_{jd} \X_{H_j} +\delta_{kd}\X_{H_k} )\,\unity_3\, ,\\
    c_{e}=& (\delta_{ie} \X_{H_i}+\delta_{je} \X_{H_j} +\delta_{ke}\X_{H_k} )\,\unity_3\,, 
\end{aligned}
\ee
where $\delta_{if}$ is the Kronecker delta and allows us to interpolate between the different DFSZ models. Similarly, 
the gauge couplings can be written as 
\be
\label{eq:ALPbosons_DFSZgeneric}
\begin{aligned}
    c_W &= 0 \, ,\\
    c_B &= c_\gamma = -3\times\left[\X_{H_i}\left(-\frac{4}{3}\delta_{ui}+\frac{1}{3}\delta_{di}+\delta_{ei}\right)+\X_{H_j}\left(-\frac{4}{3}\delta_{uj}+\frac{1}{3}\delta_{dj}+\delta_{ej}\right)+ \right. \\ &\left. \hspace{5cm}+\X_{H_k}\left(-\frac{4}{3}\delta_{uk}+\frac{1}{3}\delta_{dk}+\delta_{ek}\right)\right] \, ,\\
    c_G &=-\frac{3}{2}\times \left[\X_{H_i}\left(-\delta_{iu}+\delta_{id}\right)+\X_{H_j}\left(-\delta_{ju}+\delta_{jd}\right) +\X_{H_k}\left(-\delta_{ku}+\delta_{kd}\right)\right] \, .
\end{aligned}
\ee

The couplings for the two Higgs doublets DFSZ scenarios can be recovered, for example, by sending first one of the vevs to zero and then $n\to m $, $k\to j$ 
in Eqs.~\eqref{eq:Higgs_charges}. In particular, the usual DFSZ model is obtained by setting $H_i = H_u$, $H_j = H_d = H_e$ 
and $m=n=2$, yielding to the following coefficients:
\begin{equation}
   H_u^\dagger H_d\phi^2\, : \quad  
   c_u = -2s_\beta^2\,\unity_3 \, , \quad 
   c_d = c_e = -2c_\beta^2\,\unity_3 \,, \quad
   c_\gamma = -8 \, , \quad 
   c_G = -3 \, .
\end{equation}

Since our goal is to explore a broader range of phenomenological possibilities where the scale $f_a$ can be experimentally 
testable, we propose three UV alternatives with specific characterising features. More specifically, we define a two 
Higgs doublet DFSZ model with no anomalous coupling to gluons but with anomalous couplings under QED, in the following 
referred to as the QED-DFSZ~\cite{DiLuzio:2024jip}. This setup is achieved by choosing $H_i = H_u = H_d$, $H_j = H_e$ and $n=m=-2$, 
resulting in the following couplings:
\begin{equation}\label{eq:model_QEDDFSZ}
    H_q^\dagger H_e (\phi^\ast)^2: \quad 
    c_{u} = -c_{d} = 2 s_\beta^2 \,\unity_3\, , \quad 
    c_{e} = 2c_\beta^2 \,\unity_3\, , \quad 
    c_{\gamma} = -6 \, , \quad
    c_G = 0 \, .
\end{equation}
In this scenario, where there is no ALP--gluon interaction, all quarks couple to the ALP in the same manner but 
differently from leptons. 

Other interesting benchmark models, where the ALP couplings to a specific fermion specie can be set to zero, 
can be derived in the more general context of three Higgs doublets DFSZ scenario. For this to happen, we need to 
impose a vanishing PQ charge to one of the Higgses. There are several ways of achieving this condition, for example 
by choosing $n\neq m$ and $v_i^2=nv_j^2/(m-n) $ if $(m-n)/n>0$ or $v_j^2=-n \,v_k^2/m$ if instead $n/m<0$. 
For example, one can set:
\begin{align}
\label{eq:3HDM_benchmark1}
    & m=2,\quad n=1,\quad v_j=v_i\implies \X_{H_i}=1,\quad \X_{H_j} =-1,\quad \X_{H_k} = 0 \, , \\
\label{eq:3HDM_benchmark2}   
    & m=-2,\quad n=1,\quad v_j=v_k/\sqrt{2}\implies \X_{H_i}=0,\quad \X_{H_j} =2,\quad \X_{H_k} = -1 \, ,
\end{align}
Notice, however, that none of Higgs doublets should have a vanishing vev to avoid a massless fermion species. One can also 
get a large suppression on the PQ charge of one of the doublets (e.g. $H_i$) by taking the other two vevs very small (e.g.  $v_j,v_k 
\ll v_i$). However, bounds on the perturbativity of the Yukawa couplings can set lower limits on the different vevs~\cite{Bjorkeroth:2019jtx}. 
Using the conditions described in Eqs.~\eqref{eq:3HDM_benchmark1}-\eqref{eq:3HDM_benchmark2} one can therefore decouple 
the up-type quarks (and leptons) from the down-type ones. This can be achieved by setting $i = u$, $j = d$ and $k=e$, 
leading to the following couplings:

\begin{eqnarray}
& & H_u^\dagger H_d \phi^2,\, H^\dagger_u H_e \phi:\quad 
    c_{u} = -\unity_3 \, , \, 
    c_{d} = -\unity_3 \, , \, c_{e} = 0 \, , \, 
    c_{\gamma} = 5 \, , \quad 
    c_{G} = 3 \,, \label{eq:uDFSZ} \\ 
& & H_u^\dagger H_d (\phi^*)^2,\, H^\dagger_u H_e \phi :\quad  
    c_{u} = 0 \, , \, 
    c_{d} = 2\cdot\unity_3 \, , \,
    c_{e} = -\unity_3 \, , \, 
    c_{\gamma} =  1 \, , \, 
    c_{G} = -3 \, . \qquad
\label{eq:dDFSZ}
\end{eqnarray}
The benchmark models of Eqs.~\eqref{eq:uDFSZ} and \eqref{eq:dDFSZ} will be dubbed in the following as $e$-DFSZ and $u$-DFSZ 
respectively.

At one loop, the effects of running and matching give rise to off-diagonal couplings of the ALP with fermions, 
inducing flavour violating effects. The dominant contributions are typically proportional to the ALP--top 
quark coupling, when present. While our numerical analysis is performed considering the full set of running 
equations (see App.~\ref{app:Running}), it is instructive to derive analytical expressions, retaining only the 
the top Yukawa coupling contributions. Within the leading log approximation, they read
\begin{align}
\label{eq:cmu_1loop}
c^A_{\ell\ell'}(\mu_\textrm{ew})&=\left(c_e(\Lambda) -c_u(\Lambda)\frac{3y_t^2\log (\mu_\mathrm{ew}/\Lambda)}{8\pi^2} \right) 
   \delta_{\ell\ell'} \,,\\
\label{eq:cVA_FV}
c_{ij}^{V,A}(\mu_\textrm{ew}) &= c_{d}(\Lambda)\,\delta_{ij}\, \pm  \,c_u(\Lambda)\frac{y_t^2}{16\pi^2}V_{ti}^*V_{tj}\left[\frac{1}{2}\log
\frac{\mu^2_\mathrm{ew}}{m_t^2}-\frac{1}{4}-\frac{3}{2}\frac{1-x_t+\log x_t}{(1-x_t)^2}\right]\,,
\end{align}

for $i,j=d,s,b$, while off-diagonal ALP--up quark couplings are neglected being suppressed by a $y^2_b/y^2_t$ factor, compared to the down-quark sector. 
Note that off-diagonal couplings do not run below the electroweak scale 
\cite{Bauer:2020jbp,Bauer:2021mvw}.

\subsubsection*{KSVZ-like ALPs}
In the previous class of models, ALP couplings to SM fermions are induced at tree level due to the mixing between the 
singlet and Higgs doublet scalars. In contrast, the traditional KSVZ axion model~\cite{Kim:1979if,Shifman:1979if} 
introduces heavy exotic fermions that couple directly to the scalar field. Restricting the following analysis to 
the case of Vector-Like Fermions (VLFs), that is fermions transforming vectorially under the gauged SM symmetries, 
but chirally under the PQ one, the generic Lagrangian including the PQ scalar singlet can be written as
\begin{equation}
    -\mathcal{L}^\textrm{KSVZ} = \sum_\psi y_\psi \bar\psi_L \psi_R \phi + \text{h.c.}\,.
\end{equation}
Once the PQ symmetry is broken, 
\be
\phi\supset \dfrac{f_a}{\sqrt2}e^{ia/f_a}\,,
\label{ALPKSVZ}
\ee
anomalous ALP couplings with the SM gauge bosons are generated. Dubbing the transformation properties of the generic VLF under  $\mathrm{SU}(3)_c \times \mathrm{SU}(2)_L \times \mathrm{U}(1)_Y \times \mathrm{U}(1)_\mathrm{PQ}$ as $(\mathcal{C}_\psi, \mathcal{I}_\psi, \mathcal{Y}_\psi, \mathcal{X}_\psi)$, the resulting anomalous coefficients are:
\be
\label{eq:match_KSVZ}
\begin{aligned}
    c_G &= -\sum_\psi \mathcal{X}_\psi\,d(\mathcal{I}_\psi)\,T(\mathcal{C}_\psi)\,,\\
    c_W &= -\sum_\psi \mathcal{X}_\psi\,d(\mathcal{C}_\psi)\,T(\mathcal{I}_\psi)\,,\\
    c_B &= -\sum_\psi \mathcal{X}_\psi\,d(\mathcal{C}_\psi)\,d(\mathcal{I}_\psi)\,\mathcal{Y}_\psi^2\,,
\end{aligned}
\ee
where $d$ is the dimension of the representation and $T$ its Dynkin index. In particular, for the case of QCD 
representations the Dynkin index for the singlet, triplet and octer is respectively $T(\boldsymbol{1}) = 0$,  
$T(\boldsymbol{3}) = 1/2$ and $T(\boldsymbol{8}) = 3$.

In specific KSVZ scenarios, tree-level interactions of the VLFs with the SM fields may arise depending on the VLF 
transformation properties introduced (see, for instance, Ref.~\cite{Alonso-Alvarez:2023wig}). These interactions 
would necessarily lead to tree-level ALP couplings to SM fermions. However, in the following, we will stick to the 
``standard" KSVZ framework in which SM fermions have no PQ charge, and therefore (universal) vanishing leading order 
coupling with the ALP, i.e. $c_f=0$ in Eq.~\eqref{eq:UVLag}.

In what follows we consider three different benchmark cases where only one VLF is introduced at a time: the 
transformation properties and the corresponding anomalous ALP--gauge boson couplings are given by 
\begin{align}
\textrm{Q-KSVZ}\quad&\rightarrow\quad  Q_{L,R}\sim\left(\mathbf{3},\, \mathbf{1}, \, 0, \, \pm\frac{1}{2}\right)\,,\quad         c_G=-\frac{1}{2}\, , 
\quad c_f=c_B=c_W=0 \, ,\\ 
\textrm{L-KSVZ}\quad&\rightarrow\quad L_{L,R}\sim\left(\mathbf{1},\, \mathbf{2}, \, 0, \, \pm\frac{1}{2}\right)\,,\quad 
c_W=-\frac{1}{2}\, , \quad
c_f=c_B=c_G=0  \, ,\\ 
\textrm{Y-KSVZ}\quad&\rightarrow\quad \mathcal{Y}_{L,R}\sim\left(\mathbf{1},\, \mathbf{1}, \, \frac{1}{2}, \, \pm\frac{1}{2}\right)\,, \quad c_B=-\frac{1}{8}\, , \quad 
c_f=c_G=c_W=0 \, .
\end{align}

In these scenarios, ALP couplings to SM fermions are generated only at the 2-loop order and therefore much more suppressed 
compared with the DFSZ setup discussed in the previous subsection. Depending on the specific KSVZ model considered, a 
different gauge boson will generate contributions to the ALP--lepton coupling, as well as the ALP flavour violating 
couplings in the down-sector. In the following, we specify three possible cases.

\begin{itemize}
    \item{\bf \boldmath Couplings to $W^\pm$ (L-KSVZ):}
    \begin{eqnarray}
        c_{\ell\ell'}^A (\mu_\textrm{ew}) &=&   \frac{9 \alpha_\textrm{em}^2 \log (\mu_\mathrm{ew}/\Lambda)}{32\pi^2 s_w^4}\, \delta_{\ell\ell'},\\
        c_{ij}^V(\mu_\textrm{ew})& = & -c_{ij}^A(\mu_\textrm{ew}) = -\frac{3 \alpha_{\mathrm{em}} y_t^2}{64\pi^3 s_w^2} V_{ti}^* V_{tj} \frac{1 - x_t + x_t \log x_t}{(1 - x_t)^2}\,. \hspace{4.3cm} \nn
    \end{eqnarray}
    \item{\bf \boldmath Couplings to $B$ (Y-KSVZ):}
    \begin{align}
        c_{\ell\ell'}^A(\mu_\textrm{ew}) &=  \frac{9 \alpha_\textrm{em}^2 \log (\mu_\mathrm{ew}/\Lambda)}{128\pi^2 c_w^4}\,\delta_{\ell\ell'},\\
        c_{ij}^V(\mu_\textrm{ew})\! &=\!-c_{ij}^A(\mu_\textrm{ew}) \!= \!\frac{17 \alpha_\textrm{em}^2 y_t^2 \log (\mu_\mathrm{ew}/\Lambda)}{6144\pi^4 c_w^4}  V_{ti}^* V_{tj} \!\left[\frac{1}{2} \log \frac{\mu_\mathrm{ew}^2}{m_t^2} - \frac{1}{4} - \frac{3}{2} \frac{1 - x_t + \log x_t}{(1 - x_t)^2}\right]\,.\nn
    \end{align}
    \item \textbf{Couplings to gluons (Q-KSVZ):}
    \begin{align}
    c_{\ell\ell'}^A(\mu_\textrm{ew}) &= - \frac{3 \alpha_s^2 y_t^2\log^2 (\mu_\mathrm{ew}/\Lambda)}{8\pi^4}\,\delta_{\ell\ell'},\\
    c_{ij}^V(\mu_\textrm{ew}) &= -c_{ij}^A(\mu_\textrm{ew}) = \frac{\alpha_s^2 y_t^2 \log (\mu_\mathrm{ew}/\Lambda)}{16\pi^4}  V_{ti}^* V_{tj} \left[\frac{1}{2} \log \frac{\mu_\mathrm{ew}^2}{m_t^2} - \frac{1}{4} - \frac{3}{2} \frac{1 - x_t + \log x_t}{(1 - x_t)^2}\right]\,.\nonumber
    \end{align}
\end{itemize}
In the previous equations the indexes $i,j$ run over all the down type quarks, i.e. $i,j=d,s,b$, while again ALP--up quarks couplings are not shown as extra suppressed by a $y^2_b/y^2_t$ factor.

A comment is in order for the ALP--lepton couplings for the Q-KSVZ scenario. The ALP coupling to gluons does not generate 
one-loop contributions to $c_{\ell\ell}$. However, it induces a coupling to top quarks, which subsequently generates 
a coupling to leptons at next order. By approximating the solution for $c_{tt}(\mu)\equiv(c_{u})_{33}(\mu)$ using the 
leading-log and substituting it into the RGE for $c_{\ell\ell}$, we obtain the result shown above.

\subsection{Flavour Non-Universal Frameworks}
\label{sec:nonuniversal}

In all the models described in the previous subsection flavour violating couplings of the ALP to SM fermions are 
only induced at NLO. On the other hand, constructions with LO flavour-violating couplings already at tree level 
have been also widely discussed in the literature. Therefore in this subsection, we discuss for completeness 
also three non-universal tree-level ALP--fermion couplings frameworks.

\subsubsection*{Top-philic ALP}

A common setup, particularly in the context of collider phenomenology, is the so-called top-philic ALP 
scenario\footnote{Not to be confused with other studies that focus on top-physics with an ALP \cite{Carmona:2022jid,
Bruggisser:2023npd,Phan:2023dqw,Anuar:2024qsz,Inan:2025bdw}.} \cite{Esser:2023fdo,Rygaard:2023dlx,Blasi:2023hvb}. 
In this framework, all ALP couplings except those to the top quark are set to zero at high scales. To impose this 
condition in the UV, one can introduce on a non-universal scenario where only the third quark generation couples 
to the ALP. Moreover, in order to avoid tree-level couplings to the $b$-quark, which is a highly constrained sector, 
the ALP couples exclusively to RH quarks through the $(c_u)_{33}$ coupling in Eq.~\eqref{eq:UVLag}. This may appear 
a straightforward setup at first sight. However, while on the one hand it cannot be consistently implemented as a 
well-defined UV model (see the discussion in App.~\ref{appendix:top-philic}), on the other hand the effect of 
fermion mixing inevitably switches on ALP couplings with all the other fermions at NLO, that have to be accounted 
for in a global phenomenological analysis.

Moving ahead, assuming to have been able to find a mechanism that (approximately) justifies the top-philic scenario, that is, that the only non-vanishing ALP--fermion coupling at the UV is $c_{tt}(\Lambda)\equiv(c_{u})_{33}(\Lambda)$ and that at the same time fermion masses can be correctly described. Then, ALP couplings to the other fermions, generated by 
running effects, have to be accounted for. To exemplify this, working with the leading log approximation and retaining only the contributions proportional to the gauge couplings and the top Yukawa coupling one obtains (adapting the results of Refs.~\cite{Bauer:2020jbp,Bauer:2021mvw}):

\begin{align}
    (c_q)_{ij}(\mu_\mathrm{ew}) &= c_{tt}(\Lambda)\frac{\log (\mu_\mathrm{ew}/\Lambda)}{16\pi^2}\left[-\left(\frac{4}{9}\frac{\alpha^2_\textrm{em}}{c_w^4}+8\alpha_s^2\right)\delta_{ij}+y_t^2\delta_{i3}\delta_{j3}\right]\,, \nonumber\\
    (c_u)_{ij}(\mu_\mathrm{ew}) &= c_{tt}(\Lambda)\delta_{i3}\delta_{j3} + c_{tt}(\Lambda)\frac{\log (\mu_\mathrm{ew}/\Lambda)}{16\pi^2}\left[\left(\frac{64}{9}\frac{\alpha^2_\textrm{em}}{c_w^4}+8\alpha_s^2-6 y_t^2\right)\delta_{ij}+2y_t^2\delta_{i3}\delta_{j3}\right]\,,\nn\\
    (c_d)_{ij}(\mu_\mathrm{ew}) &= c_{tt}(\Lambda)\frac{\log (\mu_\mathrm{ew}/\Lambda)}{16\pi^2}\left[\frac{16}{9}\frac{\alpha^2_\textrm{em}}{c_w^4} + 8\alpha_s^2- 6   y_t^2\right]\delta_{ij}\,, \\
    (c_\ell)_{ij}(\mu_\mathrm{ew}) &= c_{tt}(\Lambda)\frac{\log (\mu_\mathrm{ew}/\Lambda)}{16\pi^2}\left[-4\frac{\alpha^2_\textrm{em}}{c_w^4}\right]\delta_{ij}\,, \nonumber\\
    (c_e)_{ij}(\mu_\mathrm{ew}) &= c_{tt}(\Lambda)\frac{\log (\mu_\mathrm{ew}/\Lambda)}{16\pi^2}\left[16\frac{\alpha^2_\textrm{em}}{c_w^4}-6 y_t^2\right]\delta_{ij}\,.\nn
\end{align}
Note that all the couplings generated at one loop are universal, except those associated with the third-quark generation in $c_q$ and $c_u$. As a consequence, moving to the fermion mass basis, ALP flavour-violating couplings are induced. For example, if we assume the LH rotation matrix to be proportional to the CKM, $L_d\sim V_\textrm{CKM}$, the LH coupling 
$c'_{d_L}$ develops non-diagonal entries. According to Eq.~\eqref{PrimedCouplingsPhysicalBasis}, this coupling can indeed be written as $c'_{d_L} = V^\dagger_\mathrm{CKM} c_q V_\mathrm{CKM}$, that is
\begin{equation}
    (c'_{d_L})_{ij}(\mu_\mathrm{ew}) = 
    c_{tt}(\Lambda)\frac{\log (\mu_\mathrm{ew}/\Lambda)}{16\pi^2}\left[-\left(\frac{4}{9}\frac{\alpha^2_\textrm{em}}{c_w^4}+ 
    8\alpha_s^2+6 \beta_qy_t^2\right)\delta_{ij}+y_t^2 V_{ti}^*V_{tj}\right]\, .
\end{equation}
Additionally, the integration of the top quark produces a second non-diagonal contribution to $c'_{d_L}$ in the matching conditions,
\begin{equation}
    \Delta (c'_{d_L})_{ij} = 
    c_{tt}(\Lambda)\frac{y_t^2}{16\pi^2}V_{ti}^*V_{tj}\left[\frac{1}{2}\log\frac{\mu^2_\mathrm{ew}}{m_t^2}-\frac{1}{4}-
    \frac{3}{2}\frac{1-x_t+\log x_t}{(1-x_t)^2}\right]\,,\qquad i\neq j\,,
\end{equation}
where $x_t = m_t^2/m_W^2$. Below the EW scale, the flavour-violating couplings do not run at one loop, while the 
running of the flavour-conserving ones is negligible. 

All in all, the final low-energy expressions for the couplings relevant for our phenomenological analysis are given by
\be
\begin{aligned}
    c_{\ell\ell'}^A(\mu_\textrm{EW})  &\simeq c_{tt}(\Lambda)\frac{\log (\mu_\mathrm{ew}/\Lambda)}
    {16\pi^2}\left[20\frac{\alpha_\textrm{em}^2}{c_w^4}-6 y_t^2\right]\, \delta_{\ell\ell'}\,, \\
    c_{ij}^V (\mu_\textrm{EW}) &= -c_{ij}^A(\mu_\textrm{EW}) = c_{tt}(\Lambda)\frac{y_t^2}{16\pi^2}V_{ti}^*V_{tj}
    \left[\frac{1}{2}\log\frac{\Lambda^2}{m_t^2}-\frac{1}{4}-\frac{3}{2}\frac{1-x_t+\log x_t}{(1-x_t)^2}\right]\, .
\end{aligned}
\ee
One important conclusion is that both ALP--lepton couplings and flavour-violating couplings to down-type quarks are 
generated at $\mathcal{O}(y_t^2/(16\pi^2))$, with the quark ones also experiencing additional CKM suppressions. Both 
types of couplings should be consistently taken into consideration. Moreover, in contrast with what one may naively 
expect, in top-philic scenarios ALP couplings to quarks different to the top one can also be strongly constrained, 
despite of their loop and CKM suppression. We will explore these phenomenological aspects in more detail in the 
following sections.

\subsubsection*{ALPs from the Flavour Puzzle}

A common effective description to address the flavour puzzle is the Froggatt-Nielsen (FN) mechanism \cite{Froggatt:1978nt}. 
In this framework, the Yukawa interactions are forbidden at the renormalizable level by a global $\mathrm{U}(1)_F$ flavour symmetry. 
Fermion masses and mixings thus arise as effective operators obtained from the traditional SM Yukawa terms introducing 
insertions of an additional scalar field $\phi$, transforming under $\mathrm{U}(1)_F$. Once $\phi$ acquires a vev, spontaneously 
breaking the flavour symmetry, each entry of the Yukawa matrices is then suppressed by powers of $\epsilon \equiv 
\vev{\phi}/(\sqrt2\Lambda)$, being $\Lambda$ the cut-off scale. 

Specifically, the Lagrangian for the FN mechanism can be written as
\begin{equation}
    \mathcal{L}_\textrm{FN}=\bar{q}_{Li}\,y_{u,ij}\tilde{H}\,u_{Rj}\left(\frac{\phi}{\Lambda}\right)^{\X_{q_L}^i-\X_{u_R}^j}\!\!\!+
    \bar{q}_{Li} \, y_{d,ij} H \, d_{Rj} \left(\frac{\phi}{\Lambda}\right)^{\X_{q_L}^i - \X^i_{d_R}}\!\!\!+
    \bar{\ell}_{Li} \, y_{e,ij} H \, e_{Rj} \left(\frac{\phi}{\Lambda}\right)^{\X_{\ell^i_L} - \X^j_{e_R}} \,, \nn
\end{equation}
where $\X_{q_L}^i$, $\X_{u_R}^j$, $\X_{d_R}^j$, $\X_{\ell_L}^i$, and $\X_{e_R}^j$ are the flavour charges of the different 
fermion fields and it is assumed that $\X_{\phi}=1$. Here, $y_{u,ij}$, $y_{d,ij}$, and $y_{e,ij}$ are dimensionless 
$\mathcal{O}(1)$ couplings.
After spontaneous breaking of the $\mathrm{U}(1)_F$ symmetry, the effective Yukawa couplings are generated, taking the form
\begin{equation}
\label{YukawaFNModels}
    (Y_f)_{ij}= y_{f,ij} \, \epsilon^{\X_{f_L}^i - \X_{f_R}^j} \,,
\end{equation}
leading to the masses and mixings of the SM fermions. By choosing appropriate flavour charges for a given value of 
$\epsilon$, the CKM mixing matrix as well as the mass hierarchies among the fermions can be naturally explained. In full 
generality, one can parameterise the Yukawa matrices as products of two unitary matrices and a diagonal matrix: the 
diagonal matrix contains the individual fermion Yukawa couplings in the mass basis, and the unitary matrices correspond 
to the transformation to move from the mass and the interaction bases: 
\begin{align}
\label{YukawasFN}
    &Y_f = L_f \hat{Y}_f R^\dag_f \sim  \\  
    &\resizebox{\textwidth}{!}{$
    \begin{pmatrix}
    1 & \epsilon^{\X_{f_L^1} - \X_{f_L^2}} & \epsilon^{\X_{f_L^1} - \X_{f_L^3}} \\
    \epsilon^{\X_{f_L^1} - \X_{f_L^2}} & 1 & \epsilon^{\X_{f_L^2} - \X_{f_L^3}} \\
    \epsilon^{\X_{f_L^1} - \X_{f_L^3}} & \epsilon^{\X_{f_L^2} - \X_{f_L^3}} & 1
    \end{pmatrix}\!\!
    \begin{pmatrix}
    \epsilon^{\X_{f_L^1} - \X_{f_R^1}} & 0 & 0 \\
    0 & \epsilon^{\X_{f_L^2} - \X_{f_R^2}} & 0 \\
    0 & 0 & \epsilon^{\X_{f_L^3} - \X_{f_R^3}}
    \end{pmatrix}\!\!
    \begin{pmatrix}
    1 & \epsilon^{\X_{f_R^2} - \X_{f_R^1}} & \epsilon^{\X_{f_R^3} - \X_{f_R^1}} \\
    \epsilon^{\X_{f_R^2} - \X_{f_R^1}} & 1 & \epsilon^{\X_{f_R^3} - \X_{f_R^2}} \\
    \epsilon^{\X_{f_R^3} - \X_{f_R^1}} & \epsilon^{\X_{f_R^3} - \X_{f_R^2}} & 1
    \end{pmatrix}
    $} \, .\nonumber
\end{align}
The entries of these matrices are expressed in terms of combinations of the FN charges of the fermions, but we are 
omitting the free $\cO(1)$ parameters, typical of the FN context. This is indeed the mayor drawback of these 
constructions, as indeed the absence of any control on these free parameters translates into the lack of predictive 
power of the FN models. Despite this, it represents an improvement with respect to the SM scenario as, in general, 
the 't Hooft Naturalness principle is well respected. The freedom to choose the flavour charges and the parameter 
$\epsilon$ leads to a wide variety of models, which have been extensively studied, see for instance 
Refs.~\cite{Altarelli:2000fu,Altarelli:2002sg,Chankowski:2005qp,Buchmuller:2011tm,Altarelli:2012ia,
Bergstrom:2014owa,Smolkovic:2019jow,Cornella:2023zme,Ibe:2024cvi}. 

In this work, we focus on the Goldstone boson arising from the spontaneous breaking of the global symmetry, i.e. 
the angular mode of the scalar field $\phi$, see Eq.~\eqref{ALPKSVZ}. This particle has been investigated in 
various contexts, including axions (flaxions or axiflavons) and ALPs \cite{Ema:2016ops,Calibbi:2016hwq,
Arias-Aragon:2017eww,Bonnefoy:2019lsn,Greljo:2024evt}.
In general, the ALP of this type of models exhibits explicit flavour violating couplings at tree level. In 
particular, once the FN scalar field gets the vev according to the notation in Eq.~\eqref{ExpNotVevs}, the 
ALP--fermion couplings as defined in Eq.~\eqref{eq:Lag-chi-flipping} read,
\be
(\tilde{c}_f)_{ij}=(Y_f)_{ij}\left(\chi_{f_L}^i-\chi_{f_R}^j\right)\,,
\label{FlaxionALPCouplings}
\ee
where the Yukawa couplings are the ones in Eq.~\eqref{YukawaFNModels}.
The couplings in the physical basis can only be read after performing the unitary transformations in Eq.~\eqref{PrimedCouplingsPhysicalBasis}, and they lead to flavour violation observables at tree-level.

As a benchmark, for the following phenomenological analysis we choose the charges of the flaxion defined in Ref.~\cite{Ema:2016ops}, which are:
\begin{align}
    \X_{q_L} = 
    \begin{pmatrix}
        3 & 2 & 0 \
    \end{pmatrix},& \quad\X_{u_R} = 
    \begin{pmatrix}
        -5 & -1 & 0 \
    \end{pmatrix}, \quad
    \X_{d_R} = 
    \begin{pmatrix}
        -4 & -3 & -3
    \end{pmatrix}, \nonumber
\\
    \X_{\ell_L} &= 
    \begin{pmatrix}
        1 & 0 & 0 
    \end{pmatrix},\quad
    \X_{e_R}=\begin{pmatrix}-8 & -5 & -3
    \end{pmatrix}.
    \label{eq:flaxion_charges}
\end{align}
After adopting this choice and with the notation used in Eq.~\eqref{YukawasFN}, we can easily obtain the expressions 
for the Wilson coefficients in the mass basis in Eq.~\eqref{PrimedCouplingsPhysicalBasis} for the ALP--fermion couplings.

\subsubsection*{Non-universal DFSZ}

Non-universal DFSZ models have been used for different reasons: to avoid domain wall problems, as certain constructions 
allow for  $N_\textrm{DW}=2N=1$ \cite{Davidson:1984ik,Cox:2023squ}, or leading to astrophobic axion models, i.e.~models with suppressed couplings to nucleons and electrons \cite{DiLuzio:2017ogq,Bjorkeroth:2019jtx,DiLuzio:2022tyc}. A generic consequence of non-universal DFSZ models is flavour violation, arising both from the heavy scalar sector and the ALP \cite{DiLuzio:2023ndz}. However, unlike the flaxion scenario, the couplings in these models depend on the vev ratio $\tan\beta = v_2 / v_1$, providing additional freedom to accommodate experimental constraints and fit observables.

The construction of these models is done allowing the 2HDM Higgs to couple to different generations of the same family 
of fermion\footnote{In the 2HDM context, this breaks the typical $\mathbbm{Z}_2$ symmetry imposed to avoid tree-level FCNC, 
see Ref.~\cite{deGiorgi:2023wjh}. However, there exist models where these FCNC are not too dangerous \cite{Branco:1996bq}.}. 
This construction leads to non-universal charges, and the violation of some of the conditions in 
Eq.~\eqref{eq:LHFlavourCondition}-\eqref{eq:RHFlavourCondition2}. The benchmark model considered here is given by 
\begin{align} 
\label{eq:Lyuk2HDM}
\mathcal{L}_Y^{\rm PQ\text{-}2HDM} = 
&- \bar q_{L, i} Y_{ij}^u\left[  (\delta_{i1}+\delta_{i2}) \tilde{H}_1 +\delta_{i3} \tilde{H}_2\right] u_{R,j}
+ \bar q_{L, i} Y_{ij}^d \left[  \delta_{i3} H_1+ (\delta_{i1}+\delta_{i2})H_2 \right]d_{R,j} \nonumber \\
&+ \bar \ell_{L,i} Y_{ij}^e \left[ \delta_{i3} H_1 +  (\delta_{i1}+\delta_{i2}) H_2\right]e_{R,j} 
+H_2^\dagger H_1 \phi + \text{h.c.}  \, , 
\end{align}
which corresponds to the UV non-universal Wilson coefficients:
\begin{align}
\label{eq:couplingsM1}
c_{q}  = c_{\ell}= &\,\mathrm{diag} (0,0,1)  \, ,  \quad  c_{u} = s^2_\beta \unity_3   \, , \quad 
c_{d} = c_{e}= c^2_\beta\unity_3 \, , \nonumber \\
c_G &= -\frac{1}{2}  , \quad c_W=2 , \quad c_B = -\frac{10}{3} \, .
\end{align}

\subsection{Summary}

\begin{table}[t!]
    \centering
    \begin{tabular}{ c| c c c|c|c c|c}
         Universal & $c_u$ & $c_d$ & $c_e$ & $c_G$ & $c_W$ & $c_B$ & $c_\gamma$    \\
         \hline
         &&&&&&&\\[-4mm]
          QED-DFSZ 
         & $2s_\beta^2$ & $-2s_\beta^2$&$2c_\beta^2$& 0& 0& $-6$& $-6$  \\[1mm]
          $u$-DFSZ 
         & $0$ & $-2$ & $1$ &  $-3$ &0 & $-1$& $-1$ \\[1mm]
         \ $e$-DFSZ 
         & $-1$& $1$& $0$&  $-3$ &0 & $5$& $5$  \\[1mm]
         \hline
         &&&&&&&\\[-4mm]
         L-KSVZ 
         & 0& 0& 0& 0 & $-\frac12$ & 0& $-\frac12$  \\[1mm]
          Y-KSVZ 
         & 0 & 0 & 0& 0 & $ 0 $ & $-\frac18$& $-\frac18$ \\[1mm]
          Q-KSVZ 
         & 0& 0& 0& $-\frac12$ & 0& 0& 0  \\[1mm]
         \hline
    \end{tabular}\vspace{5mm}
\resizebox{\linewidth}{!}{
\begin{tabular}{@{\hspace{-1cm}}c@{\hspace{-.1cm}}|@{\hspace{-.2cm}}c@{\hspace{-.3cm}} c@{\hspace{-.3cm}} c@{\hspace{-.3cm}} c@{\hspace{-.3cm}} c@{\hspace{-.3cm}}} 
\multicolumn{1}{c|@{\hspace{-.2cm}}}{Non-universal} & $c_q$ & $c_u$ & $c_d$ & $c_\ell$ & $c_e$ \\[1mm]
\hline
&&&&&\\[-3mm]
 Top-philic 
& 0& $\begin{pmatrix} 0 & 0 & c_{tt} \end{pmatrix}$  & 0 & 0 & 0 \\
 Flaxion 
& $\begin{pmatrix} -3 & -2 & 0 \end{pmatrix}$ &
  $\begin{pmatrix} 5 & 1 & 0 \end{pmatrix}$ &
  $\begin{pmatrix} 4 & 3 & 3 \end{pmatrix}$ &
  $\begin{pmatrix} -1 & 0 & 0 \end{pmatrix}$ &
  $\begin{pmatrix} 8 & 5 & 3 \end{pmatrix}$ \\
 \multirow{2}{7em}{\centering{Non-Universal DFSZ}} 
& \multirow{2}{7em}{\centering{$\begin{pmatrix} 0 & 0 & 1 \end{pmatrix}$}} &
  \multirow{2}{7em}{\centering{$\begin{pmatrix} c_\beta^2 & c_\beta^2 & c_\beta^2 \end{pmatrix}$}} &
  \multirow{2}{7em}{\centering{$\begin{pmatrix} s_\beta^2 & s_\beta^2 & s_\beta^2 \end{pmatrix}$}} &
  \multirow{2}{7em}{\centering{$\begin{pmatrix} 0 & 0 & 1 \end{pmatrix}$}} &
  \multirow{2}{7em}{\centering{$\begin{pmatrix} c_\beta^2 & c_\beta^2 & c_\beta^2 \end{pmatrix}$}} \\[6mm]
  \hline
  \hline
&  $c_G$ & $c_W$ & $c_B$ & $c_\gamma$ & \\[1mm]
\hline
&&&&&\\[-3mm]
 Top-philic 
& 0 & 0 & 0 & 0 & \\[1mm]
 Flaxion 
& $-13$ & $-8$ & $-\frac{86}{3}$ & $-\frac{110}{3}$ & \\
 \multirow{2}{7em}{\centering{Non-Universal DFSZ}} 
& \multirow{2}{7em}{\centering{$-\frac{1}{2}$}} & \multirow{2}{7em}{\centering{$ 2$}} & \multirow{2}{7em}{\centering{$-\frac{10}{3}$}} & \multirow{2}{7em}{\centering{$-\frac{4}{3}$}} & \\[6mm]
\end{tabular}}
\caption{\em Matching of benchmark models to the ALP effective Lagrangian. Fermionic and anomalous couplings are grouped separately for clarity. Matching follows Eq.~\eqref{eq:UVLag} for all the models.}
\label{tab:ModelCouplings}
\end{table}

To conclude this benchmark models section we provide a summary table that facilitates the comparison among the different UV realisations considered. In particular, in Table~\ref{tab:ModelCouplings} we show the expressions of the  tree-level ALP couplings to SM fermions. The DFSZ- and KSVZ-type models, in the upper part of the table, present universal couplings, but the top-philic, flaxion models and the non-universal DFSZ construction, in the lower part, present a non trivial flavour structure. The matching is performed with the ALP effective Lagrangian in the derivative basis in Eq.~\eqref{eq:UVLag}. For all models,  it requires to move from the chirality-flipping basis to the derivative one. Moreover, the same table shows the corresponding anomalous couplings induced at one-loop level. For the DFSZ-descriptions, we take advantage of the freedom to redefine the PQ charges and we show the couplings in the basis where only the RH fields transform under PQ, which implies that the $c_W$ coupling vanishes. Notice that for the KSVZ-descriptions, these anomalous contributions arise from the ALP couplings to the exotic fermions. 

In Table~\ref{tab:FlavourCoefficientEstimates}, we introduce the na\"ive expectation, including loop-contributions, of the flavour-changing ALP couplings to down-type quarks, relevant to our phenomenological study. We have only shown here, for shortness, the dominant top quark contribution and the first term in the Cabibbo angle expansion. As the dependence on the running scale is very much similar among the different realisations, we omit it in the table to simplify the comparison among the contexts. 

\begin{table}[tb]
    \centering
    \begin{tabular}{c| c c c }
        Construction & $s\to d$ &  $b\to s$ &  $b\to d$ \\
        \hline
        &&&\\[-4mm]
        QED-DFSZ 
        & $c_u\frac{y_t^2}{16\pi^2} \lambda^5 $ & $c_u\frac{y_t^2}{16\pi^2} \lambda^2 $ 
        & $c_u\frac{y_t^2}{16\pi^2} \lambda^3 $  \\[1mm]
        $u$-DFSZ 
        & $0 $ 
        & $0 $ 
        & $0 $  \\[1mm]
        $e$-DFSZ 
        & $c_u\frac{y_t^2}{16\pi^2} \lambda^5 $ 
        & $c_u\frac{y_t^2}{16\pi^2} \lambda^2 $ 
        & $c_u\frac{y_t^2}{16\pi^2} \lambda^3 $  \\[1mm]
        \hline  
        &&&\\[-4mm]
        L-KSVZ 
        & $\frac{3\alpha_\text{em} y_t^2  }{64 \pi^3 s_w^2} \lambda^5$ 
        & $\frac{3 \alpha_\text{em} y_t^2}{64 \pi^3 s_w^2}  \lambda^2$ 
        & $\frac{3 \alpha_\text{em} y_t^2}{64 \pi^3 s_w^2} \lambda^3$ \\[1mm]
        Y-KSVZ 
        & $\frac{17 \alpha^2_\text{em} y_t^2 }{6144\pi^4 c_w^4}\lambda^5 $ 
        & $\frac{17 \alpha^2_\text{em} y_t^2 }{6144\pi^4 c_w^4}\lambda^2 $
        &  $\frac{17 \alpha^2_\text{em} y_t^2 }{6144\pi^4 c_w^4}\lambda^3 $\\[1mm]
        Q-KSVZ 
        & $\frac{\alpha_s^2 y_t^2 }{16\pi^4}\lambda^5$ 
        & $\frac{\alpha_s^2 y_t^2 }{16\pi^4}\lambda^2$ 
        & $\frac{\alpha_s^2 y_t^2 }{16\pi^4}\lambda^3$ \\[1mm]
        \hline
        &&&\\[-4mm]
        Top-philic 
        &$\frac{y_t^2 c_t}{16\pi^2} \lambda^5 $ 
        & $\frac{y_t^2 c_t}{16\pi^2} \lambda^2 $ 
        & $\frac{y_t^2 c_t}{16\pi^2} \lambda^3 $ \\[1mm]
        Flaxion 
        & $5\lambda$ 
        & $2\lambda^2$ 
        & $5\lambda^3$ \\
        Non-Universal DFSZ 
        & $\lambda^5$ 
        & $\lambda^2$ 
        & $\lambda^3$ \\
    \end{tabular}
    \caption{\em Estimation of flavour changing ALP couplings to down-type quarks at low-energy in the different models, in the fermion mass basis. The dependence on the running scale is not explicitly shown. Only the first term in the expansion in terms of the Cabibbo angle $\lambda$ is reported.}
    \label{tab:FlavourCoefficientEstimates}
\end{table}

In the next sections, we will show a detailed phenomenological analysis of both the generic effective Lagrangian and these specific constructions.

\section{ALP decay channels}
\label{sec:decays}
This section we study the different ALP decay channels. Although this topic has been presented in depth in previous 
works (see Refs.~\cite{Bauer:2021mvw}), recent efforts have focused on improving the formulation and modeling of 
hadronic decays (see Refs.~\cite{Aloni:2018vki,Bauer:2021wjo,Bai:2024lpq,Ovchynnikov:2025gpx,Bai:2025fvl,Balkin:2025enj}). These hadronic 
contributions also include ALP--meson mixing, which leads to a much richer phenomenology; further details can be 
found in App.~\ref{sec:alpmixing}. 

The ALP decay modes can be broadly categorised into three main classes: decays into photons, hadrons, and leptons.

\subsection{Decays into photons}

According to the discussion in Ref.~\cite{Aloni:2018vki}, it is possible to write the decay width as
\begin{equation}
\label{eq:decay-photon}
\Gamma \left(a\rightarrow \gamma\gamma\right)=\frac{\alpha_\text{em}^2 m_a^3}{(4\pi)^3f_a^2}\left|c_\gamma + 
\mathcal{C}_\gamma^\chi+\mathcal{C}_\gamma^\text{VMD}+ \mathcal{C}_\gamma^\ell+\mathcal{C}_\gamma^q+
\mathcal{C}_\gamma^W\right|^2\,,
\end{equation}
where $c_\gamma$ is the tree-level contribution (present in the Lagrangian of Eq.~\eqref{eq:LagafterEW}), 
$\mathcal{C}_\gamma^\chi$ stands for the contribution coming from the chiral rotation in Eq.~\eqref{eq:effCgamma}, 
$\mathcal{C}_\gamma^\text{VMD}$ for the contribution stemming from the Vector-Meson Dominance (VMD) in 
Eq.~\eqref{eq:VMDphotons} and  $\mathcal{C}_\gamma^\ell$, $\mathcal{C}_\gamma^q$ and $\mathcal{C}_\gamma^W$ 
is the contribution coming from the loops involving leptons, quarks in the perturbative region of QCD and 
$W^\pm$ bosons if they have not been integrated out, respectively. The one-loop expression for these last 
contributions is~\cite{Bauer:2017ris}:
\begin{align}
\label{eq:full-contribution}
    \mathcal{C}_{\gamma}^\ell &=  \frac{1}{2} \sum_{\ell} N_{\ell} c_{\ell}^A B_1\left(\frac{4m_\ell^2}{m_a^2}\right),\\
    \mathcal{C}_{\gamma}^q &=  \frac{1}{2} \sum_{q\neq t} N_{q} Q^2_q c_{q}^A B_1\left(\frac{4m_q^2}{m_a^2}\right),\\
    \mathcal{C}_{\gamma}^W&= 2\frac{\alpha_\text{em}}{\pi}\frac{c_W}{s_w}B_2\left(\frac{4m_W^2}{m_a^2}\right),
\end{align}
where $N_q=3$ $(N_{\ell}=1)$ is the number of fermion colours and  
\begin{equation}\label{eq:B1B2}
\begin{split}
    B_1\left(\tau \right) &= 1 - \tau f(\tau)^2\,, \qquad B_2\left(\tau \right) = 1-(\tau-1)f(\tau)^2\,,\\
    f(\tau) &= \begin{cases}
        \arcsin\left(\frac{1}{\sqrt{\tau}}\right)\qquad&\text{for}\quad \tau\geq1 \\
        \frac{\pi}{2} + \frac{i}{2} \log{\frac{1 + \sqrt{1 - \tau}}{1 - \sqrt{1 - \tau}}}\qquad\qquad&\text{for}\quad \tau<1\,.
    \end{cases}
\end{split}
\end{equation}

In our numerical analysis, the contributions from the chiral transformation and the VMD region are adapted from 
Ref.~\cite{Aloni:2018vki} (see App.~\ref{appendix: ALP_chiPT_details} for details). In the regions where QCD perturbativity 
applies, the expressions are extracted from Ref.~\cite{Bauer:2021mvw}.

\subsection{Decays to hadrons}
In discussing ALP decays to hadrons one has to distinguish two different mass regimes: in the low ALP mass region 
($m_a \lesssim 1$ GeV) chiral perturbation theory ($\chi$PT) can be safely employed, while for $m_a$ above 3 GeV, 
perturbative QCD (pQCD) gives instead sufficiently accurate results. The transition between non-perturbative and perturbative regimes is performed at the energy scale where both descriptions predict the same hadronic partial decay widths.

In the low mass region we follow the analysis of Ref.~\cite{Ovchynnikov:2025gpx}, where reparameterisations invariance 
under chiral rotations is accounted for. In the case of ALP decays into three pseudoscalar mesons ($a\to 3\pi$, $a\to \eta\pi\pi$ 
and $a\to\eta'\pi\pi$), there is a contact interaction obtained directly from the chiral Lagrangian, as well as resonant 
contributions mediated by scalar ($a_0$, $f_0$ and $\sigma$) and tensor ($f_2$) mesons. We also consider ALP decays into two 
vector mesons, described by VMD. In particular, we include $a\to\omega\omega$ and $a\to\gamma\pi\pi$, where one of the vector 
mesons mixes with a photon and the other decays into two pions. We note that in Refs.~\cite{Aloni:2018vki,Ovchynnikov:2025gpx}  additional ALP decay channels are included. We omitted them here since they do not contribute significantly to the total decay 
width, nor seem to be phenomenologically relevant. 

\begin{figure}[t!]
\centering
{\includegraphics[width=0.9\linewidth]{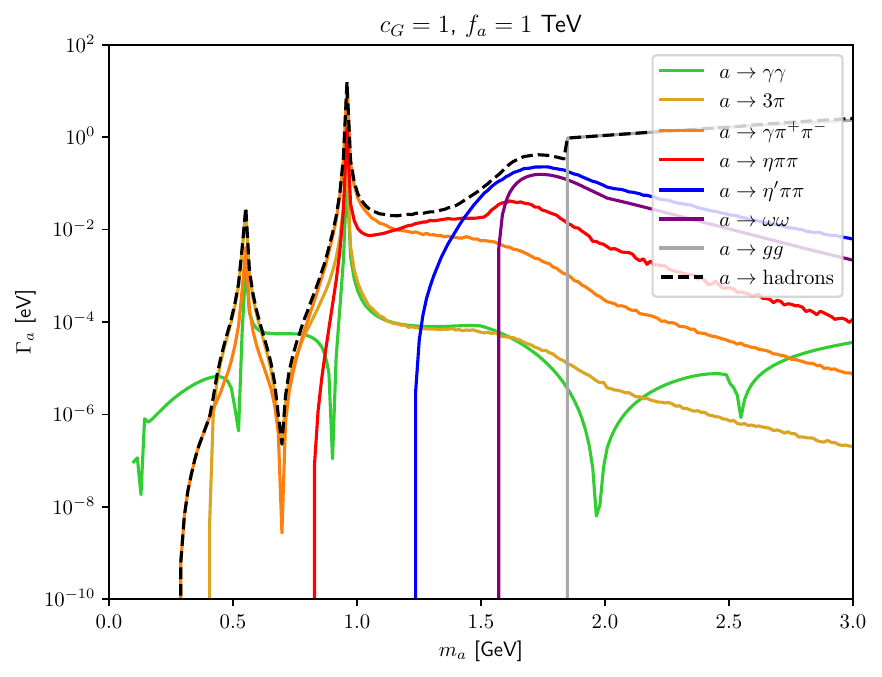}}
\, 
\caption[]{\em Decay width of different hadronic channels, as well as the decay $a\rightarrow\gamma\gamma$ for the case 
of all vanishing ALP tree-level couplings, but $c_G=1$. In grey, the pQCD $a\rightarrow gg$ channel is shown.}
\label{fig:hadron_width}
\end{figure}

In the pQCD regime, the inclusive decays into charmed and bottom-flavoured hadrons is computed via the quark-level widths $a\to c\bar{c}$ and $a\to b\bar{b}$ respectively (see next paragraph and Eq.~\eqref{eq:dw_fermions}). For the light-flavoured hadrons, we can rely on the quark--hadron duality~\cite{Poggio:1975af,Shifman:2000jv}, obtaining~\cite{Spira:1995rr,Bauer:2017ris}
\begin{equation}
    \Gamma(a\to\mathrm{light\ hadrons}) \approx \Gamma(a\to gg) = \frac{\alpha_s^2}{8\pi^3}\frac{m_a^3}{f_a^2}|c_G^\mathrm{eff}|^2 \left(1 +\frac{\alpha_s}{4\pi}\frac{291-14n_q}{12}\right)\,,
\end{equation}
where $n_q$ is the number of active quark flavours, and the effective coupling to gluons, including one-loop quark contributions is given by
\begin{equation}\label{eq:cG_eff}
    c_G^\mathrm{eff} = c_G + \frac{1}{2} \sum_{q\neq t} c_{q} B_1\left(\frac{4m_q^2}{m_a^2}\right)\,. 
\end{equation}

In the intermediate mass regime, that is 1 GeV $\lesssim m_a \lesssim $ 3 GeV, we follow the data-driven 
approach of Ref.~\cite{Aloni:2018vki,Balkin:2025enj}, where the validity range of the $\chi$PT expressions 
is extended by adding phenomenological hadronic form factors $\mathcal{F}(m_a)$, which are obtained by fitting 
to $e^+e^-\to \rho\pi, \omega\pi,K^*K,\phi\eta$ data.

In order to verify that our formulae are correct, we study our expressions in the limit in which all ALP couplings vanish 
except for the ALP coupling to gluons, $c_G=1$, thus reproducing the results of Fig.~4 of Ref.~\cite{Ovchynnikov:2025gpx}. 
A similar analysis is presented in Fig.~\ref{fig:hadron_width} for $f_a = 1\,\mathrm{TeV}$ fixing $\Lambda=4\pi f_a$ and  
including the $a \to \gamma\gamma$ decay. The peaks in the widths are located at $m_a\approx m_\pi$ (only visible in 
the $a\to \gamma \gamma$ channel), $m_\eta, m_{\eta^\prime}$. The grey line is the $a\to gg $ decay width, which for 
this specific setup, will clearly dominate  in the pQCD region.

\subsection{Decays to fermions}

The ALP decay to SM fermions in the perturbative regime has been derived in Refs.~\cite{Bauer:2020jbp,Bauer:2021mvw}. The decay rate of $a\rightarrow f\overline{f}$, when kinematically allowed, can be written as
 \be
 \Gamma\left(a\rightarrow f\ov{f}\right)=N_c^f\frac{m_a m_f^2}{8\pi f_a^2}|c^\mathrm{eff}_f|^2
 \sqrt{1-\frac{4 m_f^2}{m_a^2}}\,.
 \label{eq:dw_fermions}
 \ee
The effective coupling to fermions, $c_f^\text{eff}$, includes the tree-level contribution from the axial coupling $c_{f}^A$, as well as one-loop contributions generated by the couplings to gauge bosons, that is $c_\gamma$, $c_{\gamma Z}$, $c_Z$, $c_W$, and, in the case of  decays to quarks, also $c_G$~\cite{Bauer:2017ris}.

\subsection{Comparison of benchmark models}

\begin{figure}
     \centering
     \includegraphics[width=0.74\textwidth]{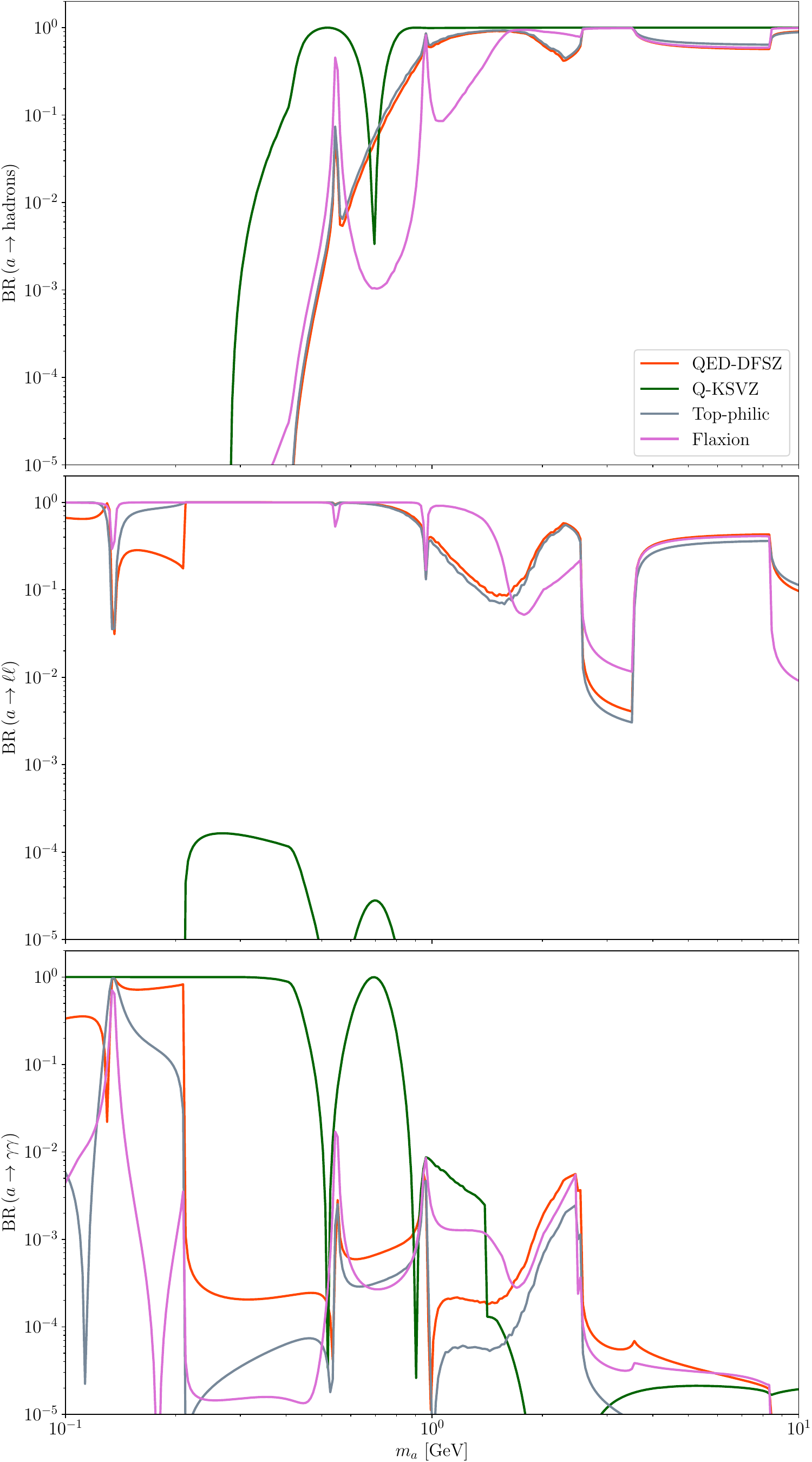}
\caption{\em Comparison of the contribution to the branching ratio to leptons (top), hadrons (middle) and photons (bottom) of four different benchmark models: QED-DFSZ with $\tan{\beta}=1$ (in red), Q-KSVZ (in green), top-philic with $c_{tt}=1$ (in grey) and flaxion (in pink). }
\label{fig:br-comparison}
\end{figure}

The models explained in Section~\ref{sec:models} exhibit distinct features that result in a different pattern of couplings, 
which clearly affects the channels in which the ALP can decay. Fig.~\ref{fig:br-comparison} illustrates the respective contributions to the branching ratio of the decay channels to leptons (top), hadrons (middle) and photons 
(bottom) for the QED-DFSZ, Q-KSVZ, top-philic and flaxion scenarios. For $m_a<1$ GeV, the decay to leptons is dominant 
for both the QED-DFSZ, top-philic and flaxion cases; whereas the predominant decay channel in the case of Q-KSVZ is to 
photons. For masses $m_a>1$ GeV, the hadronic decays will be dominant for all four scenarios, especially for the Q-KSVZ 
model. In all four cases, the photonic channel is subdominant with respect to the leptonic one in this mass regime. 
The peaks present at $m_a\sim 130, 550, 960$ MeV are a consequence of the mixing of the ALP with $\pi^0,\,\eta,\,
\eta^\prime$, respectively. 

We emphasise the relevance of hadronic channels in the intermediate mass region. These channels play a significant role 
not only in models like Q-KSVZ, but also in other scenarios where hadronic decay modes suppress the branching ratio into 
leptons. This suppression affects the resulting bounds: although decays into leptons —primarily muons— continue to provide 
the most stringent constraints in this region, the branching ratio can be reduced by up to an order of magnitude due to 
hadronic contributions.

\begin{figure}[t]
	\centering
{\includegraphics[width=\linewidth]{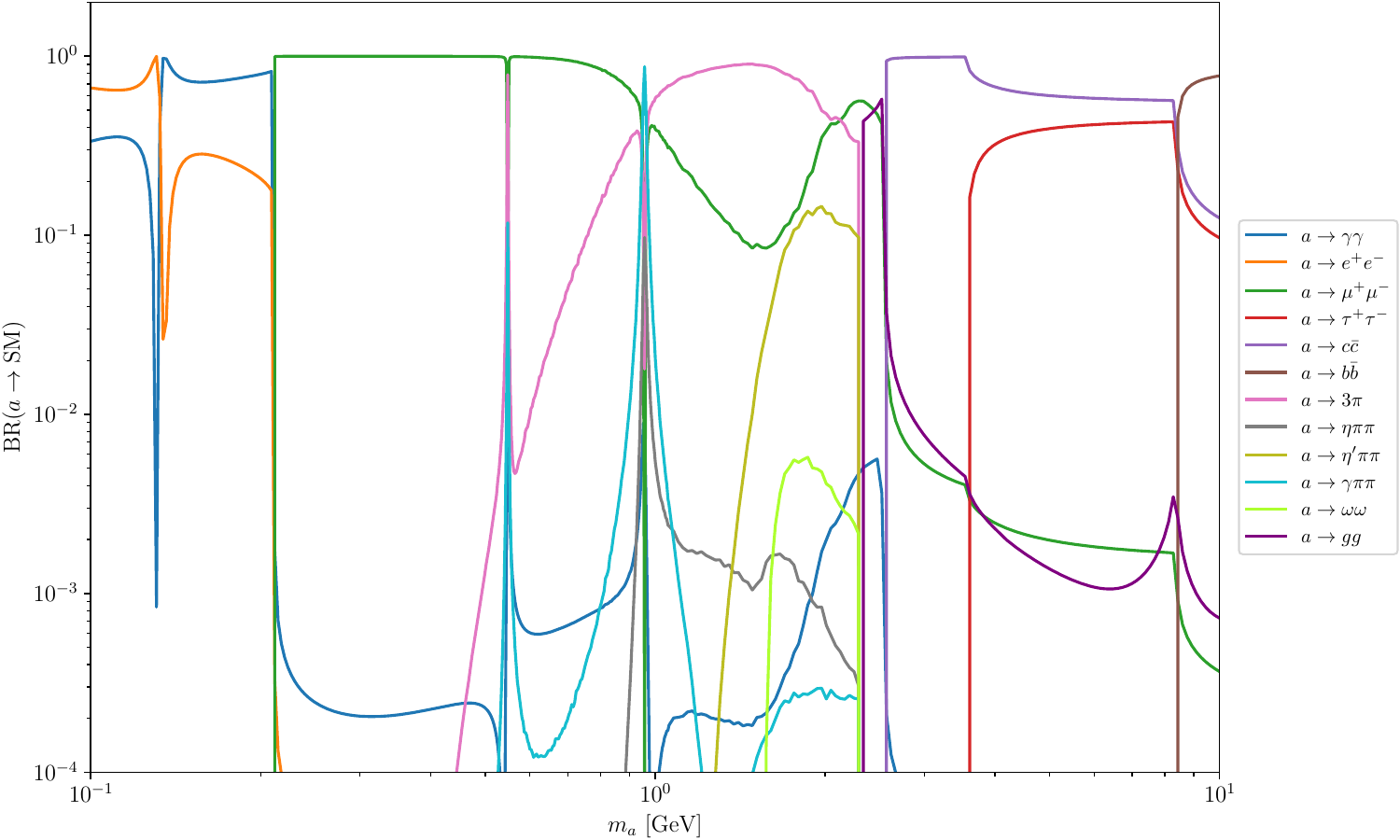}}
     \,
     \caption[]{\em Branching ratios of all studied channels for the QED-DFSZ model with $\tan{\beta}=1$.}
	\label{fig:br_QED-DFSZ}
\end{figure}

Once we know the general behaviour of each model, it is also enlightening to see the individual evolution of the branching 
ratios with respect to the mass. Fig.~\ref{fig:br_QED-DFSZ} shows the branching ratios for different channels in the case 
of the QED-DFSZ model. The decays to electrons and photons are dominant until $m_a=2m_\mu$. Once the muon channel is opened 
and until $m_a\sim 1-2\GeV$, where hadronic effects become relevant, this will represent the most preferred channel, except 
for masses $m_a\sim 1.5\GeV$, where the decay to $3\pi$ becomes relevant. For higher values of the ALP mass, where 
pQCD applies, the most relevant channels are the decays to quarks, first to charm and then to bottom, once this channel 
is kinematically allowed.

\begin{figure}[t]
	\centering
{\includegraphics[width=\linewidth]{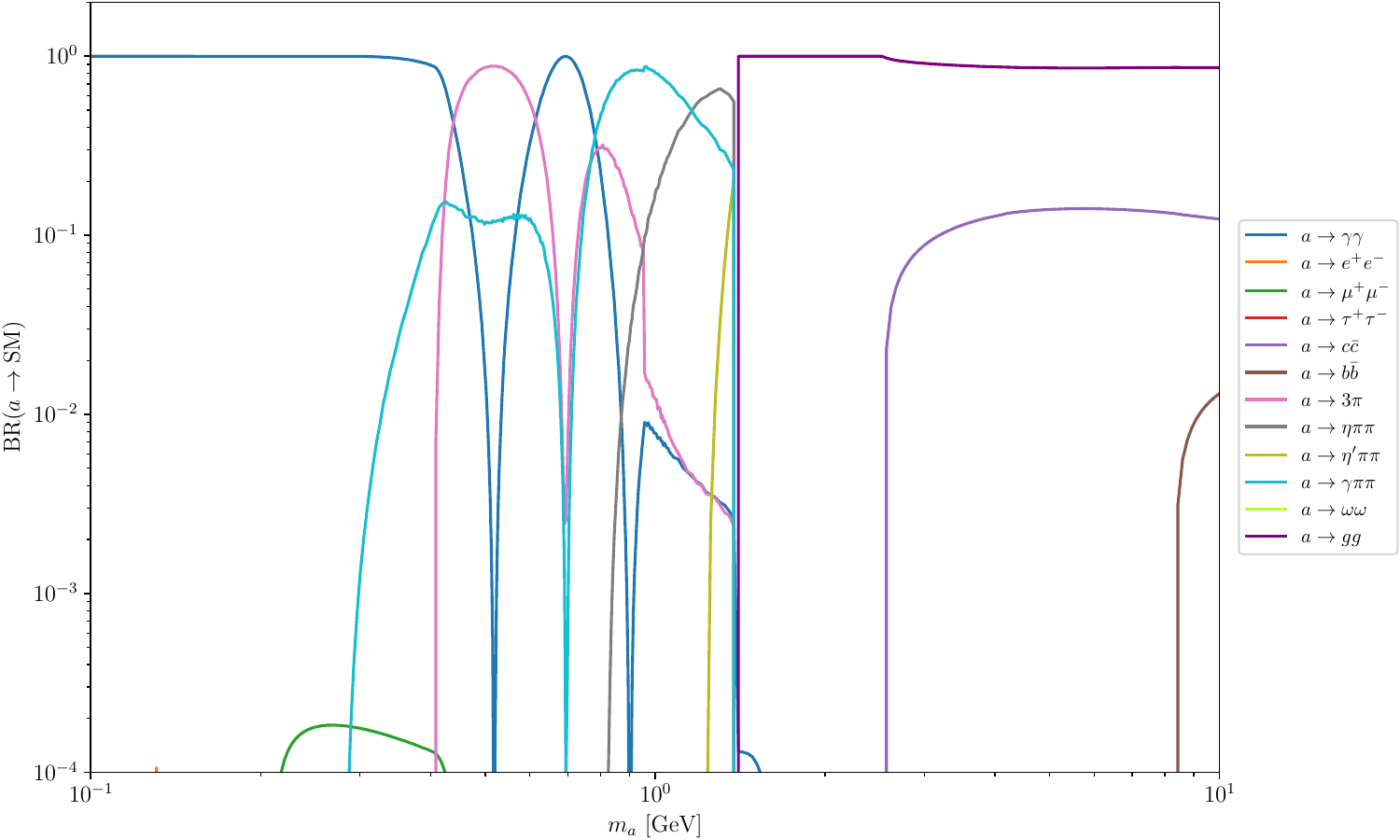}}
     \,
     \caption[]{\em Branching ratios of all studied channels for the Q-KSVZ model.}
	\label{fig:br_KSVZ-Q}
\end{figure}

The behaviour of the Q-KSVZ benchmark model, shown in Fig.~\ref{fig:br_KSVZ-Q}, shows completely different characteristics. 
As in this model only the coupling to gluons is present at tree level, the decays to leptons are highly suppressed and 
barely seen in the plot. 
For lower masses, the decay to photons is dominant, then, in the region $0.4\lesssim  m_a\lesssim 1$ GeV, the decay to 
photons competes in relevance with the hadronic channels $3\pi$ and $\gamma \pi\pi$. For ALP masses $m_a>1$ GeV the dominant 
channels are $\eta^{(\prime)} \pi\pi$ first and then the inclusive decay to all hadrons is the most relevant, as couplings 
to quarks and leptons are generated at one- and two-loop respectively. 

These two contrasting models present the most relevant phenomenological channels that will bound the ALP parameter space in 
future sections, however, note that in particular the flaxion also exhibits flavour-violating decays, which have been also 
implemented in \alpaca\!\!. These channels, in particular those containing LFV final states do not modify substantially 
the picture, since they are proportional to the masses of these fermions $m_f m_{f'}$, and hence, once the ALP can decay 
to 2 of the heaviest of the fermions, this decay tends to dominate. However, there are relevant LFV searches in mesons 
such as $B^0\to K^{0*} \mu^\pm\tau^\mp$ in LHCb \cite{LHCb:2022wrs}.

\section{ALP processes}
\label{sec:signatures}
For an ALP mass in the $[0.01,10]$ GeV range, some of the most promising ALP production channels are Flavour-Changing Neutral Current (FCNC) meson decays, like $B\to K^{(*)}a$ and $K\to \pi a$, and quarkonia decays, as 
$\Upsilon(nS)\to\gamma a$ and $J/\psi\to\gamma a$.

In most of the FCNC meson decays, the dominant contribution comes from the effective flavour-violating coupling 
$q_1 \to q_2 a$ (where $q_1$ and $q_2$ are the initial and final state quark, respectively). As seen in the 
previous section, in a realistic UV model this effective coupling gets an irreducible contribution both by 
the rotation to the mass basis and the integration of the top and $W^\pm$ fields, in addition to any possible 
model-dependent tree-level or one-loop sources of flavour violation. 

If both the initial and final mesons are pseudoscalars or vectors, the decay is mediated by the vector coupling 
$c_{q_1q_2}^V$, while pseudoscalar to vector or vector to pseudoscalar decays are mediated by the axial coupling 
$c_{q_1 q_2}^A$. The initial state radiation and final state radiation of the ALP, mediated by the flavour-conserving 
couplings $c_{qq}^A$, are typically more suppressed~\cite{Guerrera:2022ykl}. The decay rates of $B$ mesons into 
lighter hadrons $P=K, \pi$ or $V=K^*, \rho$ are given by~\cite{Izaguirre:2016dfi}
\begin{equation}
\begin{split}
\Gamma(B^+\to P^+a) &= \frac{m_B^3}{64\pi f_a^2}  |c^V_{ib}|^2 |F_0^{B\to P}(m_a^2)|^2\sqrt{1-\frac{m_P^2}{m_B^2}} \,
\lambda^{1/2}\!\!\left(1,\frac{m_P^2}{m_B^2}, \frac{m_a^2}{m_B^2} \right)\,,\\ 
\Gamma(B^+\to V^+a)&=\frac{m_B^3}{64\pi f_a^2}  |c^A_{ib}|^2 |A_0^{B\to V}(m_a^2)|^2 \,
\lambda^{3/2}\!\!\left(1,\frac{m_V^2}{m_B^2}, \frac{m_a^2}{m_B^2} \right)\,,
\end{split}
\end{equation}
where $F_0$ and $A_0$ are the scalar form factors for the corresponding hadronic transition and $\lambda(a,b,c) = 
a^2+b^2+c^2-2(ab+ac+bc)$ is the K\"all\'en function. The $K\to\pi a$ decays depend not only on the flavour-violating   
coupling $c_{ds}^A$, but also on $c_G$ and on the flavour-conserving couplings to light quarks~\cite{Bauer:2021wjo,
Guerrera:2021yss,Cornella:2023kjq}. 

Quarkonia decays are sensitive to different combinations of the flavour-conserving coupling $c_{qq}^A$ ($q=c$ for 
$J/\psi\to\gamma a$ and $q=b$ for $\Upsilon(nS)\to\gamma a$) and the photon coupling $c_\gamma$, depending on whether 
the decay proceeds resonantly, non-resonantly or mixed~\cite{Merlo:2019anv,DiLuzio:2024jip}. The decay width of the 
quarkonia meson, relevant for resonant searches, is\footnote{For $J/\Psi\to a\gamma$ we use lattice input to include hadronic corrections to the second term of the expression, see Ref.~\cite{Colquhoun:2025xlx}}
\begin{equation}
    \Gamma(V\to\gamma a) \simeq \frac{\alpha_\mathrm{em} Q_q^2 m_V}{24} \frac{f_V^2}{f_a^2}\left|
    \frac{\alpha_\mathrm{em} c_\gamma}{\pi} \left(1-\frac{m_a^2}{m_V^2}\right)-2c_{qq}^A\right|^2\,,
\end{equation}
while the non-resonant production cross-section $e^+e^-\to\gamma a$ is
\begin{equation}
\sigma_\mathrm{NR}(s) = \frac{\alpha_\mathrm{em}^3}{24\pi^3}\frac{|c_\gamma|^2}{f_a^2}\left(1-\frac{m_a^2}{s}\right)^3\,.
\end{equation}

\subsection{Experimental signatures of on-shell ALPs}
\label{Sect:ExpSigOnShellAlps}

As seen in Section~\ref{sec:decays}, the ALP is an unstable particle, and it will eventually decay. In our set-up, the 
condition $\Gamma_a \ll m_a$ is always fulfilled, and consequently we can work in the Narrow Width Approximation (NWA), 
where the ALP is produced on-shell and the branching ratio for the complete process of ALP production in $M \to F_1 a$ 
decays and subsequent ALP decay $a\to F_2$ (where $M$ is the initial meson, $F_1$ and $F_2$ can be multi-particle states) 
is given by
\begin{eqnarray}
    \label{eq:NWA}
    \mathrm{BR}(M \to F_1 F_2)_\mathrm{NWA} \approx \mathrm{BR}(M\to F_1 a) \times \mathrm{BR}(a\to F_2)\,.
\end{eqnarray}

Depending on the decay length of the ALP $c\tau_a = c \hbar/\Gamma_a$ and the characteristics of the experiment, 
there are three possible signatures for the ALP processes: if the ALP is very short-lived, the decay $M\to F_1 F_2$ 
will be prompt, with the decay products of the ALP appearing to originate at the primary interaction point. If the 
decay length of the ALP is comparable with the spatial resolution of the experiment, the decay $a\to F_2$ will 
happen at a displaced vertex separated from the primary vertex, provided the experiment can distinguish a secondary 
vertex. Finally, if the ALP is long-lived, it will escape the detector before decaying, resulting in an invisible 
decay $M\to F_1 + \mathrm{missing\ energy}$. A further, model-dependent scenario, is that the ALP decays mainly 
into a dark sector, which results in invisible decays even if the decay length is small. In the present work we 
will treat the branching ratio into the dark sector as a free parameter, without dwelling in any of the possible  
realisations.

If the ALP has velocity $\beta_a$ and Lorentz boost factor $\gamma_a$ in the LAB frame, the probability of finding 
an ALP that decays after traveling a distance between $r_1$ and $r_2$ is\cite{Ferber:2022rsf,Bruggisser:2023npd}
\begin{equation}
    P(r_1,r_2) = \exp\left(-\frac{r_1}{c \tau_a \beta_a \gamma_a}\right)-\exp\left(-\frac{r_2}{c \tau_a \beta_a \gamma_a}\right)\,.
\end{equation}
If the detector is able to resolve secondary vertices located between $r_\mathrm{min}$ and $r_\textrm{max}$ of the 
collision point, the probability of a prompt decay is $P_\textrm{prompt} = P(0, r_\textrm{min})$, for a displaced-vertex 
decay is $P_\textrm{d.v.} = P(r_\textrm{min}, r_\textrm{max})$, and finally for a decay outside the detector is 
$P_\textrm{out} = P(r_\textrm{max}, \infty)$. The explicit expressions for these two decays are
\begin{eqnarray}
    \label{eq:prompt_inv}
    P_\textrm{prompt}= 1-\exp\left(-\frac{r_\textrm{min}}{c\tau_a \, \beta_a \gamma_a}\right)\, , \quad  
    P_\textrm{out}= \exp\left(-\frac{r_\textrm{max}}{c\tau_a \, \beta_a \gamma_a}\right)\,.
\end{eqnarray}

\begin{figure}[t]
    \centering
\includegraphics[width=\linewidth]{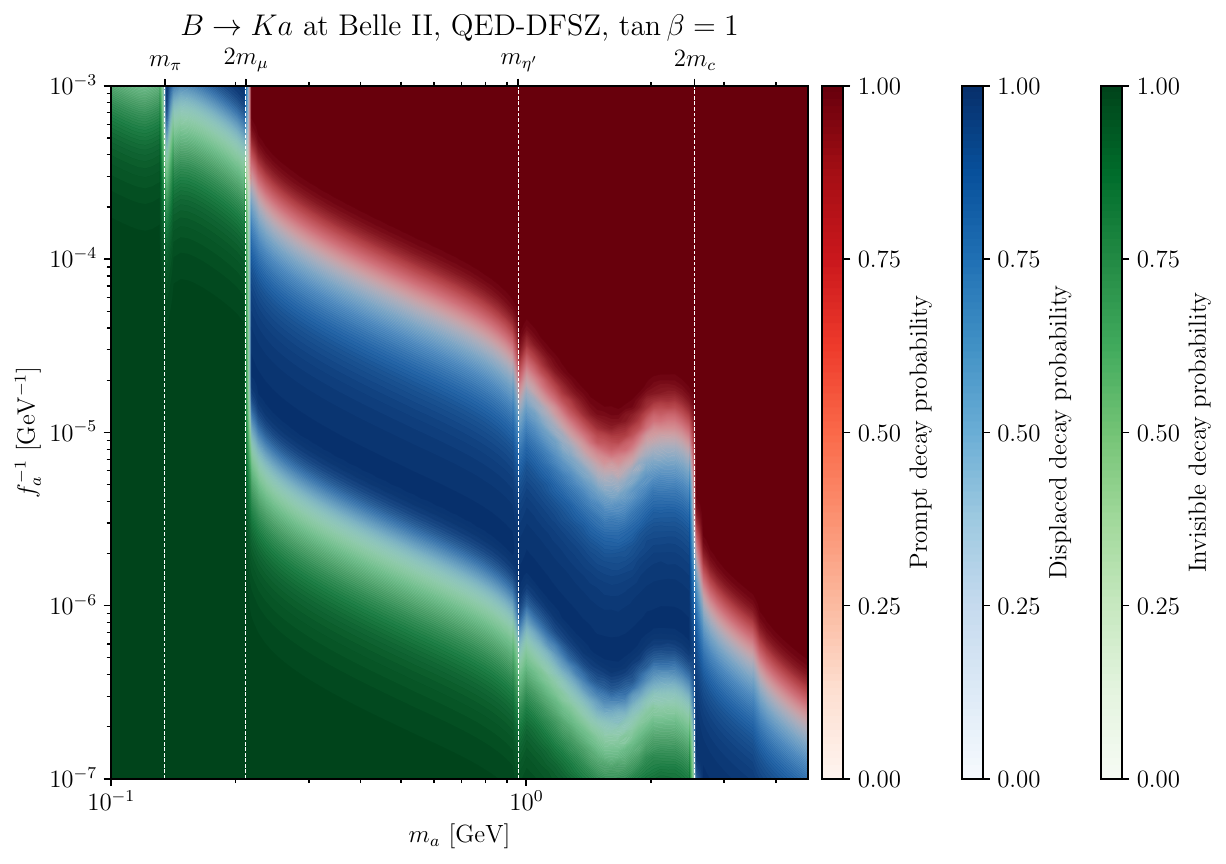}
\caption{\em Probability distribution of a QED-DFSZ ALP of decaying promptly (red), as a secondary displaced 
vertex (blue) and invisibly (green) at Belle II. This model does not include dark sector decay modes, and hence the only 
invisible decays are those where the ALP escapes outside the detector.}
    \label{fig:ProbDecaysBelle2}
\end{figure}

\begin{figure}[ht]
    \centering
    \includegraphics[width=0.8\textwidth]{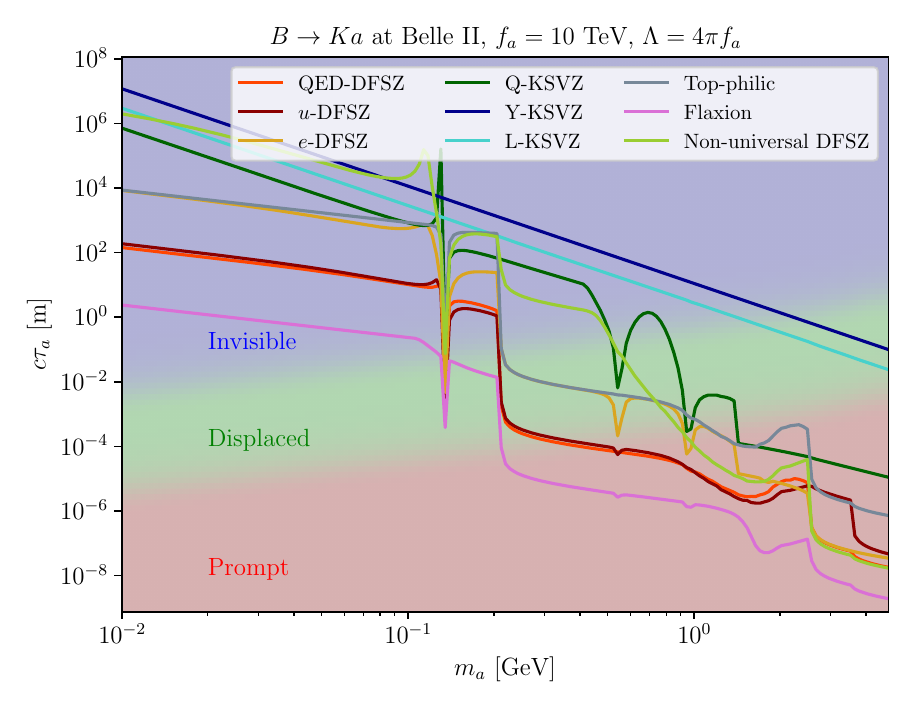}
    \caption{\em The lines represent the proper decay length $c\tau_a$ for ALPs of mass $m_a$ in the UV benchmarks discussed in this work. For DFSZ-like benchmarks, the lines correspond to $\tan \beta=1$, except for the non-universal DFSZ, where $\tan\beta = 10$ is used instead (see discussion at the end of Section~\ref{sec:Belle2} for a motivation). The background colours represent the main signature (prompt decay in red, decay with displaced vertex in green and invisible decay in blue) of an ALP of the given $m_a$ and $c\tau_a$ that was produced in a $B\to K a$ process at Belle II ($\gamma_B\beta_B = 0.28$, $r_\mathrm{min} = 1\,\mathrm{mm}$, $r_\mathrm{max} = 1\,\mathrm{m}$). The value of $f_a=10\,\mathrm{TeV}$ has been chosen to better illustrate the different regimes.}
    \label{fig:regions_BKa_BelleII}
\end{figure}

As a study case, let us consider the QED-DFSZ model defined in Eq.~\eqref{eq:model_QEDDFSZ}. In Fig.~\ref{fig:ProbDecaysBelle2}
we plot the probability distribution as a function of $m_a$ and $f_a$, for the chosen value $\tan\beta=1$. In this plot, 
the interplay between the mass and the scale of the ALP in producing a prompt decay, a displaced decay, or an invisible decay
is shown. As this model does not include dark sector decay modes, the only invisible decays are those where the ALP 
escapes outside the detector. We see, that after the $m_a=2m_\mu$ and $m_a=2m_c$ thresholds the prompt decay region 
get enhanced as $\Gamma_a$ increases when any new channel opens up. We also observe how in the $1-2\,$GeV region, the 
hadronic channels, mainly $a\to3\pi$, are crucial to understand not only the allowed decay channels, but also the lifetime. 
For large enough $f_a$ the ALP decay gets delayed, there is a higher probability to detect a displaced vertex. Then, 
either for light ALP masses, in particular below the $\mu$ production threshold (where the decays are suppressed by the 
electron mass), or sufficiently large $f_a$, the ALP travels outside the detector before decaying, showing up as an 
invisible decay channel. Finally, the spikes at $m_a=m_\pi$ and $m_a=m_{\eta'}$ are due to the ALP mixing with the 
$\pi^0$ and the $\eta'$.

In Fig.~\ref{fig:regions_BKa_BelleII} we show a comparison of the proper ALP decay lengths as a function of the ALP mass 
for the different benchmark models presented previously. In the cases where tree-level leptonic decays are possible, 
ALPs produced by Belle II cease to be invisible once the muon threshold ($m_a > 2 m_\mu$) is crossed. For ALPs that 
couple predominantly to gluons, as in the Q-KSVZ model, the decay becomes visible once the channel $a\to \eta \pi\pi$ 
is open. Finally, in the Y-KSVZ and L-KSVZ benchmarks, the only tree-level decay available in the mass range considered 
is $a\to\gamma\gamma$, which results in decays outside the detector for $m_a \lesssim 3\,\mathrm{GeV}$.

For our analysis, we will interpret the experimental measurements as follows:
\begin{itemize}
    \item In \textbf{prompt decays} $M\to F_1 a(\to F_2)$  the quantity $P_\mathrm{prompt} \times \mathrm{BR}(M\to F_1 a)\times\mathrm{BR}(a\to F_2)$ will be confronted in the analysis with the experimental determination of $\mathrm{BR}(M\to F_1F_2)$.
    \item For \textbf{displaced vertex analyses}, the experimental collaborations report bounds on $\mathrm{BR}(M\to F_1 a)\times\mathrm{BR}(a\to F_2)$ (in several experimental searches this particle is labelled as $A$, an unspecified light particle, that in our case we identify with the ALP) as functions of $m_a$ and $c\tau_a$. Therefore, when $(c\tau_a)_\mathrm{min} < c\tau_a < (c\tau_a)_\mathrm{max}$, we directly compare $\mathrm{BR}(M\to F_1 a)\times\mathrm{BR}(a\to F_2)$ to the reported bounds. If $c\tau_a < (c\tau_a)_\mathrm{min}$, i.e. in the prompt region, we assume that the bounds are equal to those at $(c\tau_A)_\mathrm{min}$ and that $P_\mathrm{prompt}+P_\mathrm{d.v.} = 1$. If $c\tau_a > (c\tau_a)_\mathrm{max}$, we take the bounds at $(c\tau_A)_\mathrm{max}$  weighted by $P_\mathrm{d.v.}$. This approach is consistent with the tool provided by LHCb to obtain the bounds on $\mathrm{BR}(B^0 \to K^{*0}a) \times \mathrm{BR}(a\to \mu^+\mu^-)$ in~\cite{LHCb:2015nkv}.
    \item In the case of \textbf{invisible decays}, if the ALP only decays into SM particles, we have to confront $P_\mathrm{out} \times \mathrm{BR}(M\to F_1 a)$ with the measurements of $\mathrm{BR}(M\to F_1 + \mathrm{inv.})$. If the ALP can decay also into a dark sector, then our prediction is given by $\mathrm{BR}(M\to F_1 a) \times[P_\mathrm{out} \times\mathrm{BR}(a\to \mathrm{SM}) + \mathrm{BR}(a\to\mathrm{dark})]$.
\end{itemize}

Table~\ref{tab:experiments_decaytypes} reports useful information about the experimental setup used in this paper. 
In particular, we list the region where we expect the ALP to promptly decays and the distance to consider it as 
missing energy. We also show the boost factor $\beta\gamma$ necessary to compute the different probabilities. 
These input data are implemented in the \alpaca routines and are considered in the phenomenological analysis 
that follows. Notice that if one wants to study all possible decay modes, and cover the complete parameter space, 
different experiments sensitive to different lifetime (or distance) ranges that allow to investigate all types of 
signatures are needed. In particular, notice how the CHARM~\cite{CHARM:1985anb} and MicroBooNE~\cite{MicroBooNE:2021sov} 
experiment in Table~\ref{tab:experiments_decaytypes} have the longest detection lengths, but since they study different 
meson decays, the two searches are relevant in different portions of the parameter space. Future projections of the 
SHiP experiment~\cite{SHiP:2015vad,SHiP:2021nfo} show that it will be able to look for long particles with a similar 
sensitivity. Contrary to this, LHCb~\cite{LHCb:2016awg} is able to track secondary vertices with higher resolution, 
allowing to have better constraints on displaced signatures. 

\begin{table}[t]
    \centering
    \begin{tabular}{c|c|c|c|c}
        $P_1$ & Experiment & Prompt region & Invisible region & $\beta_{P_1}\gamma_{P_1}$ \\\hline
        $K$ & NA62 & $\tau_a < 100\,\mathrm{ps}$ & $\tau_a > 5\,\mathrm{ns}$ & --\\
        $K$ & MicroBooNE & $c\tau_a < 0.3\,\mathrm{m}$ & $c\tau_a > 3\times10^7\,\mathrm{m}$ & --\\
        $B$ & BaBar & $r_\mathrm{min} = 2\,\mathrm{cm}$  & $r_\mathrm{max} = 3\,\mathrm{m}$ & $0.42$ \\
        $B$ & Belle & $r_\mathrm{min} = 4\,\mathrm{cm}$  & $r_\mathrm{max} = 1\,\mathrm{m}$ & $0.42$ \\
        $B$ & Belle II & $c\tau_a < 1\,\mu\mathrm{m}$  & $c\tau_a > 4\,\mathrm{m}$ & $0.28$ \\
        $B$ & LHCb & $\tau_a < 0.1\,\mathrm{ps}$ & $\tau_a > 1\,\mathrm{ns}$ & --\\
        $B$ & CHARM & $\tau_a <250\,\mathrm{ps}$ & $\tau_a>1\,\mu\mathrm{s}$ & -- \\
        $B$ & SHiP & $\tau_a <10\,\mathrm{ps}$ & $\tau_a>1\,\mu\mathrm{s}$ & -- \\
        $\Upsilon(1S) $ & Belle & &$r_\mathrm{max} =1\,\mathrm{m}$ & 0.0\tablefootnote{The $\Upsilon(1S)$ at Belle are generated in the three-body decay $\Upsilon(2S)\to\Upsilon(1S) \pi^+\pi^-$, and therefore its boost follows a probability distribution. We have checked that in the ranges of interest $|\vec{p}_a|\gg|\vec{p}_{\Upsilon(1S)}|$, and as a simplifying assumption we set the $\Upsilon(1S)$ approximately at rest.} \\
    \end{tabular}
    \caption{\em List of the relevant parameters that mark the different search regions, depending on the different process and detector characteristics. Whenever is given, we used the experimental sensitivity to the ALP lifetime $\tau_a$, as reported in this table, otherwise we estimate by roughly taking the lengths of the detector.}
\label{tab:experiments_decaytypes}
\end{table}

\section{Phenomenological analysis}
\label{sec:pheno}
This section is mostly devoted to highlighting the potential of \alpaca for ALP analyses. We will 
start in Section~\ref{sec:phenoUV} by presenting a phenomenological study of the allowed parameter space 
for the UV benchmark models presented in Section~\ref{sec:models}. Then, in Section 
\ref{sec:phenoEFT}, we will consider a more agnostic scenario by showing how the different ALP-EFT Wilson coefficients get 
constrained by the most recent experimental data. These studies will allow us to highlight the complementarity 
between all the different experimental searches, as well as the different information achievable within these 
two orthogonal approaches, i.e. UV vs EFT. Then, in Section \ref{sec:Belle2}, as a study case, we focus 
on the Belle II anomaly in the $B^+\rightarrow K^+ \nu \overline{\nu}$ channel, pointing out, for the first 
time, the importance of fully accounting for the decay length information when combining measurements from 
different experiments. Finally, we show how the Belle II anomaly could be accounted for in 
one of the specific non-universal ALP--fermion model, previously discussed.

\subsection{UV models: A bound on the scale of the ALP}
\label{sec:phenoUV}

In Section~\ref{sec:models}, we have introduced a selection of UV models to highlight the variety of ALP 
behaviours. As it is well known, the most stringent constraints on ALP couplings arise from flavour-changing 
transitions, involving either visible or invisible channels. To illustrate the phenomenological differences in this 
type of processes, let us focus on the QED-DFSZ, Q-KSVZ, top-philic and flaxion frameworks as they provide 
a representative set of benchmarks to explore the different suppression mechanisms of flavour transitions: in the flaxion 
scenario they appear already at tree level, in the QED-DFSZ and top-philic models flavour violating couplings are induced 
at one loop, whereas in the Q-KSVZ model only at two loops. Owing to this hierarchy in flavour-violating couplings, the 
bounds on the corresponding scales differ by several orders of magnitude, making both visible and invisible searches 
essential to constrain these models.

\begin{figure}[tb!]
	\centering
\includegraphics[width=\linewidth]{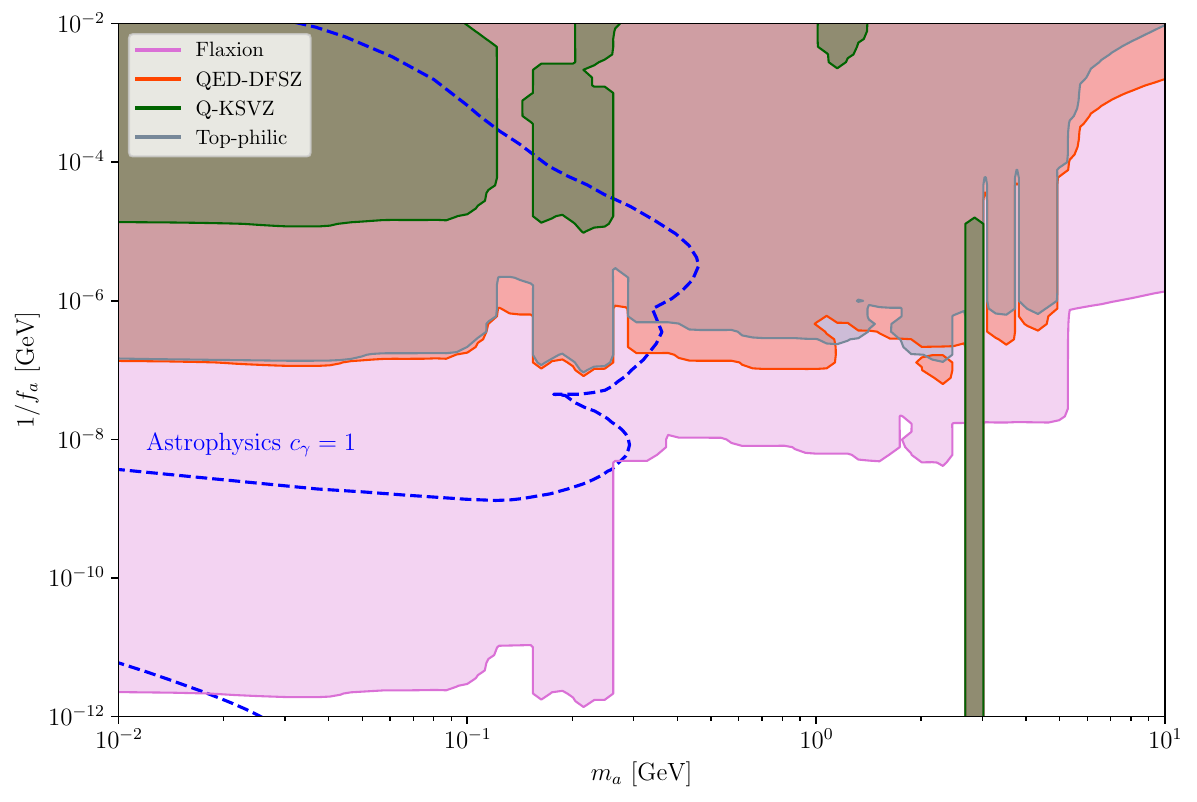}
\caption[]{\em $3\,\sigma$ excluded parameter space for the different models studied in Section~\ref{sec:models}. 
The red region corresponds to the QED-DFSZ model for $\tan\beta = 1$, the green region to the Q-KSVZ model, the grey region to the top-philic model and the pink region to the flaxion model. The dashed blue line delimit, for reference, the region of parameter space tested by astrophysics, for $c_\gamma=1$~\cite{AxionLimits}.}
\label{fig:model-exclusion}
\end{figure}

In Fig.~\ref{fig:model-exclusion}, we present the excluded regions in the $(m_a, 1/f_a)$ parameter space for the various 
models considered. For the QED-DFSZ case $\tan\beta = 1$ has been chosen. 

At low ALP masses, the dominant constraint across all models arises from the flavour-changing transition $K \to \pi\, a$ 
measurement at NA62~\cite{NA62:2021bji,Guadagnoli:2025xnt}. In this case we have used the reanalysis of the $K\to \pi \nu\bar{\nu}$ 
measured at NA62~\cite{NA62:2024pjp} performed in Ref.~\cite{Guadagnoli:2025xnt}, which shows good agreement with previous 
analyses performed by the experimental collaboration. In this regime, the ALP is effectively invisible, since for 
$m_a \leq 2m_\mu$ the couplings to photons and electrons are so strongly suppressed, by $\alpha_\textrm{em}$ and $m_e$ 
respectively, that the ALP fails to decay within the detector. A small region around $m_a \sim 100\,\textrm{MeV}$ is 
additionally constrained by the MicroBooNE experiment, which searches for long-lived ALPs via the decay $K \to \pi\, 
a(\to e^+ e^-)$ (see Ref.~\cite{Coloma:2022hlv}); in this case, the long baseline is particularly sensitive to such 
displaced signals~\cite{MicroBooNE:2021sov}.

For $m_a \gtrsim 2m_\mu$, the models exhibit distinctive phenomenological signatures. On the one hand, the flaxion model can be probed via both displaced and invisible signals up to the $B$ meson mass threshold, as its flavour-violating coupling is only mildly suppressed by two powers of the Cabibbo angle, $\lambda^2$. The main constraint arises from $B^+ \to \rho^+ a$~\cite{Belle:2017oht}, though channels such as $B^0 \to \pi^0 a(\to\mu^+ \mu^-)$ from BaBar~\cite{BaBar:2013qaj}, and $B^+ \to K^+ a(\to\mu^+ \mu^-)$ from LHCb~\cite{LHCb:2016awg} and CHARM~\cite{CHARM:1985anb} are also competitive. Although this model allows for lepton flavour-violating decays, which we have included in the analysis, the strongest bounds still come from decays into muons or missing energy.
On the other hand, models in which flavour-violating couplings arise at one loop, such as QED-DFSZ and top-philic, are constrained in the region where displaced muon searches via $B^+ \to K^+\, a(\to\mu^+ \mu^-)$ become sensitive, extending up to $m_a \simeq m_B - m_K$. These bounds are set by CHARM~\cite{CHARM:1985anb}, LHCb \cite{LHCb:2016awg}, and Belle II \cite{Belle-II:2023ueh}. In the case of invisible ALP decay at Belle II, we use the reinterpretation of $B^+ \to K^+\, \nu\bar{\nu}$ from Ref.~\cite{Altmannshofer:2023hkn}, which reports a $2.8\,\sigma$ excess. This excess translates into a lower bound on these models, as they cannot accommodate the observed anomaly. This issue will be discussed in detail in Section~\ref{sec:Belle2}. The distinction between the QED-DFSZ and top-philic benchmarks arises from the origin of the muon coupling in the latter, which is also generated at one loop, resulting in a slightly longer ALP lifetime, as shown in Fig.~\ref{fig:regions_BKa_BelleII}.

For the Q-KSVZ model, the two-loop suppression of flavour violation pushes the scale lower, and the ALP is predominantly visible. However, the leading decay channels are hadronic, as illustrated in Fig.~\ref{fig:br_KSVZ-Q}. In this case, the lack of a comprehensive inclusive measurement for $B \to K \, a\,(\to\text{hadrons})$, along with the absence of precise theoretical predictions for individual branching ratios (for $m_a\gtrsim 2\,$GeV), limits the sensitivity of meson searches in this region. Only the intermediate mass range is constrained via decays such as $B \to K\, a(\to \eta \pi\pi)$ and $B \to K\, a(\to3\pi)$ from BaBar~\cite{
   BaBar:2008rth,BaBar:2011kau,Chakraborty:2021wda}, which we can estimate theoretically, as discussed in Section~\ref{sec:decays}.

The heavier ALP region of the parameter space is constrained by several complementary observables. For the flaxion benchmark, the measurement of $\Delta m_{B}$ at Belle II~\cite{Belle-II:2023bps} imposes a strong bound due to the large flavour-violating couplings. For the QED-DFSZ and top-philic models, a combination of $\Upsilon(1S) \to \gamma\, a(\to\tau^+ \tau^-)$, from BaBar~\cite{BaBar:2009oxm}  (see Ref.~\cite{DiLuzio:2024jip} for a detailed analysis), and the decay $B_s \to \mu^+ \mu^-$, from CMS~\cite{CMS:2022mgd}, constrain the final portion of the parameter space considered here.

In Fig.~\ref{fig:model-exclusion} the blue dashed line shows the parameter space that can be covered by astrophysical 
observables. As a reference value we have set $c_\gamma =1$. A detailed analysis of all astrophysical bounds is 
beyond the scope of this paper. The main reason to depict them here is to highlight their complementarity with meson 
decays, at least in the sub-GeV ALP mass range.
 
\begin{figure}[tb!]
    \centering
    \includegraphics[width=\linewidth]{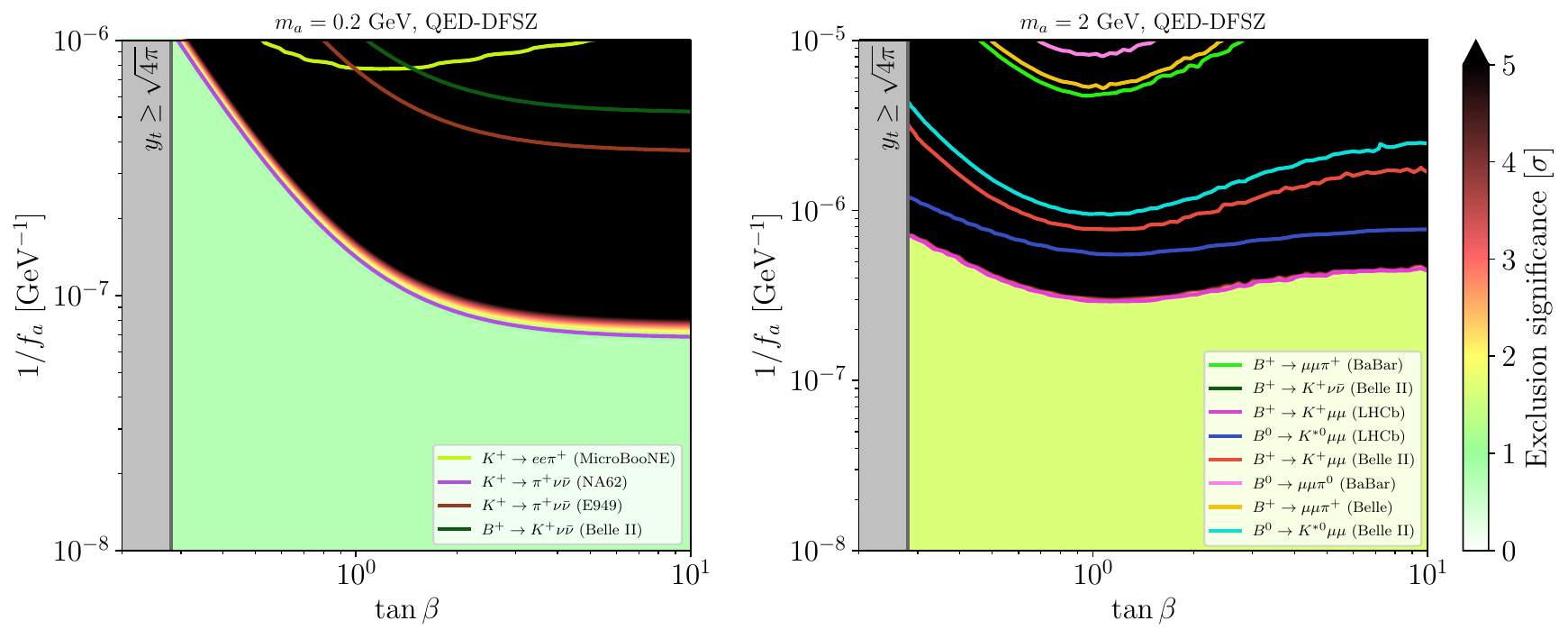}
    \caption{\em Bounds on the $\tan\beta-f_a$ plane for $m_a=0.2\,\mathrm{GeV}$ (left) and $m_a=2\,\mathrm{GeV}$ (right) for the QED-DFSZ model detailed in Section~\ref{sec:models}. In the second case, the green-yellow background indicates a slight tension with Belle II measurement.
    }
    \label{fig:QED_DFSZ}
\end{figure}

To conclude, in Fig.~\ref{fig:QED_DFSZ} we show the $\tan\beta$ dependence of the QED-DFSZ model bounds for an ALP mass $m_a = 0.2\,\mathrm{GeV}$ ($2\,\mathrm{GeV}$) in the left (right) panel. The different ALP masses lead to distinct decay patterns: for small $m_a$, the ALP is long-lived and effectively invisible in the observables under consideration; for large $m_a$, by contrast, the ALP may either decay promptly or produce a displaced vertex. As a general comment, increasing $\tan\beta$ enhances the quark couplings ($c_{u,d}\propto s_\beta^2$) and suppresses the lepton couplings ($c_e\propto c_\beta^2$). In the left panel of Fig.~\ref{fig:QED_DFSZ}, this results in a longer ALP lifetime, thereby strengthening the bound from $K \to \pi \nu\bar{\nu}$. Similarly, in the right panel, a large $\tan\beta$ slightly reduces the branching ratio $\mathrm{BR}(a \to \mu\mu)$, thus weakening the bound. On the other hand, for small $\tan\beta$ values, the flavour-violating couplings, induced by the top quark through the running, are suppressed, and consequently, the ALP production probability is reduced.

\subsection{ALP EFT}
\label{sec:phenoEFT}

\begin{figure}[t]
    \centering
    \includegraphics[width=1\linewidth]{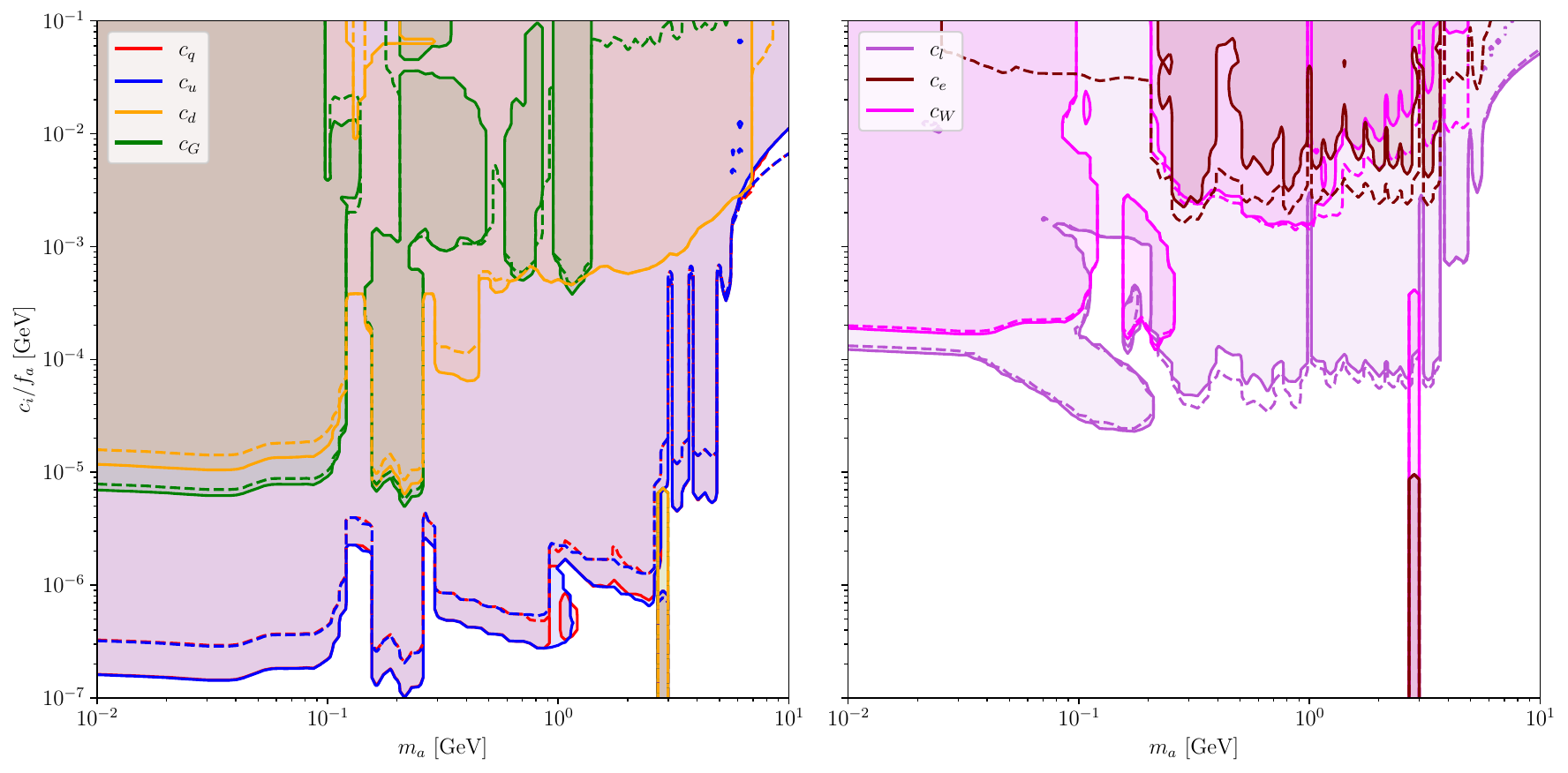}
    \caption{\em Bounds on $c_i/f_a$ for the individual Wilson coefficients of the theory above the electroweak scale (Eq.~\eqref{eq:UVLag}), for the couplings to quarks and gluons (left) and to leptons and $W^\pm$ bosons (right). Solid lines correspond to $c_i=1$ and $f_a$ as free parameter, while dashed lines to the opposite case with $f_a=1\,\mathrm{TeV}$ and free Wilson coefficients. See the text for details.}
    \label{fig:OneWilsonCoef}
\end{figure}

Instead of defining a specific model, and therefore assuming \textit{ab initio} a definite value for the ALP couplings at the 
UV scale $\Lambda$, one can assume an agnostic approach and make use of the ALP-EFT Lagrangian of Eq.~\eqref{eq:UVLag}, 
where all the coefficient are considered as arbitrary parameters at the scale $\Lambda$, that is the scale up to which 
the effective theory is meaningful. Being all the Wilson coefficient completely arbitrary, the simplest approach is to 
give bounds separately to each of them, setting to zero all the other coefficients. For the sake of simplicity, in the 
phenomenological analysis of this subsection all the couplings defined in Eq.~\eqref{eq:UVLag} will be assumed flavour universal at the scale $\Lambda$, thus flavour non-universality will enter only through loop effects and matching conditions. In Fig.~\ref{fig:OneWilsonCoef} we show the corresponding bounds for the hadronic (left) and weak-leptonic sector (right), respectively. 

As all the observables depend on $c_i/f_a$ ratios, two slightly different approaches can be pursued that correspond, 
respectively, to the bounds delimited by the solid and dashed lines. The first approach consists in fixing all the 
coefficients $c_i(\Lambda) \equiv 1$ and letting $f_a$ as a free parameter. The running effect, which is performed 
from the high scale $\Lambda=4\pi f_a$ down to the physical scale of the process under consideration, may greatly 
differ in distinct points of the parameter space: clearly, the smaller the value $1/f_a$, the larger the effects 
of the running contributions. This case is shown in Fig.~\ref{fig:OneWilsonCoef} as solid lines, that therefore 
provide an exclusion bound directly on the scale $f_a$ associated to the corresponding ALP--SM particle couplings 
to which each observable is sensitive to. In the complementary approach, instead, the Wilson coefficients are treated 
as free parameters and $f_a$ is fixed and chosen here, for definiteness, as $f_a=1\TeV$. The running is then performed 
from the same high scale $\Lambda=4\pi f_a$ in the whole parameter space. The corresponding result is given by the 
dashed lines, that then provide a direct bound on the Wilson coefficients $c_i(\Lambda)$. 

As one can see in Fig.~\ref{fig:OneWilsonCoef}, while the two approaches give similar bounds in much of the $(m_a, f_a)$ parameter space, significant discrepancies appear in parts of the $m_a$ range considered. Indeed, the main responsible for these differences is the running, that indirectly depends on the ALP mass, through the $B_{1,2}$ functions defined in Eq.~\eqref{eq:B1B2}. The differences start to appear moving apart from $c_i/f_a\approx 1 \, \mathrm{TeV}^{-1}$, as the higher the scale, the larger the running contributions. Indeed, we can see that the bound is weaker for a constant $c_i$ (solid lines) than for constant $f_a$ (dashed lines), when $c_i/f_a\gtrsim 1 \, \mathrm{TeV}^{-1}$: in this region of the parameter space, the solid lines show the results of the scan with $f_a\lesssim 1\TeV$; on the contrary, the dashed lines represent the case with $f_a= 1\TeV$ and thus a stronger running takes place. On the contrary, the constant $c_i$ approach provides stronger bounds below this threshold value, as the running starts from a scale larger than $1\TeV$. 

In the left plot of Fig.~\ref{fig:OneWilsonCoef} we study the effect on flavour-universal quark and gluons operators, and the results are very similar to what we see in the top-philic and Q-KSVZ model of the previous section, see Fig.~\ref{fig:model-exclusion}. However, with $c_{d_R}$ we start finding some differences with respect to the previous benchmarks. Again, at low masses NA62 \cite{NA62:2020pwi} and E949 \cite{BNL-E949:2009dza} invisible searches set strongest bounds before the muon threshold, while CHARM $B\to K \mu\mu$ \cite{CHARM:1985anb}, bounds the intermediate region. However, as hadronic decays start dominating over the one-loop-generated lepton couplings, the BaBar bound coming from $\Upsilon(3S)\to \gamma \,a (\to\textrm{hadrons})$ dominates until large masses \cite{BaBar:2011kau}. 

In the right plot of Fig.~\ref{fig:OneWilsonCoef}, we find the leptons and EW anomalous couplings, where $c_B$ does not yield bounds in this region of parameter space. Contrary to quarks, we see a large difference between starting with the LH lepton operator or the RH one. For the Wilson coefficient $c_{\ell_L}$, a combination of NA62 \cite{NA62:2020pwi} and E949 \cite{E949:2014gsn} invisible search bound up to $f_a\sim 10^3$, but MicroBooNE \cite{MicroBooNE:2021sov} search constrains further the parameter space, until the $m_a=2m_\mu$ threshold. Then for both $c_{\ell_L}$ and $c_{e_R}$, LHCb $B^+\to K^+ \mu\mu$ \cite{LHCb:2016awg} becomes the most constraining bound, however with more than one order of magnitude difference. This difference is due to the EW running of the LH and RH operators \cite{Bauer:2021mvw}. In the first case flavour violation is induced at two-loops via $W^\pm$ gauge bosons, while in the RH case is due to the $\tilde{c}_B$ coefficient induced by the running. Finally the high mass range of this plot is bounded by the $B_s\to \mu\mu$ measurement from CMS~\cite{CMS:2022mgd}. The bounds on the $c_W$ coefficient follow the same story as for $c_{\ell_L}$, but with the lepton operators generated at one-loop, hence, we see no bound coming from displaced $a\to ee$ MicrooBooNE searches. Then again bounds coming from LHCb set the most constraining bounds, but the one-loop suppression on the decay to muons sets the bounds weaker than for the LH lepton coupling.

Considering two non-vanishing Wilson coefficients at the same time, can highlight the existence of possible  correlations. In particular we will focus to study correlation in the ALP--fermion sector. As a benchmark point for the ALP mass, we choose $m_a=0.3$ GeV, since for this mass value $s \to d$ transitions still play a role, even if not anymore the dominant ones, and we can compare different flavour-violating channels, as well as different decay modes. Moreover, in all the plot presented in this subsection, we are going to fix the scale $\Lambda=4\pi f_a$, with $f_a=1\,\mathrm{TeV}$. 

\begin{figure}[t!]
\centering
\includegraphics[width=1\linewidth]{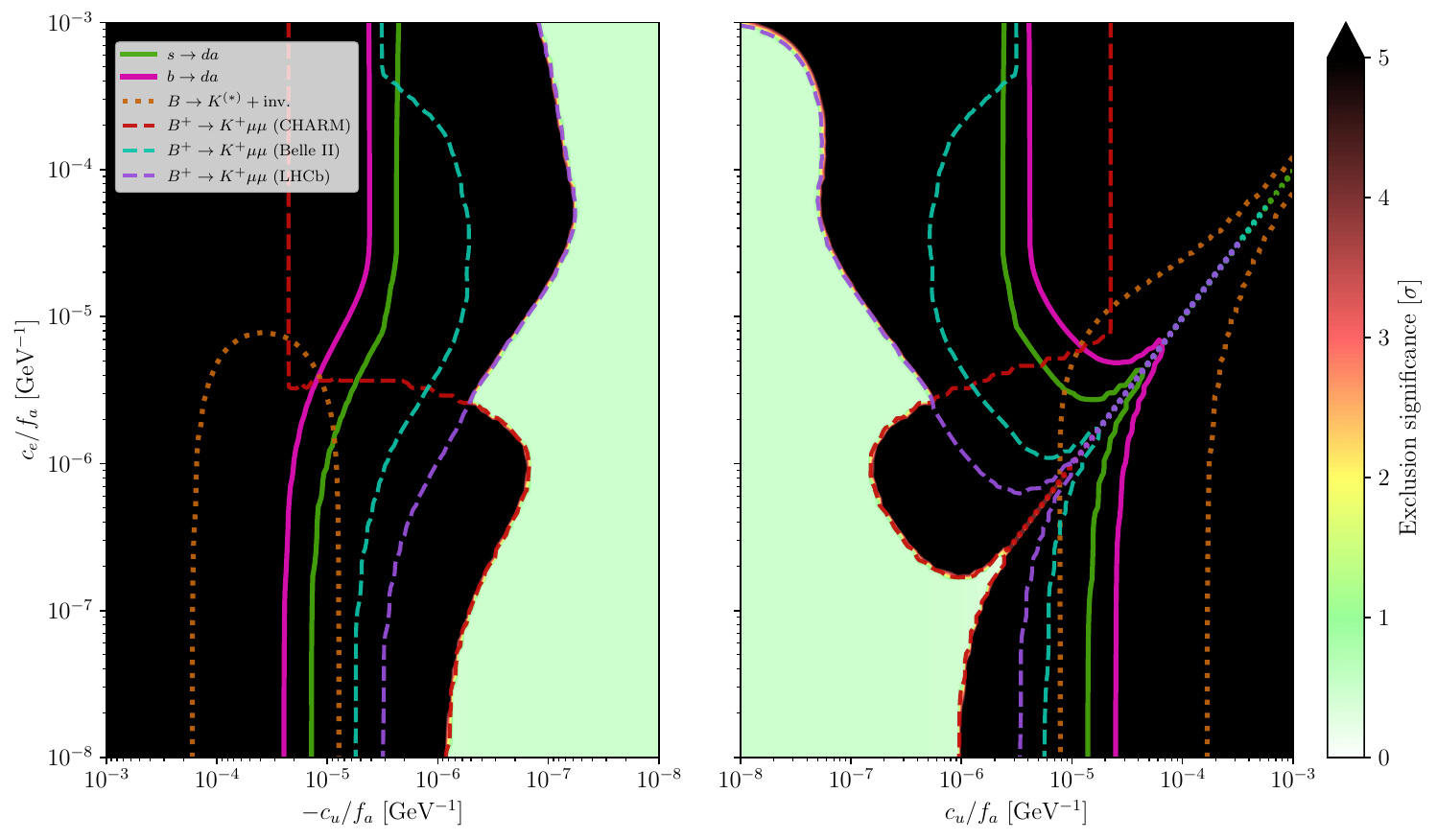}
\caption{\em Parameter space $-c_u/f_a$ vs $c_\ell/f_a$ (left) and $c_u/f_a$ vs $c_\ell/f_a$ (right) for an ALP mass 
$m_a=0.3\,\mathrm{GeV}$ and $f_a=1\,\mathrm{TeV}$. The dark region represents the portion of parameter space that is experimentally 
excluded, while the light green is the one allowed. The solid green (pink) lines represent the $2\,\sigma$ exclusion limit 
placed by processes involving $s\to d a$ ($b\to d a$) transitions, while the dotted orange line represents 
$B\to K^{(\ast)}$ processes. The dashed lines represent the $2\,\sigma$ limit of $B^+\to K^+ \mu^+ \mu^-$ measured 
in CHARM\cite{CHARM:1985anb,Dobrich:2018jyi} (red), Belle II\cite{Belle-II:2023ueh} (cyan) and LHCb\cite{LHCb:2016awg} 
(violet).}
\label{fig:exc_cevscu}
\end{figure}

\paragraph{\bf $\boldsymbol{c_u}$ vs. $\boldsymbol{c_\ell \, (c_d=0)}$}
In Fig.~\ref{fig:exc_cevscu} we show the excluded regions in the $(c_u, c_\ell)$ plane for the two possible 
independent sign choices. We can observe that $b\to s$ transitions give the strongest constraints. In particular, the bounds are dominated by the $B^+\to K^+ \mu^+ \mu^-$ decay, fact that can be understood considering that the coupling to leptons is generated at tree-level, while the quark flavour violating one at one loop. Indeed, the latter is so suppressed with respect to the leptonic coupling, that once an ALP is produced, it subsequently decays into a $\mu^-\mu^+$ pair with a displace vertex topology. We also include the exclusion limits of transitions of the type $b\to d a$ (solid pink line) and $s\to d a$ (solid green line) to show the relative strength with respect to $b\to s$ processes. The full list of observables included in each sector is detailed in Ref.~\cite{ALPACA}. The constraints are subdominant with respect to $b\to s$ due to CKM matrix and quark mass suppressions.

In the upper part of the plot in Fig.~\ref{fig:exc_cevscu} (larger values of $c_\ell/f_a$) the $B^+\to K^+ \mu^+ \mu^-$ decay by CHARM\cite{Dobrich:2018jyi,
CHARM:1985anb} gives the most stringent bounds while in the lower part of the plot (smaller values of $c_\ell/f_a$) is the LHCb \cite{LHCb:2016awg} experiment that dominates, highlighting the complementarity between these two experiments, due to having different detection lengths. Therefore, CHARM and LHCb experiments are able to put constrains on bigger (longer decay lengths) and smaller (shorter decay lengths) values of $f_a$, respectively.

In the right plot of Fig.~\ref{fig:exc_cevscu}, a flat direction can be identified along the diagonal of the parameter space, corresponding to a vanishing ALP--lepton effective coupling, see Eq.~\eqref{eq:dw_fermions}. This follows from a cancellation between the tree-level lepton coupling and the corresponding 2-loop-induced contribution: the latter arises from the ALP--top coupling, which induces a $a\gamma\gamma$ vertex, subsequent generating an $a\mu^+\mu^-$ contribution, with a the top mass enhancement. This flat direction is almost completely excluded by the invisible search of $B\to K^{(*)}\nu\bar{\nu}$.

\begin{figure}[t]
\centering
\includegraphics[width=1\linewidth]{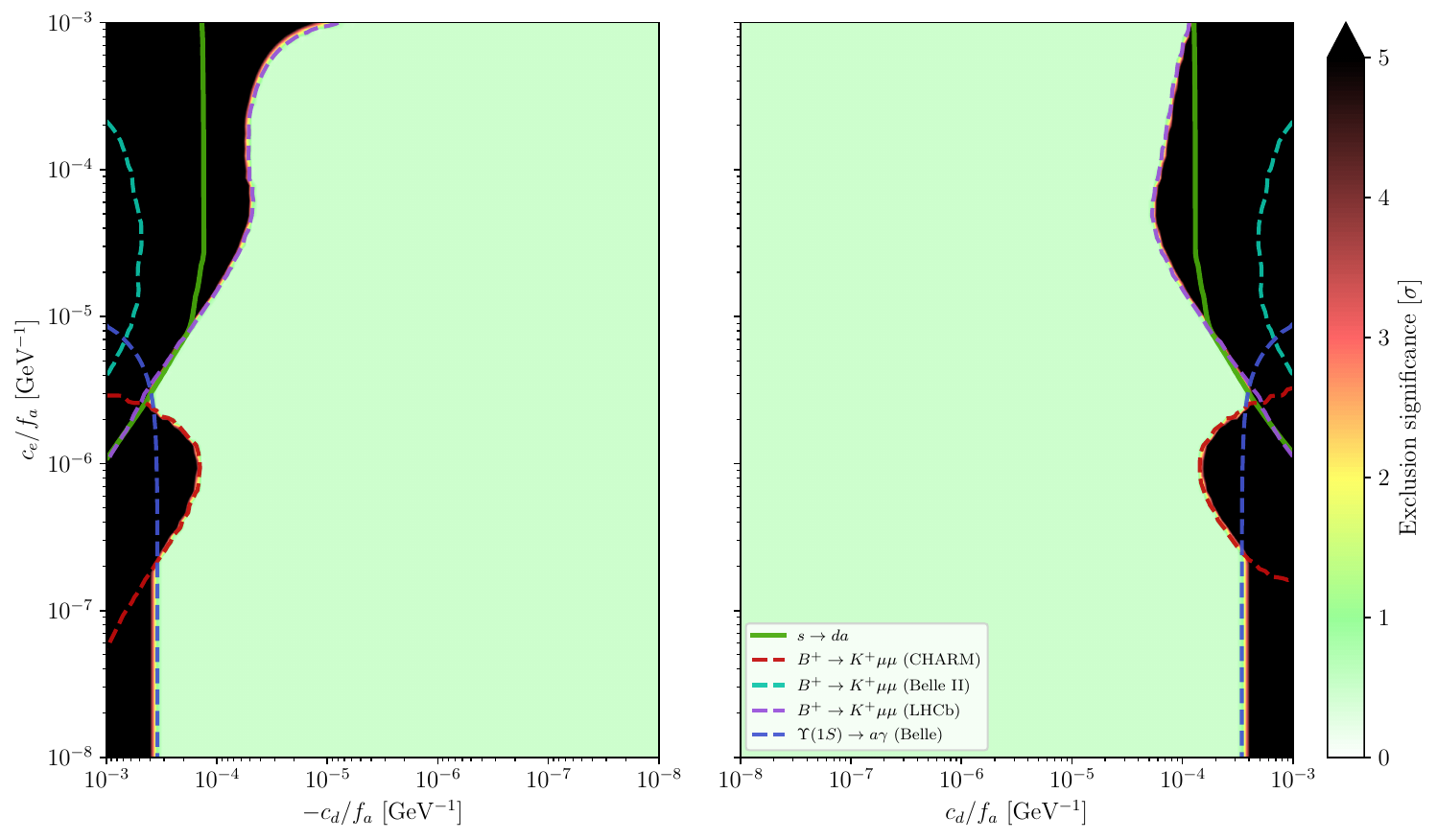}
\caption{\em Parameter space $-c_d/f_a$ vs $c_\ell/f_a$ (left) and $c_d/f_a$ vs $c_\ell/f_a$ (right) for an ALP mass $m_a=0.3\,\mathrm{GeV}$ and $f_a=1\,\mathrm{TeV}$. The dark region represents the portion of parameter space that is experimentally excluded, while the light green is the one allowed. The solid green line represents the $2\,\sigma$ exclusion limit placed by processes involving $s\to d a$ transitions. The dashed lines represent the $2\,\sigma$ limit of $B^+\to K^+ \mu^+ \mu^-$ measured in CHARM\cite{CHARM:1985anb,Dobrich:2018jyi} (red), Belle II\cite{Belle-II:2023ueh} (cyan) and LHCb\cite{LHCb:2016awg} (violet), and $\Upsilon(1S)\to \gamma a$ measured in Belle~\cite{Belle:2018pzt} (dark blue).}
\label{fig:exc_cevscd}
\end{figure}

\paragraph{\bf $\boldsymbol{c_d}$ vs. $\boldsymbol{c_\ell \, (c_u=0)}$}
To continue, we focus on the relations between the down-quark sector and the leptons, setting $c_u=0$. The exclusion plots obtained, again assuming the running of the couplings at a scale $\Lambda=4\pi f_a$, with $f_a=1\,\mathrm{TeV}$ can be seen in Fig. \ref{fig:exc_cevscd}. In contrast to the previous case, the $s \to d$ sector is now comparable to the $b\to s$ one, in the region of large lepton couplings, where the $K^+\to \pi^+ \mu\mu$ channel~\cite{NA482:2016sfh} is comparable to the 
LHCb $B^+\to K^+ a (\to\mu^+ \mu^-)$ decay \cite{LHCb:2016awg}. The complementarity between different experiments can also be seen here, with the $\Upsilon(1S)\to \gamma a$ decay searches at Belle~\cite{Belle:2018pzt} bounding the smaller lepton couplings region. It is important to highlight the huge difference between the bounds obtained for $c_d/f_a$ and the one of $c_u/f_a$ shown in the previous plot. The main reason is the fact that the top-coupling, that 
typically gives the dominant contribution in the down sector FCNC, is here generated only by loop effects, thus 
explaining the three order of magnitude difference. 

\begin{figure}[t]
\centering
\includegraphics[width=1\linewidth]{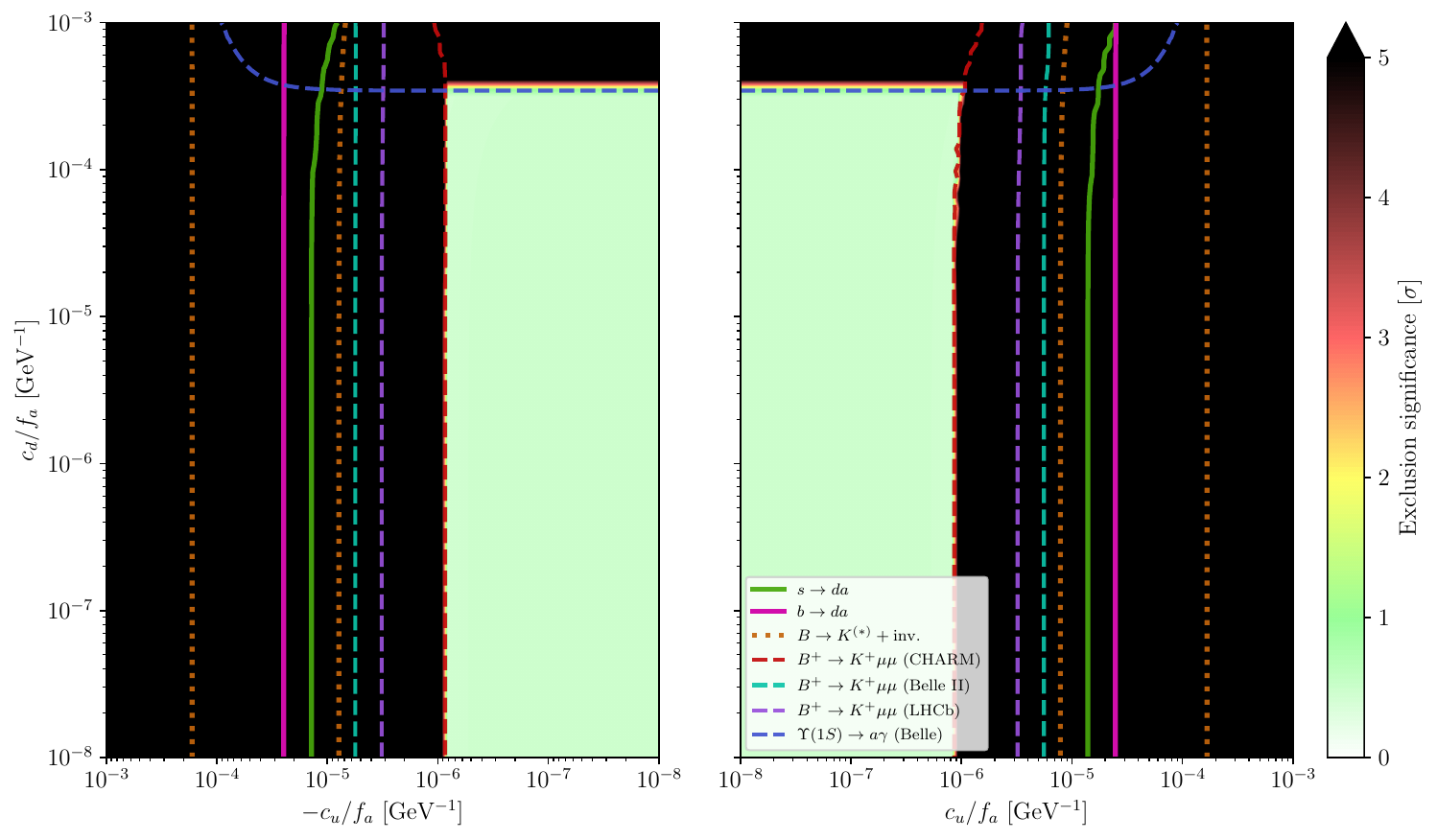}
\caption{\em Parameter space $-c_u/f_a$ vs $c_d/f_a$ (left) and $c_u/f_a$ vs $c_d/f_a$ (right) for an ALP mass 
$m_a=0.3\,\mathrm{GeV}$ and $f_a=1\,\mathrm{TeV}$. The dark region represents the portion of parameter space that is experimentally 
excluded, while the light green is the one allowed. The solid green (pink) lines represent the $2\,\sigma$ exclusion limit 
placed by processes involving $s\to d a$ ($b\to d a$) transitions, while the dotted orange line represents $B\to K^{(\ast)}$ 
processes. The dashed red lines represent the $2\,\sigma$ limit of $B^+\to K^+ \mu^+ \mu^-$ measured by CHARM 
\cite{CHARM:1985anb,Dobrich:2018jyi} (red), Belle II\cite{Belle-II:2023ueh} (cyan) and LHCb\cite{LHCb:2016awg} (violet), 
and $\Upsilon(1S)\to \gamma a$ measured in Belle~\cite{Belle:2018pzt} (dark blue).}
\label{fig:exc_cdvscu}
\end{figure}

\paragraph{\bf $\boldsymbol{c_u}$ vs. $\boldsymbol{c_d \, (c_\ell=0)}$}
Finally, we consider the interplay between the up- and down-sector quarks by setting $c_\ell=0$, obtaining the exclusion plots in Fig. \ref{fig:exc_cdvscu}. Once again, the most constraining process for the chosen values of $(m_a, f_a)$ is the $B^+\to K^+ \mu^+ \mu^-$ decay by CHARM~\cite{Dobrich:2018jyi, CHARM:1985anb}, even though in this scenario there are no tree level coupling to leptons. In this case only one experiment is needed to constrain the parameter space, as in this scenario the lifetime of the ALP is longer, and CHARM experiment had more sensitivity to longer lifetimes. For larger values of $c_d/f_a$, the quarkonia process $\Upsilon(1S)\to \gamma a$~\cite{Belle:2018pzt} constrains the parameter space.\\

As we have seen in the previous three examples, for the chosen value of $m_a=0.3\GeV$, and assuming flavour universal ALP 
fermion couplings, the majority of the bounds come from the $b\to s$ sector, with only subdominant contributions from the 
$s\to d$ and $b\to d$ observables. No relevant constraints come, instead, from $D$ decays, as up-quarks FCNC transitions 
do not benefit from the large top mass enhancement. While a better experimental precision from the $D$ sector would not 
heavily alter our findings here, it could have, however, a noticeable impact in specific benchmark models, like for example 
a charm-philic ALP, see for instance Ref.~\cite{Carmona:2021seb}. For this reason, it is important that the experimental effort done in charm physics continues. 

Although we are not plotting all possible masses, these cases are instructive enough to see that complementary searches 
are necessary to constrain all the available EFT parameter space. The same logic would apply to different ALP mass, 
keeping in mind, however, as a general guideline, that going to smaller masses invisible channels will start to dominate, 
as less decay channels are available while moving to larger ALP masses prompt or displaced searches will become dominant.

\subsection{The Belle II anomaly: a case study}
\label{sec:Belle2}

In the previous subsections we have discussed the two main alternative ways of analysing the ALP phenomenology, both 
considering explicit UV realisations and the EFT description, for an ALP mass in the range $[0.01\,, 10]\GeV$ 
and obtained constraints on the relevant coefficients by considering several flavour observables, implemented 
in \alpaca with the latest experimental results. 

We now focus on a particularly interesting case of study, that is the recent Belle II $2.8\,\sigma$ excess in the 
$B^+\rightarrow K^+ \nu \overline{\nu}$ decay\cite{Belle-II:2023esi}. This excess is localised in few bins of the 
$q^2$ distribution, and although a specific optimised analysis for two body decay would shed some light on the possible 
interpretation of this anomaly, Ref.~\cite{Altmannshofer:2023hkn} showed that it may be compatible with an ALP with 
$m_a=2\,$GeV. The authors made their claim combining Belle~II and BaBar data and assuming the following non-universal 
Vector and Axial ALP--quark couplings at the low energy scale $\mu_b \sim m_b$:
\begin{equation}
\label{eq:Belle2_lagrangian}
\mathcal{L}^\mathrm{LE}_\mathrm{ALP}(\mu_b) \subset \frac{\partial_\mu a}{f_a}\left(c_{sb}^V\ov{s}\gamma^\mu b +
          c_{sb}^A\ov{s}\gamma^\mu \gamma_5 {b}\right) +\hc \, .
\end{equation}

Retaining only the ALP--quark couplings in Eq.~\eqref{eq:Belle2_lagrangian} leads to a long-lived ALP that decays 
outside the Belle II and BaBar detectors, thus allowing to combine the two measurements directly. 
However, from a more general point of view, the Lagrangian in Eq.~\eqref{eq:Belle2_lagrangian} does not capture 
all the low-energy coefficients of the operators of the EFT Lagrangian in Eq.~\eqref{eq:UVLag}, that are inevitably 
generated at higher order by the running and matching procedures. Therefore, when considering UV models or 
EFT-benchmarks at high energies, with kinematically allowed decays (which are to be expected also in this case 
as $m_a\sim 2\GeV$), the finite lifetime of the ALP always needs to be taken into account in combining different 
measurements. Even more so when experiments have different setups, and hence, what is invisible for one detector 
may be visible for another.

\begin{figure}[t!]
\centering
\includegraphics[width=0.8\linewidth]{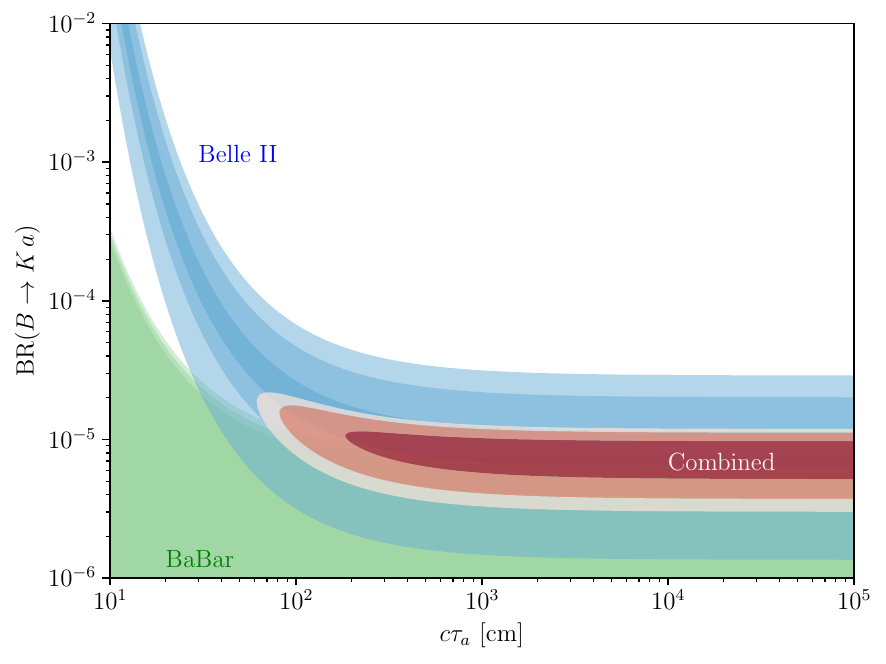}
\caption{\em Fit to $B^+\to K^+ a$ from Belle II~\cite{Belle-II:2023esi} and BaBar~\cite{BaBar:2013npw} for $m_a = 2\,\mathrm{GeV}$. 
The region shaded in blue is allowed by Belle II, in green by BaBar and in red by the combined fit. Darker colours correspond 
to $1\,\sigma$, medium colours to $2\,\sigma$ and lighter colours to $3\,\sigma$ significances. }
\label{fig:combined_belle2_babar}
\end{figure}

In the case at hand, the combination of the Belle II~\cite{Belle-II:2023esi} and BaBar~\cite{BaBar:2013npw}  measurements of $\mathrm{BR}(B^+\to K^+\,\nu 
\bar{\nu})$  can be performed using the procedure described in Section~\ref{sec:signatures}, Eqs.~\eqref{eq:NWA}
-\eqref{eq:prompt_inv}. After fixing the experimental setups, the two free parameters are the branching ratio 
$\mathrm{BR}(B^+\to K^+\,a)$ and the ALP lifetime $\tau_a$. We use the results from Ref.~\cite{Altmannshofer:2023hkn} but we extend their analysis taking into consideration the finite ALP lifetime. The result of combining BaBar and Belle II measurements in this two-dimensional parameter space can be seen in Fig.~\ref{fig:combined_belle2_babar}, where for definiteness we have fixed 
$m_a = 2\,\mathrm{GeV}$. Since the characteristic length of the detectors and the boost factor of the $B$ are different, the branching ratio must be weighted with the probability that the ALP is long-lived, $P_\textrm{inv}(\tau_a)$, and it follows that the two data sets can constrain quite different values of $\mathrm{BR}(B^+\to K^+\,a)$ in the small lifetimes region. At longer lifetimes, instead, $P_\textrm{inv}(\tau_a) \longrightarrow 1$ and Belle II and BaBar data point to regions of the parameter space that partially 
overlap. All in all, the result of the combined fit, depicted in red, allows us to set a lower bound on the ALP proper decay length 
$c\tau_a \gtrsim 80\,\mathrm{cm}$ (95\% C.L.). 

\begin{figure}[t!]
\centering
\includegraphics[width=\textwidth]{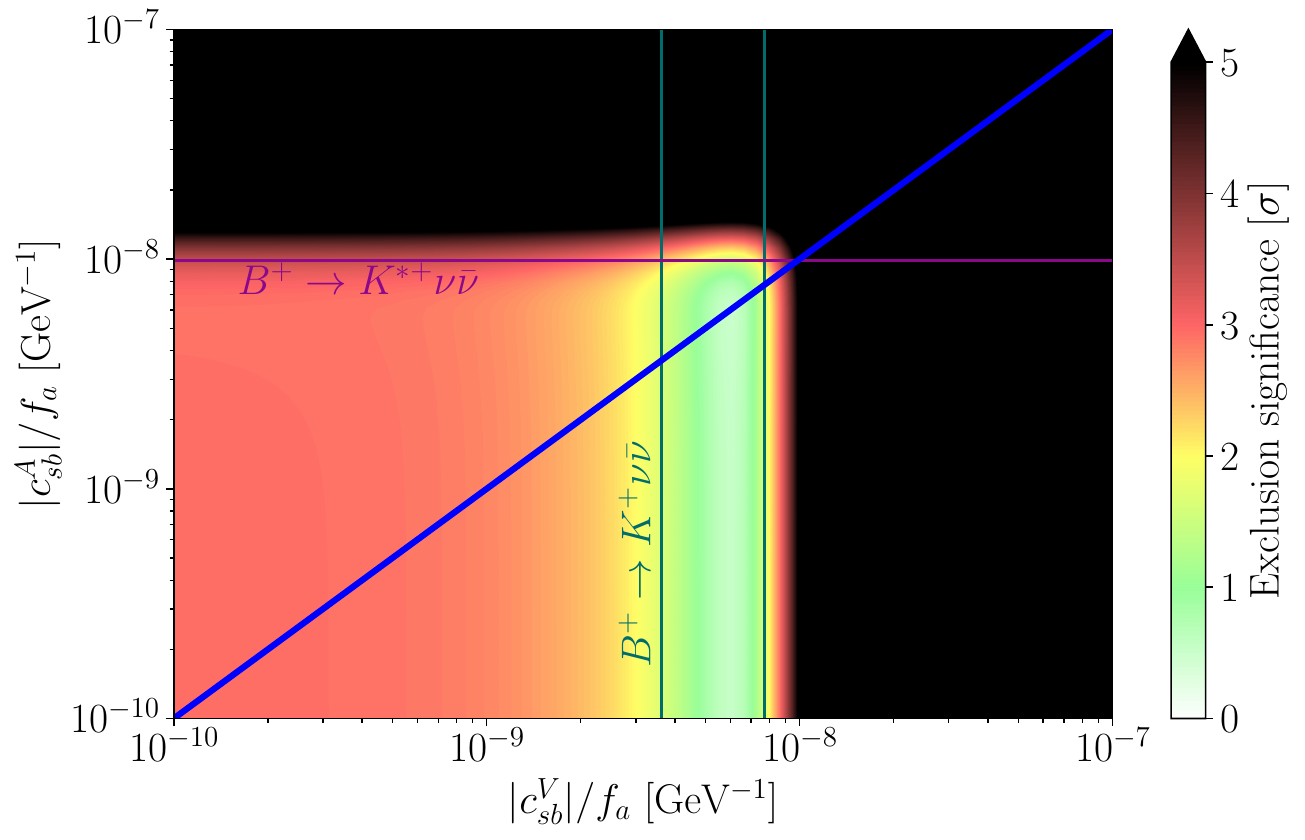}
\caption{\em Preferred region of parameter space $c_{sb}^V/f_a$ {\it vs.} $c_{sb}^A/f_a$ to explain the Belle II anomaly. The line in blue highlights where the couplings generated by the benchmarks lie. The mass of the ALP is set to $m_a=2\,\mathrm{GeV}$.}
\label{fig:belle-eft-cv-ca}
\end{figure}

\begin{figure}[th!]
\centering
\includegraphics[width=\textwidth]{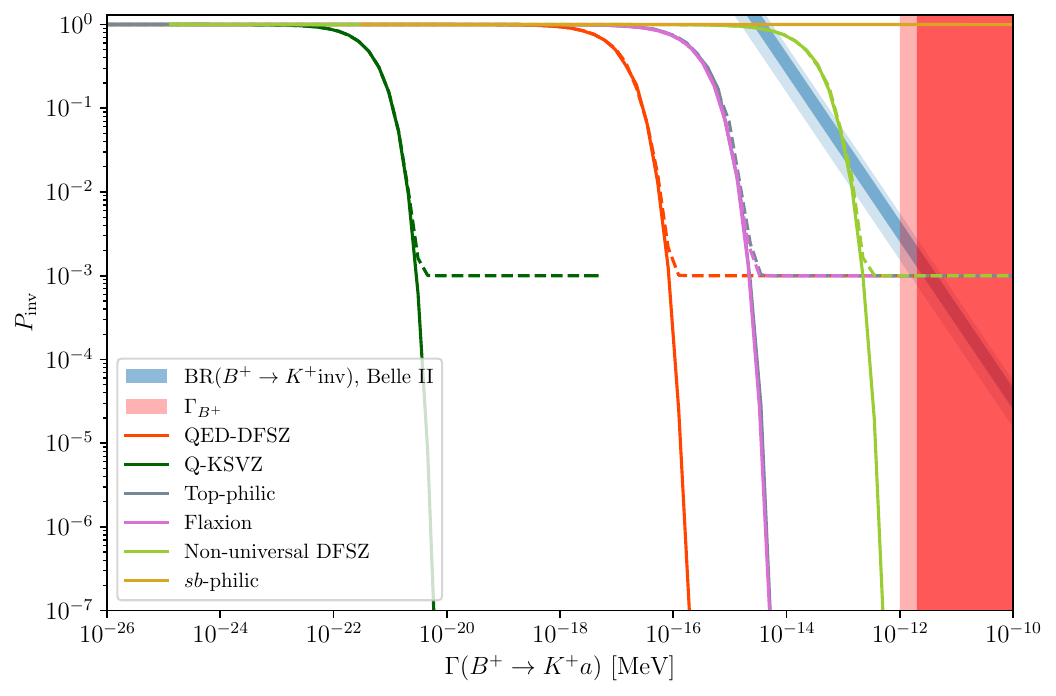}
\caption{\em Probability of invisible decay as a function of the decay width of $B^+\to K^+ a$ for the four models studied for a branching ratio to a dark sector of $0$ 
$(\, 10^{-3})$ for the continuous (dashed) lines. We set $m_a=2\,\mathrm{GeV}$. The blue-shaded region represents the branching ratio obtained experimentally at Belle II, while the red-shaded region is excluded by the decay width of $B^+$.}
\label{fig:belle-pinv}
\end{figure}

We now follow the same procedure as in the previous sections, attempting to explain this anomaly first in terms of 
an EFT description and then via a specific UV model. In Fig.~\ref{fig:belle-eft-cv-ca}, we plot the preferred parameter 
space in terms of the low-energy EFT Lagrangian of Eq.~\eqref{eq:Belle2_lagrangian}, where the measurement of 
$B^+\rightarrow K^{\ast +} \nu \ov{\nu}$ \cite{BaBar:2013npw} places a direct constraint on the axial coupling, 
$|c_{sb}^A|/f_a \lesssim 1\times10^{-8}$\GeV$^{-1}$, while the combined measurements of $B^+\to K^+\nu\bar{\nu}$ 
point to a value for the vector coupling of $3\times 10^{-9}\,\mathrm{GeV}^{-1} \lesssim |c_{sb}^V|/f_a \lesssim 8 
\times10^{-8}\,\mathrm{GeV}^{-1}$ in the $2\,\sigma$ region. The values for the couplings agree with the ones reported 
in Ref.~\cite{Altmannshofer:2023hkn}, thus being a cross-check of our results. 

In the same plot in Fig.~\ref{fig:belle-eft-cv-ca}, we also show that the predictions of the different models lie along the diagonal, that is the blue line. For the models with no flavour-violating couplings at tree-level, such as QED-DFSZ, Q-KSVZ and top-philic, this follows from the fact that flavour violation is loop generated in the LH quark sector, thus implying $c^V_{sb}=-c^A_{sb}$. For the non-universal DFSZ benchmark, Eqs.~\eqref{eq:Lyuk2HDM}-\eqref{eq:couplingsM1}, where the RH fermions have universal couplings, while those of the LH down-type quarks are diagonal but non-universal, flavour violation  is generated at tree-level through the rotations into the mass basis (Eq.~\eqref{PrimedCouplingsPhysicalBasis}), but the relation $c^V_{sb} = - c_{sb}^A$ still holds. 
For the benchmark flaxion model considered, Eq.\eqref{eq:flaxion_charges}, this relation is a numerical accident. As can be seen, these models can easily generate the flavour-violating couplings at low-energy necessary to describe the Belle II anomaly. 

However, this is not enough to explain an excess corresponding to an ``invisible'' channel of an ALP. Apart from a suitable flavour-violating coupling, the ALP needs to be long-lived enough to escape the detector. 

To emphasise this point, we plot in Fig.~\ref{fig:belle-pinv} the partial width $\Gamma(B^+ \to K^+\,a)$ and $P_\text{inv}$, running implicitly as a function of $f_a$, $f_a \in [100, 10^{10}]\,\mathrm{GeV}$. The region that would explain Belle II anomaly is the blue-shaded one. The red-shaded area is excluded by the requirement that $\Gamma(B^+\to K^+ a)$ must be smaller than the total decay width of the $B$ meson~\cite{ParticleDataGroup:2024cfk}.
As we can see, the only benchmark model that has an overlap with this region is the non-universal DFSZ with $\tan\beta \gtrsim 10$ (solid light-green line). On the contrary, the QED-DFSZ, Q-KSVZ and top-philic models fail as the loop-induced coupling requires a scale $f_a$ incompatible with an invisible ALP. Moreover, in the flaxion model studied here, flavour violation is slightly suppressed due to Cabibbo factors, making again the scale too small for the ALP to be long-lived. Notice that it is possible to reconcile these models with the Belle II measurement, assuming that the ALP can decay into a Dark Sector (DS) with $\mathrm{BR}(a\to \mathrm{DS})\gtrsim \mathcal{O}(10^{-3})$ (dashed lines).

Finally, the solid orange line represents the prediction of a specific benchmark EFT scenario, labelled $sb$-philic, where the only non vanishing coupling in Eq.~\eqref{eq:UVLag} is $(c_d)_{23}(\Lambda)$, that is the flavour violating parameter in the $sb$ sector. In this scenario, the ALP is long-lived in both BaBar and Belle II detectors, as the dominant decay $a\to\eta\pi\pi$ is very much suppressed, receiving contributions at one EW loop level.

\begin{figure}[t]
    \centering
    \includegraphics[width=0.98\linewidth]{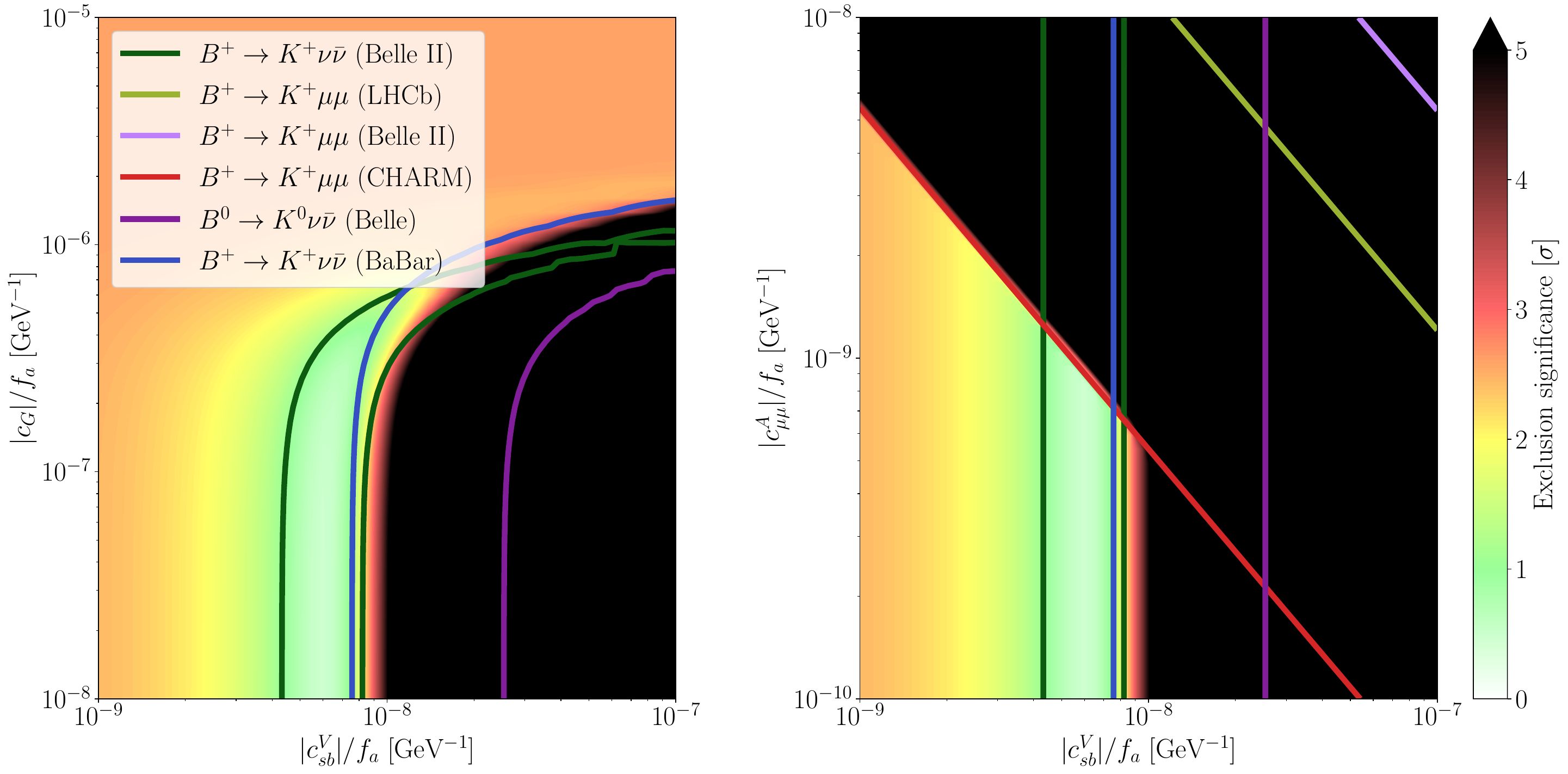}
    \caption{\em Bounds on the low energy couplings $c_G$ (left) and $c^A_{\mu\mu}$ with a generic vector coupling $c^V_{sb}$, for $m_a=2\,\mathrm{GeV}$, showing the upper limit needed to explain the Belle II anomaly with different additional couplings.}
    \label{fig:Belle_cg_cmu_csbV}
\end{figure}

From these simplified benchmark models we learn that if we see a missing energy signal compatible with an ALP, the model must either exhibit flavour violation relatively unsuppressed, or certain channels must be cancelled at low energies to make the particle invisible. In many cases, both conditions are necessary. About the last point we can generalise this argument further in the EFT framework, studying to what degree one must cancel the different ALP decay channels. In particular, muon and hadronic channels, as discussed in Section~\ref{Sect:ExpSigOnShellAlps} and specifically in Figs.~\ref{fig:ProbDecaysBelle2} and \ref{fig:regions_BKa_BelleII}, need to be tamed if one wants to obtain an invisible ALP with masses of $m_a=2\GeV$. In Fig.~\ref{fig:Belle_cg_cmu_csbV} we see how a gluon coupling (left) at low energies must satisfy $|c_G|/f_a\lesssim 6\times10^{-7}\,\GeV^{-1}$ such that the ALP is long-lived enough. Similar bound applies to $|c_{u}|/f_a$. On the other hand, in the case of the muon-coupling (right), experiments with longer detection length such as CHARM~\cite{CHARM:1985anb,Dobrich:2018jyi} could observe the ALP decay into muons, leading to a stronger constraint of $|c_{\mu
\mu}^A|/f_a \lesssim 10^{-9}\,\mathrm{GeV}^{-1}$. Other couplings at low energies, such as the photon coupling, do not impose a relevant constraint on the scale. 

\begin{figure}[t!]
    \centering
    \includegraphics[width=1\linewidth]{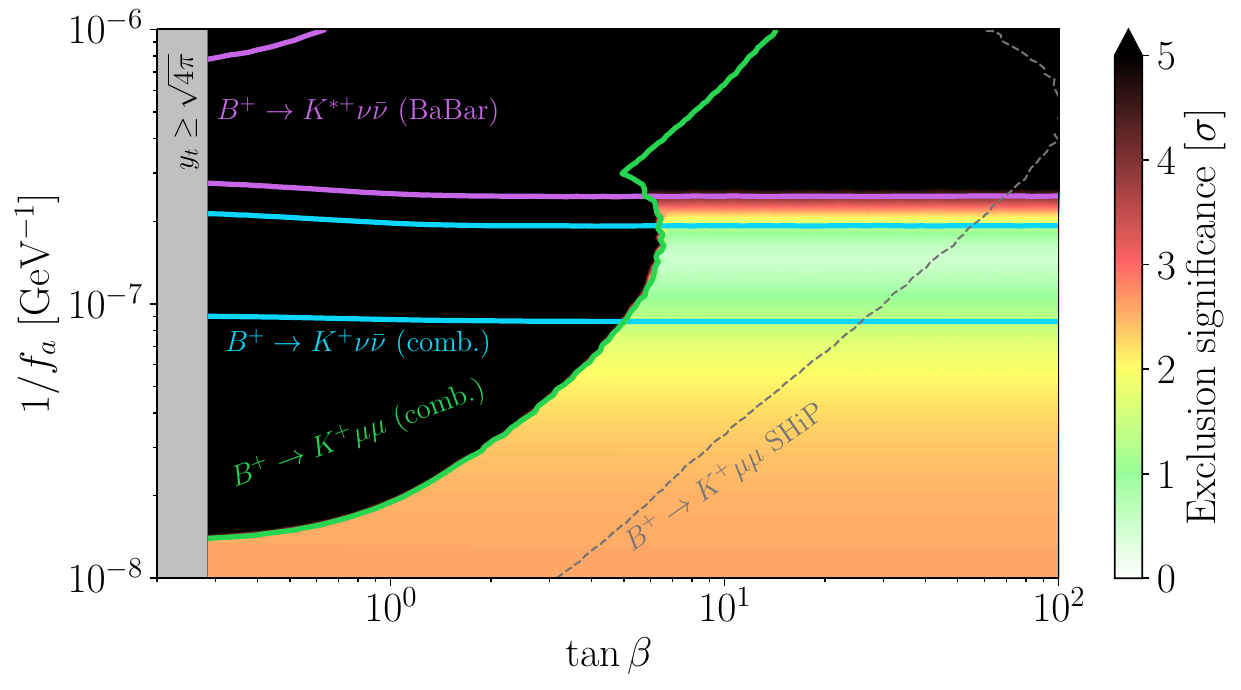}
    \caption{\em Parameter space of the non-universal DFSZ model presented in Section~\ref{sec:models}, including invisible $B^+\to K^+ \nu\bar{\nu}$  (light-blue) and $B^+\to K^{*+} \nu\bar{\nu}$ (pink) $2\,\sigma$ contours and $2\,\sigma$ bounds from the visible decay channel $B^+\to K^+ \mu\mu$ (green). The light green shades correspond to the global fit regions where it this model is allowed. }
    \label{fig:2HDM_fit}
\end{figure}

The non-universal DFSZ model presented in Section~\ref{sec:models}, with the couplings in the UV defined in Eq.~\eqref{eq:couplingsM1}, has flavour violation at tree-level generated by non-universal structure of $c_{q}$ coupling and fulfils the long-lived condition. The lepton coupling still depends on $\tan\beta$, which can take any value allowed by perturbativity of the $t$ and $b$ Yukawa couplings, that is $ \tan\beta \in [0.28, \, 146]$. In Fig.~\ref{fig:2HDM_fit} we scan over the values of $f_a$ and $\tan\beta$, showing in light blue the combined  $2\,\sigma$ regions of Belle II (reinterpreted) \cite{Belle-II:2023esi,Altmannshofer:2023hkn} and BaBar \cite{BaBar:2013npw} $B^+\to K^+ \nu\bar{\nu}$ searches respectively, while in pink we plot the bound set by BaBar on $B^+\to K^{*+} \nu\bar{\nu}$ \cite{BaBar:2013npw,Altmannshofer:2023hkn}. The green line correspond to the $2\,\sigma$ limit of CHARM~\cite{CHARM:1985anb,Dobrich:2018jyi} and LCHb \cite{LHCb:2016awg}, which cut the regions allowed by the combined fit of $B^+\to K^+ \nu\bar{\nu}$, constraining the model to larger $\tan\beta$, if we want to explain the Belle II anomaly. The light green region corresponds to the global fit performed taking all these observables into account, showing how there exists a region of parameters, consistent with perturbativity limits, which is not ruled out by any present experiment, and could be further tested by the SHiP experiment \cite{Dobrich:2018jyi}, shown in grey. This light green area contains the small region in Fig.~\ref{fig:belle-pinv} characterised by $f_a \sim 10^7\,\mathrm{GeV}$ and $P_\mathrm{inv}\approx 1$ where the model line crosses the blue-shaded band; the other intersection corresponding to $f_a \sim 2\times 10^5\,\mathrm{GeV}$ and $P_\mathrm{inv}\approx 10^{-2}$ is experimentally excluded.

The reason why this model works is just a combination of the results obtained in the analysis of this section:\\
i) The model exhibits flavour violation at tree level (with some suppression from the CKM~$\sim\lambda^2$), making the scale compatible with displaced-invisible ALP-searches. This can also be seen in Fig.~\ref{fig:Belle_cg_cmu_csbV} (left), where the constraint on the gluon coupling is automatically satisfied.\\
ii) We can fix the values of $\tan\beta$ to suppress the branching ratio $\mathrm{BR}(a\to \mu\mu)$ evading experimental constraints, see Fig.~\ref{fig:2HDM_fit}. 

As a final comment, this model-building exercise is important on its own, as it shows that models unavoidably require suppression on the most relevant decay channels in order to guarantee an ``invisible-ALP'' signal at masses $m_a\gtrsim\mathcal{O}(\text{GeV})$.

\section{Conclusions}
\label{sec:Conclusions}
In this work, we have provided a thorough and systematic study of ALPs with a $\mathcal{O}(\mathrm{GeV})$ mass, combining a robust theoretical framework with the latest experimental constraints. ALPs can be considered as effective descriptions of axions solving the Strong CP problem and are nowadays intensively searched for at experimental facilities. We focus on, but not limit ourselves to, meson phenomenology and provide a new computing tool to the community, the \alpaca code, to enable a comprehensive and flexible exploration of the GeV-ALP parameter space, including both EFT descriptions and specific ultraviolet completions. This approach allows for precise predictions of ALP-induced observables across various energy scales, taking into account the effects of renormalisation group evolution and matching conditions at different thresholds, particularly from the UV down to the non-perturbative QCD regime, where $\chi$PT is the correct description.

We structured our analysis around several well-motivated benchmark scenarios, including flavour-universal and flavour non-universal models such as DFSZ-like, KSVZ-like, top-philic, and flaxion frameworks. Each of these models illustrates distinct phenomenological signatures, highlighting the importance of model-dependent studies alongside model-independent EFT analyses. Our work emphasises that flavour observables, especially rare decays of $B$ and $K$ mesons, are powerful tools for constraining ALP couplings, particularly in the mass window below $\lesssim10\GeV$, where the reach of colliders is limited.

A central feature of our study is the consistent treatment of ALP interactions in the non-perturbative regime using $\chi$PT. We carefully included ALP--meson mixing and accounted for decay widths into hadrons, photons, and fermions. Furthermore, we addressed a sometimes overlooked aspect: the classification of ALP decay signatures as prompt, displaced, or invisible on an experiment-by-experiment basis. This has crucial implications for interpreting bounds and guiding future searches.

Finally, we applied our framework to the intriguing Belle II anomaly in the $B^+\to K^+\nu\ov\nu$ channel, showing under which conditions an ALP can account for the observed excess. 

While no conclusive evidence for ALPs is currently available, our results demonstrate that current and near-future experiments, especially Belle II and NA62, will significantly narrow the viable ALP parameter space. In this context, \alpaca provides a versatile tool for ongoing and future analyses, and we expect it to play a central role in the global effort to test the properties of ALPs. 

On the other hand, despite considerable efforts to improve our understanding of hadronic ALP decays, significant challenges remain.
For instance, for ALP masses above  $m_a = 2\GeV$, the individual branching ratios of the various hadronic decay channels are still poorly known.
From an experimental perspective, channels involving hadronic final states are particularly relevant for constraining models in which the ALP predominantly couples to $aG\widetilde{G}$.
In such cases, improved knowledge of the inclusive $a \to \mathrm{hadrons}$ decay rate would lead to stronger bounds on ALPs in the higher-mass regime.
Similarly, dedicated searches for $a \to 3\pi$ and $a \to 2\pi \eta\,(\gamma)$ would enhance our understanding of this scenario.

As theoretical analyses of ALPs become more precise, it is crucial that also experimental results achieve higher precision, not underestimating the relevance of presenting results in a form that can be interpreted in a model-independent way. This is preferable to providing constraints solely within the context of simplified models. In addition, it would be extremely useful to have publicly available results of the analysis performed by the experimental collaborations (for example, the upper limit bounds of the branching ratios as functions of the mass and lifetime of the particle) in machine-readable formats that can be easily incorporated into \alpaca\!\!.

\section*{Note Added}
During the final stages of this work, a new set of bounds for the processes $K^+\to\pi^+a(\to\mathrm{inv.},\ \gamma\gamma,\ \mu^+\mu^-)$ as a function of the lifetime of the ALP was reported by the NA62 collaboration~\cite{NA62:2025upx}, obtaining the strongest constraints for couplings to gluons and to fermions assuming universality, for $m_a < 250\,\mathrm{MeV}$. These bounds will be added in future versions of {{\fontfamily{cmss}\selectfont ALPaca}}.

\section*{Acknowledgements}
We thank Maksym Ovchynnikov and Maria Ramos for useful discussions. JA kindly thanks the hospitality of Universidad Aut\'onoma de Madrid/IFT during the completion of this work. MFZ kindly thanks the University of Padova for the hospitality and the COST Action COSMIC WISPers (CA21106) for financial support during the early stages of this work.

JA has received funding from the Fundaci\'on Ram\'on Areces ``Beca para ampliaci\'on de estudios en el extranjero en el campo de las Ciencias de la Vida y de la Materia'', and acknowledges support by the grants PID2021-126078NB-C21 funded by MCIN/\hspace{0pt}AEI/\hspace{0pt}10.13039/\hspace{0pt}501100011033 and ``ERDF A way of making Europe''.  The work of MFZ is supported by the Spanish MIU through the National Program FPU (grant number FPU22/\hspace{0pt}03625). XPD has partially received funding from the Swiss National Science Foundation (SNF) through the Eccellenza Professorial Fellowship ``Flavor Physics at the High Energy Frontier,'' project number 186866.
We acknowledge partial financial support by the European Union's Horizon 2020 research and innovation programme under the Marie Sk\l odowska-Curie grant agreement No.~101086085-ASYMMETRY and by the Spanish Research Agency (Agencia Estatal de Investigaci\'on) through the grant IFT Centro de Excelencia Severo Ochoa No CEX2020-001007-S and by the grant PID2022-137127NB-I00 funded by MCIN/\hspace{0pt}AEI/\hspace{0pt}10.13039/501100011033.
This article is based upon work from COST Action COSMIC WISPers CA21106, supported by COST (European Cooperation in Science and Technology).


\appendix

\section{Running and matching of the ALP couplings}
\label{app:Running}
The running of the ALP parameter is governed by the Renormalisation Group Equations, that can be written as
\begin{equation}
    \frac{d}{d \log(\mu)} c_i(\mu) = \frac{1}{16\pi^2}\gamma_{ij}(\mu) c_j(\mu) = \frac{1}{16\pi^2}\beta_i(\mu)\,.
\end{equation}

The anomalous dimension matrix $\gamma_{ij}$ depends on the energy scale through the SM parameters. The running of the gauge constants $\alpha_s(\mu)$, $\alpha_1(\mu)$ and $\alpha_2(\mu)$ is affected by the ALPs, although this effect is suppressed by $(m_a/f_a)^2$, and consequently we will ignore it. In the following we reproduce the beta functions found in Ref.~\cite{Bauer:2020jbp}.

\subsection{Running from the UV to the EW scale}
The running of the ALP couplings in the derivative basis above the electroweak scale is given by the following beta functions at one loop
\begin{align}
    \beta_{q_L} &= \frac{1}{2}\{Y_uY_u^\dagger + Y_dY_d^\dagger,\,c_{q_L}\}-Y_u c_{u_R} Y_u^\dagger - Y_d c_{d_R} Y_d^\dagger - \left(16\alpha_s^2 \tilde{c}_G + 9\alpha_2^2 \tilde{c}_W+\frac{1}{3}\alpha_1^2 \tilde{c}_B\right) \mathbb{1}\,,\nonumber\\
    \beta_{u_R} &= \{Y_u^\dagger Y_u,\,c_{u_R}\} -2 Y_u^\dagger c_{q_L} Y_u +\left(-2X + 16\alpha_s^2 \tilde{c}_G+\frac{16}{3}\alpha_1^2 \tilde{c}_B\right)\mathbb{1}\,, \nonumber\\
    \beta_{d_R} &= \{Y_d^\dagger Y_d,\,c_{d_R}\} -2 Y_d^\dagger c_{q_L} Y_d +\left(2X + 16\alpha_s^2 \tilde{c}_G+\frac{4}{3}\alpha_1^2 \tilde{c}_B\right)\mathbb{1}\,, \nonumber\\
    \beta_{\ell_L} &= \frac{1}{2}\{Y_eY_e^\dagger,\,c_{\ell_L}\} - Y_e c_{e_R} Y_e^\dagger - \left(9\alpha_2^2\tilde{c}_W+3\alpha_1^2 \tilde{c}_B\right)\mathbb{1}\,,\nonumber\\
    \beta_{e_R} &= \{Y_e^\dagger Y_e,\,c_{e_R}\}- 2Y_e^\dagger c_{\ell_L} Y_e + \left(2X+12\alpha_1^2\tilde{c}_B\right)\mathbb{1}\,,\nonumber \\
    \beta_G &=\beta_W = \beta_B = 0\,,
\end{align}
with
\begin{equation}
    X = \Tr[3 c_{q}(Y_u Y_u^\dagger-Y_dY_d^\dagger)-3c_{u}Y_u^\dagger Y_u +3c_{d}Y_d^\dagger Y_d -c_{\ell} Y_eY_e^\dagger +c_{e}Y_e^\dagger Y_e ]\,.
\end{equation}

In the chirality-flipping basis, the beta functions of the ALP couplings to fermions are
\begin{equation}
\begin{split}
    \tilde{\beta}_u =& 2\tilde{c}_u Y_u^\dagger Y_u + \frac{5}{2}Y_u Y_u^\dagger \tilde{c}_u-\frac{3}{2}Y_d Y_d^\dagger \tilde{c}_u-2 Y_d \tilde{c}_d^\dagger Y_u-\tilde{c}_d Y_d^\dagger Y_u \\&- \tilde{c}_u \left(32\pi \alpha_s + 9\pi \alpha_2 + \frac{17}{3}\pi\alpha_1 -T\right)+ Y_u \left(-2X+32\alpha_s^2 \tilde{c}_G+9\alpha_2^2\tilde{c}_W+\frac{17}{3}\alpha_1^2\tilde{c}_B\right)\,,\\
    \tilde{\beta}_d =& 2\tilde{c}_d Y_d^\dagger Y_d + \frac{5}{2}Y_d Y_d^\dagger \tilde{c}_d-\frac{3}{2}Y_u Y_u^\dagger \tilde{c}_d-2 Y_u \tilde{c}_u^\dagger Y_d-\tilde{c}_u Y_u^\dagger Y_d \\&- \tilde{c}_d \left(32\pi \alpha_s + 9\pi \alpha_2 + \frac{5}{3}\pi\alpha_1 -T\right)+ Y_d \left(2X+32\alpha_s^2 \tilde{c}_G+9\alpha_2^2\tilde{c}_W+\frac{5}{3}\alpha_1^2\tilde{c}_B\right)\,,\\
    \tilde{\beta}_e =& 2\tilde{c}_e Y_e^\dagger Y_e + \frac{5}{2}Y_e Y_e^\dagger \tilde{c}_e- \tilde{c}_e \left( 9\pi \alpha_2 + 15\pi\alpha_1 -T\right)+ Y_e \left(2X+9\alpha_2^2\tilde{c}_W+45\alpha_1^2\tilde{c}_B\right)\,,\\
\end{split}
\end{equation}
with $T = \Tr(3Y_u^\dagger Y_u + 3Y_d^\dagger Y_d + Y_e^\dagger Y_e)$.

In this basis, the couplings to bosons run, according to the beta functions
\begin{equation}
\begin{split}
    \tilde{\beta}_G &=2\Tr(Y_u^\dagger \tilde{c}_u+Y_d^\dagger \tilde{c}_d) + 96\alpha_s^2 \tilde{c}_G+27\alpha_2^2 \tilde{c}_W + 11 \alpha_1^2 \tilde{c}_B\,,\\
    \tilde{\beta}_W &=\frac{1}{2}\Tr(3 Y_u^\dagger \tilde{c}_u+3Y_d^\dagger \tilde{c}_d+Y_e^\dagger \tilde{c}_e)+ 72\alpha_s^2 \tilde{c}_G+54\alpha_2^2 \tilde{c}_W + 6 \alpha_1^2 \tilde{c}_B\,,\\
    \tilde{\beta}_B &=\frac{1}{6}\Tr(17 Y_u^\dagger \tilde{c}_u+5Y_d^\dagger \tilde{c}_d+15Y_e^\dagger \tilde{c}_e)+ 88\alpha_s^2 \tilde{c}_G+18\alpha_2^2 \tilde{c}_W + \frac{190}{3} \alpha_1^2 \tilde{c}_B\,.
\end{split}
\end{equation}

\subsection{Matching at the EW scale}
\label{app:matching}
The integration of the loops containing top quarks and $W^\pm$ bosons induces a negligible matching contribution to the couplings to gauge bosons, which can be obtained from Eqs.~\eqref{eq:full-contribution} and \eqref{eq:cG_eff},
\begin{equation}
\begin{split}
    \Delta c_G &= \frac{1}{2} [c'_{u_R}\!(\mu_w)-c'_{u_L}\!(\mu_w)]_{33}  B_1\!\!\left(\frac{4 m_t^2}{m_a^2}\right) \approx -\frac{m_a^2}{24 m_t^2} [c'_{u_R}\!(\mu_w)-c'_{u_L}\!(\mu_w)]_{33}\,,\\
    \Delta c_\gamma  &= \frac{4}{3}[c'_{u_R}\!(\mu_w)-c'_{u_L}\!(\mu_w)]_{33}  B_1\!\!\left(\frac{4 m_t^2}{m_a^2}\right) + \frac{2\alpha_\mathrm{em}}{\pi s_w^2} c_W B_2\!\!\left(\frac{4 m_W^2}{m_a^2}\right) \\ &\approx  -\frac{m_a^2}{9 m_t^2} [c'_{u_R}\!(\mu_w)-c'_{u_L}\!(\mu_w)]_{33} + \frac{\alpha_\mathrm{em}}{3\pi s_w^2}\frac{m_a^2}{m_W^2} c_W\,.
\end{split}
\end{equation}

The flavour-universal matching contributions to the couplings to fermions are

\begin{equation}
\begin{split}
\Delta c_{f_R} =& -\frac{3 y_t^2 Q_f s_w^2}{8\pi^2} [c'_{u_R}\!(\mu_w)-c'_{u_L}\!(\mu_w)]_{33} \log \frac{\mu_w^2}{m_t^2} \\ &+ \frac{3\alpha_\mathrm{em}^2Q_f^2}{4\pi^2 c_w^2} c_{\gamma Z}\left(\log\frac{\mu_w^2}{m_Z^2}+\frac{3}{2}+\delta_1\right) -  \frac{3\alpha_\mathrm{em}^2Q_f^2}{8\pi^2 c_w^4} c_Z\left(\log\frac{\mu_w^2}{m_Z^2}+\frac{1}{2}+\delta_1\right)\,,\\
\Delta c_{f_L} =& \frac{3 y_t^2 }{8\pi^2} (T_{3f} - Q_f s_w^2) [c'_{u_R}\!(\mu_w)-c'_{u_L}\!(\mu_w)]_{33}\log \frac{\mu_w^2}{m_t^2}
\\ &+ \frac{3\alpha_\mathrm{em}^2}{4\pi^2 s_w^2 c_w^2}Q_f(T_{3f}-Q_fs_w^2) c_{\gamma Z}\left(\log\frac{\mu_w^2}{m_Z^2}+\frac{3}{2}+\delta_1\right)\\ &+  \frac{3\alpha_\mathrm{em}^2}{8\pi^2 s_w^4 c_w^4}(T_{3f} - Q_f s_w^2)^2c_Z\left(\log\frac{\mu_w^2}{m_Z^2}+\frac{1}{2}+\delta_1\right)\\
&+  \frac{3\alpha_\mathrm{em}^2}{16\pi^2 s_w^4}c_W\left(\log\frac{\mu_w^2}{m_Z^2}+\frac{1}{2}+\delta_1\right)\,. 
\end{split}
\end{equation}

Additionally, the coupling to left-handed down-type quarks gets an additional matching contribution that is not flavour-universal,
\begin{equation}
\begin{split}
    \Delta^{ij}_\mathrm{FV} c_{d_L} =& \frac{y_t^2}{16\pi^2}(V^*_{3i} V_{kj}[c'_{u_L}\!(\mu_w)]_{3k}\!+\!V^*_{ki}V_{3j} [c'_{u_L}\!(\mu_w)]_{k3})\!\left(-\frac{1}{4}\log\frac{\mu_w^2}{m_t^2}-\frac{3}{8}+\frac{3}{4}\frac{1-x_t+\log x_t}{(1-x_t)^2}\right) \\
    &+ \frac{y_t^2}{16\pi^2} V^*_{31}V_{3j}[c'_{u_L}\!(\mu_w)]_{33} \\&+ \frac{y_t^2}{16\pi^2} V^*_{31}V_{3j}[c'_{u_R}\!(\mu_w)]_{33}\left(\frac{1}{2}\log\frac{\mu_w^2}{m_t^2}-\frac{1}{4}-\frac{3}{2}\frac{1-x_t+\log x_t}{(1-x_t)^2}\right)\\
    &-\frac{3 \alpha_\mathrm{em} y_t^2}{32\pi^3 s_w^2}V^*_{31}V_{3j} c_W  \frac{1-x_t + x_t \log x_t}{(1-x_t)^2}\,,
\end{split}
\end{equation}
where $x_t = m_t^2/m_W^2$.

After matching, the fermionic operators are
\begin{equation}\label{eq:matching}
\begin{split}
c_{u_i u_j}^{V,A}(\mu_w) &= [c'_{u_R}(\mu_w) \pm  c'_{u_L}(\mu_w)]_{ij} + (\Delta c_{u_R} \pm \Delta c_{u_L})\delta_{ij}\,,\qquad (i,j \neq 3)\,,\\
c_{d_i d_j}^{V,A}(\mu_w) &= [c'_{d_R}(\mu_w) \pm  c'_{d_L}(\mu_w)]_{ij} + (\Delta c_{d_R} \pm \Delta c_{d_L})\delta_{ij} \pm \Delta^{ij}_\mathrm{FV} c_{d_L} \,,\\
c_{e_i e_j}^{V,A}(\mu_w) &= [c'_{e_R}(\mu_w) \pm  c'_{e_L}(\mu_w)]_{ij} + (\Delta c_{e_R} \pm \Delta c_{e_L})\delta_{ij}\,.
\end{split}
\end{equation}

\subsection{Running below the EW scale}
\label{app:runningbelow}
Below the electroweak scale, only the diagonal terms of the axial couplings are affected by RGE running,
\begin{equation}\label{eq:run_low}
\begin{split}
\beta_{d_i d_j}^A &= \left(32 \alpha_s^2 \tilde{c}_G  + \frac{8}{3}\alpha_\mathrm{em}^2 \tilde{c}_\gamma \right)\delta_{ij}\,,\\
\beta_{u_i u_j}^A &= \left(32 \alpha_s^2 \tilde{c}_G  + \frac{32}{3}\alpha_\mathrm{em}^2 \tilde{c}_\gamma \right)\delta_{ij}\,,\\
\beta_{e_i e_j}^A &= 24\alpha_\mathrm{em}^2 \tilde{c}_\gamma\delta_{ij}\,,\\
\end{split}
\end{equation}
where the effective bosonic couplings only receive contributions from the fermions that are lighter than the scale $\mu$,
\begin{equation}
\begin{split}
\tilde{c}_G(\mu) =& c_G +\frac{1}{2}\sum_q c_{q q}^A \Theta(\mu-m_{q})\,,\\
\tilde{c}_\gamma(\mu) =& c_\gamma + \sum_f N_c^f Q_f^2c_{ff}^A \Theta(\mu-m_f)\,.\\
\end{split}
\end{equation}

\begin{figure}[tb!]
    \centering
    \includegraphics[width=0.9\textwidth]{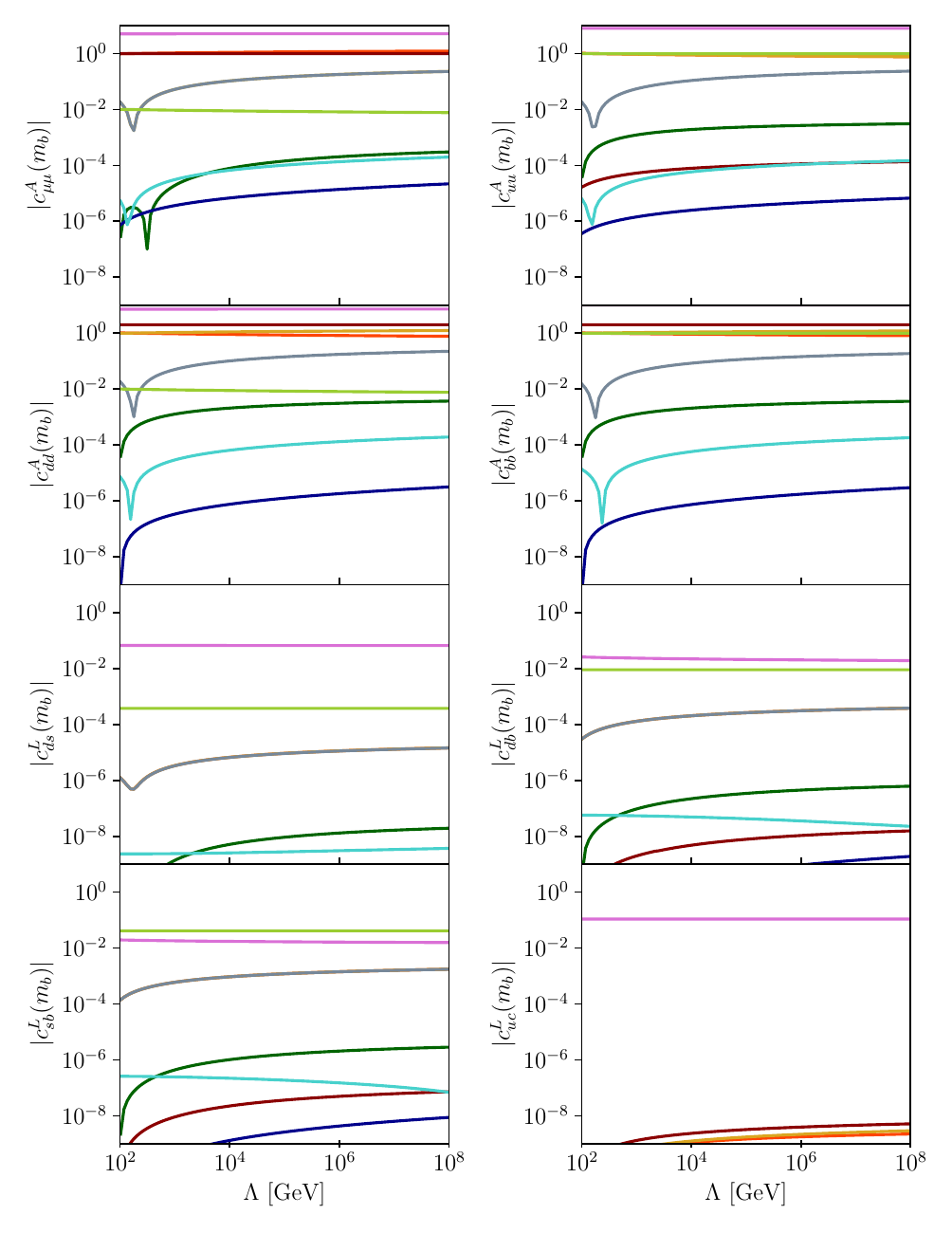}
    \caption{Fermionic couplings at the scale $\mu = m_b$ for the models considered, as function of the UV matching scale $\Lambda$: Red (\textcolor{orangered}{\rule[0.5ex]{0.3cm}{2pt}}) QED-DFSZ $\tan\beta =1$, maroon (\textcolor{darkred}{\rule[0.5ex]{0.3cm}{2pt}}) $u$-DFSZ $\tan\beta =1$, gold (\textcolor{goldenrod}{\rule[0.5ex]{0.3cm}{2pt}}) $e$-DFSZ $\tan\beta =1$, cyan (\textcolor{mediumturquoise}{\rule[0.5ex]{0.3cm}{2pt}}) L-KSVZ, blue (\textcolor{darkblue}{\rule[0.5ex]{0.3cm}{2pt}}) Y-KSVZ, green (\textcolor{darkgreen}{\rule[0.5ex]{0.3cm}{2pt}}) Q-KSVZ, grey (\textcolor{lightslategrey}{\rule[0.5ex]{0.3cm}{2pt}}) top-philic, pink (\textcolor{orchid}{\rule[0.5ex]{0.3cm}{2pt}}) flaxion, lime (\textcolor{yellowgreen}{\rule[0.5ex]{0.3cm}{2pt}}) non-universal DFSZ $\tan\beta = 10$.}
    \label{fig:running_couplings}
\end{figure}

The running of the ALP couplings relevant for the phenomenology in some of our benchmark models has been summarised in Fig.~\ref{fig:running_couplings}.

\section{Top-philic scenario}
\label{appendix:top-philic}
In the top-philic scenario, all ALP couplings except those to the top quark are set to zero at high scales. To impose this 
condition in the UV, one can introduce a non-universal scenario where only the third quark generation couples 
to the ALP. Moreover, in order to avoid tree-level couplings to the $b$ quark, which is a highly constrained sector, the ALP 
couples exclusively to RH quarks through the $(c_u)_{33}$ coupling in Eq.~\eqref{eq:UVLag}.

We notice that algebraic conditions exists that, if satisfied, assure that the ALP does not have flavour-violating couplings at tree-level. Considering the quark sector, one has
\begin{align}
    \label{eq:LHFlavourCondition}
    \left[c_{q},\, L_{u,\, d}\right] = 0 \quad &\Longleftrightarrow \quad\left[c_{q}, \, Y_{u, d} Y_{u,\, d }^\dagger \right] = 0 \,,\\
    \label{eq:RHFlavourCondition1}
    \left[c_{d},\, R_{d}\right] = 0 \quad &\Longleftrightarrow \quad\left[c_{d}, \, Y_{d}^\dagger Y_d\right] = 0 \, , \\
    \label{eq:RHFlavourCondition2}
    \left[c_{u},\, R_{u}\right] = 0 \quad &\Longleftrightarrow \quad\left[c_{u}, \, Y_{u}^\dagger Y_u\right] = 0 \,, 
\end{align}
where the conditions on both sides are equivalent. They are derived in the flavour basis, corresponding to the ALP Lagrangian in Eq.~\eqref{eq:UVLag}, but hold also for both universal couplings and non-universal but flavour conserving couplings, the latter being the case of the top-philic ALP. Interestingly, having only $(c_u)_{33}(\Lambda)\neq0$, the conditions in Eqs.~\eqref{eq:LHFlavourCondition} and \eqref{eq:RHFlavourCondition1} are trivially satisfied. However, the one in Eq.~\eqref{eq:RHFlavourCondition2} imposes a non-trivial condition on the Yukawa matrix $Y_u$:
\begin{align}
    \label{eq:conditiontophilic}
    0 \overset{!}{=} \left[c_u(\Lambda), Y_u^\dagger Y_u\right] = (c_u)_{33}(\Lambda)\begin{pmatrix}
        0 & 0 & \sum_i y^u_{i3}(y^u_{i1})^* \\
        0 & 0 & \sum_i y^u_{i3}(y^u_{i2})^* \\
        \sum_i y^u_{i1}(y^u_{i3})^* & \sum_i y^u_{i2}(y^u_{i3})^* & 0
    \end{pmatrix}\,.
\end{align}
If we demand this relation to hold exactly, the Yukawa matrix $Y_u$ must adopt one of the following textures:
\begin{align}
\label{eq:textures}
    Y_u^{(1)} \sim \begin{pmatrix}
        \cross & \cross & 0 \\
        \cross & \cross & 0 \\
        \cross & \cross & 0 \\
    \end{pmatrix}, \quad
    Y_u^{(2)} \sim \begin{pmatrix}
        0 & 0 & \cross \\
        0 & 0 & \cross \\
        0 & 0 & \cross \\
    \end{pmatrix},\quad
    Y_u^{(3)} \sim \begin{pmatrix}
        \cross & \cross & 0 \\
        \cross & \cross & 0 \\
        0 & 0 & \cross \\
    \end{pmatrix},
\end{align}
where $\cross$ denotes non-zero entries. $Y_u^{(1)}$ and $Y_u^{(2)}$ are rank-2 and rank-1 matrices, respectively, and thus cannot reproduce the observed up-type quark mass spectrum, as they result in two or one massless quarks, respectively. The texture $Y_u^{(3)}$, on the other hand, requires a UV extension involving additional symmetries, as it cannot be achieved solely with a PQ symmetry that charges only the RH top quark. The simplest such extension involves an Abelian discrete $\mathbb{Z}_2$ under which the third generation quarks are odd, while the other generations are even. However, this symmetry enforces a similar texture for $Y_d \sim Y_u^{(3)}$, which leads to two vanishing mixing angles in the CKM matrix. Consequently, this symmetry must be broken at some level to accommodate flavour data.

This analysis leads to the conclusion that a top-philic scenario can, at best, only be approximate. Indeed, assuming a hierarchical Yukawa structure consistent with the CKM matrix, specifically
\be
Y_u\sim
\begin{pmatrix}
        \lambda^5 & \lambda^4 & \lambda^3 \\
        \lambda^4 & \lambda^3 & \lambda^2 \\
        \lambda^3 & \lambda^2 & 1 \\
    \end{pmatrix}\,,
\ee
with $\lambda=0.22$ to represent the Cabibbo angle, we find from Eq.~\eqref{eq:conditiontophilic} that
\begin{equation}
    \left[c_u, Y_u^\dagger Y_u\right] =0+ \mathcal{O}(\lambda^2),
\end{equation}
that is satisfying the last condition only up to second order in terms of the Cabibbo angle expansion.

\section{Details of ALP couplings in Chiral Perturbation Theory}
\label{sec:alpmixing}
\label{appendix: ALP_chiPT_details}
The starting point is Lagrangian~\eqref{EWALP} with only the three lightest quark flavours active. The leptonic part of the Lagrangian does not play any role in this sector.
\begin{equation}
\begin{split}
    \mathcal{L} =& \bar{q}(i \slashed{D} - \boldsymbol{m}) q + \frac{1}{2} (\partial_\mu a)^2 - \frac{1}{2}m_a^2 a^2 \\ &+ \frac{\alpha_s}{4\pi f_a} c_G a G_{\mu\nu}^a \tilde{G}^{a,\mu\nu} + \frac{\alpha_\mathrm{em}}{4\pi f_a}c_\gamma a F_\mu\nu \tilde{F}^{\mu\nu} +\frac{\partial_\mu a}{f_a} \bar{q} (\boldsymbol{c_R} \gamma^\mu P_R + \boldsymbol{c_L} \gamma^\mu P_L)q\,,
\end{split}
\end{equation}
where $q=(u, d, s)^T$. The quark mass and charge matrices are $\boldsymbol{m} = \mathrm{diag}(m_u, m_d, m_s)$ and $\boldsymbol{Q} = \mathrm{diag}(2/3, -1/3, -1/3)$, respectively. If we perform a chiral rotation of the quark fields with an arbitrary matrix $\boldsymbol{\kappa}$, given by~\cite{Georgi:1986df}
\begin{equation}
    q \to \exp\!\!\left(-i \frac{a}{f_a} c_G \boldsymbol{\kappa} \gamma_5\right)q\,,
\end{equation}
the fermionic path integral measure is not invariant under this transformation, and consequently, the Jacobian shows up as a new local interaction, namely
\begin{align}
    \mathcal{L}_\mathrm{Jac} =& -\sum_q\left[ \frac{g_s^2}{8\pi^2}\frac{a}{f_a}\frac{1}{2}c_G \kappa_q  G_{\mu\nu}^a \tilde{G}^{a,\mu\nu} + \frac{e^2}{8\pi^2}\frac{a}{f_a} N_c Q_q^2 c_G \kappa_q  F_{\mu\nu} \tilde{F}^{\mu\nu}\right] \\\nonumber
    =& -\frac{\alpha_s}{4\pi f_a} 2c_G \langle\boldsymbol{\kappa}\rangle a G_{\mu\nu}^a \tilde{G}^{a,\mu\nu} - \frac{\alpha_\mathrm{em}}{4\pi f_a}\left(c_G N_c \langle\boldsymbol{\kappa}\boldsymbol{Q}^2\rangle\right) a F_{\mu\nu} \tilde{F}^{\mu\nu}\,.
\end{align}

The first term ensures that the $agg$ vertex vanishes with this transformation, as long as $2\langle \boldsymbol{\kappa}\rangle \equiv\Tr (\boldsymbol{\kappa}) =1$, while the second one provides an effective coupling to photons. The Lagrangian can be recast as
\begin{align}
    \mathcal{L} =& \bar{q}(i \slashed{D} - \boldsymbol{\hat{m}}) q + \frac{1}{2} (\partial_\mu a)^2 - \frac{1}{2}m_a^2 a^2  +\frac{\alpha_\mathrm{em}}{4\pi f_a}(c_\gamma +\mathcal{C}_\gamma^\chi) a F_{\mu\nu} \tilde{F}^{\mu\nu} \nonumber \\
    & +\frac{\partial_\mu a}{f_a} \bar{q} (\boldsymbol{\hat{c}_R} \gamma^\mu P_R + \boldsymbol{\hat{c}_L} \gamma^\mu P_L)q\,,
\end{align}
with
\begin{equation}
    \boldsymbol{\hat{m}} = \exp\!\!\left(-i \frac{a}{f_a} c_G \boldsymbol{\kappa}\right) \boldsymbol{m} \exp\!\!\left(-i \frac{a}{f_a} c_G \boldsymbol{\kappa}\right)=\boldsymbol{m}-2i\frac{a}{f_a}c_G \boldsymbol{\kappa}\boldsymbol{m}+\mathcal{O}(a/f_a)\,,
\end{equation}
\begin{equation}
   \boldsymbol{\hat{c}_R} = \exp\!\!\left(-i \frac{a}{f_a} c_G \boldsymbol{\kappa} \right)(\boldsymbol{c_R} - c_G \boldsymbol{\kappa}) \exp\!\!\left(i \frac{a}{f_a} c_G \boldsymbol{\kappa} \right) = \boldsymbol{c_R} - c_G \boldsymbol{\kappa} + \mathcal{O}(a/f_a)\,,
\end{equation}
\begin{equation}
   \boldsymbol{\hat{c}_L} = \exp\!\!\left(i \frac{a}{f_a} c_G \boldsymbol{\kappa}\right)(\boldsymbol{c_L} + c_G \boldsymbol{\kappa}) \exp\!\!\left(-i \frac{a}{f_a} c_G \boldsymbol{\kappa} \right) = \boldsymbol{c_L} + c_G \boldsymbol{\kappa} + \mathcal{O}(a/f_a)\,.
\end{equation}
\begin{equation}
    \mathcal{C}_\gamma^\chi = - c_G N_c \langle\boldsymbol{\kappa}\boldsymbol{Q}^2\rangle \approx -2 c_G\,,
\end{equation}
Including next-to-leading-order corrections in $\chi$PT, the result is $\mathcal{C}_\gamma^\chi\! =\! -(1.92\pm0.04)c_G$~\cite{GrillidiCortona:2015jxo}.

\subsection{Mixing with pseudoscalar mesons}
\label{sec:alp_meson_mix}
In the three-flavour chiral perturbation theory ($\chi$PT), the light pseudoscalar mesons are parameterised as $\mathrm{U}(3)$ matrices $\boldsymbol{U} = \exp(i \boldsymbol{\Phi}/F_0)$ with
\begin{equation}\label{eq:u3_psmesons}
    \boldsymbol{\Phi} = \sum \phi \boldsymbol{\phi} = \begin{pmatrix}
        \pi^0 + \frac{\eta_8}{\sqrt{3}}+\sqrt{\frac{2}{3}}\eta_0 & \sqrt{2}\pi^+ & \sqrt{2}K^+\\ \sqrt{2}\pi^- & -\pi^0 + \frac{\eta_8}{\sqrt{3}}+\sqrt{\frac{2}{3}}\eta_0 & \sqrt{2}K^0\\
        \sqrt{2}K^-  & \sqrt{2}\bar{K}^0 &  \frac{-2\eta_8}{\sqrt{3}}+\sqrt{\frac{2}{3}}\eta_0
    \end{pmatrix}\,,
\end{equation}
where $\boldsymbol{\phi}$ are the Gell-Mann matrices associated to the respective mesons, normalised as $\langle\boldsymbol{\phi}\boldsymbol{\varphi}^T\rangle = \delta_{\phi\varphi}$. The physical states $\eta$ and $\eta'$ are obtained as a mix of $\eta_8$ and $\eta_0$, with $\mathrm{U}(3)$ representations $\boldsymbol{\eta} = \cos \theta_{\eta\eta'} \boldsymbol{\eta_8}-\sin \theta_{\eta\eta'} \boldsymbol{\eta_0}$, $\boldsymbol{\eta'} = \sin \theta_{\eta\eta'} \boldsymbol{\eta_8}+\cos \theta_{\eta\eta'} \boldsymbol{\eta_0}$. For the calculations we will use the value of $\theta_{\eta\eta'}$ obtained using lattice methods~\cite{Christ:2010dd}, however for the explicit expressions, we will approximate $\sin \theta_{\eta\eta'} \approx-1/3$.

The $\chi$PT Lagrangian, up to $\mathcal{O}(p^2)$, is~\cite{Gasser:1983yg,Gasser:1984gg}
\begin{equation}
    \mathcal{L}_{\chi\mathrm{PT}}^{(p^2)} = \frac{F_0^2}{2}\langle D_\mu \boldsymbol{U} (D^\mu \boldsymbol{U})^\dagger\rangle + \frac{F_0^2}{2}\langle\boldsymbol{\chi} \boldsymbol{U}^\dagger + \boldsymbol{U} \boldsymbol{\chi}^\dagger\rangle-\frac{1}{3}m_{\eta_0}^2 \langle\boldsymbol{\Phi}\rangle^2\,.
\end{equation}

The last term is the contribution of the axial anomaly in QCD, with $\langle\boldsymbol{\Phi}\rangle = \sqrt{3/2}\eta_0$, and $m_{\eta_0}$ will be chosen to eliminate the mixing between $\eta$ and $\eta'$. External scalar $\boldsymbol{S}$ and pseudoscalar $\boldsymbol{P}$ currents enter the Lagrangian through $\boldsymbol{\chi}$,
\begin{equation}
    \boldsymbol{\chi} = 2 B_0 (\boldsymbol{S} + i \boldsymbol{P})\,,
\end{equation}
where $B_0 \approx m_\pi^2/(m_u+m_d)$ comes from the chiral condensate. External left-handed $\boldsymbol{L}_\mu$ and right-handed $\boldsymbol{R}_\mu$ currents are incorporated in the covariant derivative,
\begin{equation}
    D_\mu \boldsymbol{U} = \partial_\mu \boldsymbol{U} -i \boldsymbol{R}_\mu \boldsymbol{U} + i \boldsymbol{U} \boldsymbol{L}_\mu\,.
\end{equation}
The matching between the ALP Lagrangian and the $\chi$PT Lagrangian is thus defined by the conditions
\begin{equation}
    \boldsymbol{\chi} = 2 B_0 \boldsymbol{\hat{m}}\,,\qquad\qquad D_\mu \boldsymbol{U} = \partial_\mu \boldsymbol{U} -i\frac{\partial_\mu a}{f_a}(\boldsymbol{\hat{c}_R} \boldsymbol{U} - \boldsymbol{U}\boldsymbol{\hat{c}_L})\,.
\end{equation}
In the absence of weak interactions, the chiral Lagrangian only contains the combination $\boldsymbol{\hat{c}_A} = \boldsymbol{\hat{c}_R}-\boldsymbol{\hat{c}_L} = \boldsymbol{c_A} - 2 c_G \boldsymbol{\kappa}$, which results in the Lagrangian
\begin{equation}
\begin{split}
    \mathcal{L}_{\mathrm{ALP}-\chi\mathrm{PT}} =& \frac{F_0^2}{2} \langle\partial_\mu \boldsymbol{U} (\partial^\mu \boldsymbol{U})^\dagger\rangle - i \frac{F_0^2}{2 f_a} (\partial_\mu a)\langle\boldsymbol{\hat{c}_A}(\boldsymbol{U}\partial^\mu \boldsymbol{U}^\dagger-\boldsymbol{U}^\dagger \partial^\mu \boldsymbol{U})\rangle  \\
    &+ F_0^2 B_0\langle\boldsymbol{m}(\boldsymbol{U}+  \boldsymbol{U}^\dagger)\rangle+2i\frac{F_0^2 B_0}{f_a}c_G  a\langle\boldsymbol{\kappa}\boldsymbol{m}(\boldsymbol{U}-\boldsymbol{U}^\dagger)\rangle -\frac{1}{3}m_{\eta_0}^2 \langle\boldsymbol{\Phi}\rangle^2 \\
    &+ \frac{1}{2}\partial_\mu a\, \partial^\mu a - \frac{1}{2} m_a^2 a^2 + \frac{\alpha_\mathrm{em}}{4\pi f_a}(c_\gamma + \mathcal{C}_\gamma^\chi) a F_{\mu\nu} \tilde{F}^{\mu\nu}\,.
\end{split}
\end{equation}
The terms in the first line will induce a kinetic mixing between the ALP and the pseudoscalar mesons, while the terms in the second line will induce a mass mixing.

If we first focus on the real pseudoscalar mesons $\pi^0, \eta, \eta'$, their masses are obtained as
\begin{equation}\label{eq:mass_neutrmeson}
    m_\phi^2 = 2B_0\langle\boldsymbol{m}\boldsymbol{\phi}^2\rangle +\frac{2}{3} m_{\eta_0}^2\langle\boldsymbol{\phi}\rangle^2\,.
\end{equation}
Between the $\pi^0, \eta, \eta'$ mesons there is no kinetic mixing, while their mass mixing is given by the off-diagonal entries of the mass matrix,
\begin{equation}\label{eq:mixing_neutrmeson}
    M^2_{\phi \varphi} = M^2_{\varphi\phi} = 2B_0\langle \boldsymbol{m}\boldsymbol{\phi}\boldsymbol{\varphi}\rangle + \frac{2}{3}m_{\eta_0}^2\langle\boldsymbol{\phi}\rangle\langle\boldsymbol{\varphi}\rangle \,.
\end{equation}
In particular, $M^2_{\eta\eta'} = M^2_{\eta'\eta} = 0$, which in the $\sin\theta_{\eta\eta'}\approx-1/3$ approximation corresponds to $m_{\eta_0}^2 \approx \frac{3}{2} B_0 (2m_s-m_u-m_d)$.

Consequently, in the absence of ALPs, the fields are rotated as
\begin{equation}
    \phi = \hat{\phi} + \!\!\!\sum_{\varphi\in\{\pi^0, \eta, \eta' \}}\!\! \frac{M^2_{\phi\varphi}}{m_\varphi^2 - m_\phi^2}\hat{\varphi}\,.
\end{equation}
By inserting this into Eq.~\eqref{eq:u3_psmesons}, we can rewrite $\boldsymbol{\Phi}$ in terms of the rotated fields,
\begin{equation}\label{eq:u3repr_smrot}
    \boldsymbol{\Phi} = \sum_{\phi} \left(\boldsymbol{\phi} +\!\!\!\!\! \sum_{\varphi\in\{\pi^0, \eta, \eta' \}}\!\!\frac{M^2_{\varphi\phi}}{m^2_\phi - m^2_\varphi} \boldsymbol{\varphi}\right)\hat{\phi} \equiv \sum_{\phi} \hat{\phi} \boldsymbol{\hat{\phi}}\,,
\end{equation}
where the $\boldsymbol{\hat{\phi}}$ matrices are still orthonormal $\langle\boldsymbol{\hat{\phi}}\boldsymbol{\hat{\varphi}}^T\rangle = \delta_{\phi\varphi}$.

Moving on to the ALP, its kinetic mixing with the pseudoscalar mesons is
\begin{equation}
    K_{a\phi}  = -\frac{F_0}{f_a}\langle\boldsymbol{\hat{c}_A} \boldsymbol{\hat{\phi}}\rangle \,,
\end{equation}
and its mass mixing is
\begin{equation}
    M_{a\phi}^2=- 4\frac{F_0}{f_a} B_0 c_G \langle\boldsymbol{\kappa}\boldsymbol{m}\boldsymbol{\hat{\phi}}\rangle \,.
\end{equation}
The Lagrangian is then diagonalised by the field redefinitions
\begin{equation}\label{eq:redef_alp}
    a = a_\mathrm{Phys} + \sum_\phi \theta_{a\phi} \phi_\mathrm{Phys}\,,\qquad\qquad \hat{\phi} = \phi_\mathrm{Phys} + \theta_{\phi a} a_\mathrm{Phys}\,.
\end{equation}
The mixing parameters are given by
\begin{equation}\label{eq:mixing_angles}
\begin{split}
    \theta_{a\phi} &= \frac{K_{a\phi} m_\phi^2 - M_{a\phi}^2}{m_a^2-m_\phi^2}\,,\\ \theta_{\phi a} &= \frac{M_{a\phi}^2-K_{a\phi}m_a^2}{m_a^2-m_\phi^2}\,.
    \end{split}
\end{equation}
By inserting Eq.~\eqref{eq:redef_alp} into Eq.~\eqref{eq:u3repr_smrot}, we finally find the expression for $\boldsymbol{\Phi}$ in terms of the physical fields,
\begin{equation}\label{eq:phimatrix_alp}
    \boldsymbol{\Phi} = \sum_\phi \phi_\mathrm{Phys} \boldsymbol{\hat{\phi}} + a_\mathrm{Phys} \sum_\phi \theta_{\phi a} \boldsymbol{\hat{\phi}} \equiv \sum_\phi \phi_\mathrm{Phys} \boldsymbol{\hat{\phi}}  + a_\mathrm{Phys} \boldsymbol{a}\,.
\end{equation}

From now on, we will always work with the physical fields, and will drop the label ``Phys'' to ease the notation. In particular, notice that in presence of flavour-violating couplings $c_{sd}^A =(c_{ds}^A)^*$, the ALP will mix with neutral kaons, with $\theta_{K^0a} = \theta_{\bar{K}^0a}^*$ which is already $\boldsymbol{\kappa}$-independent. Flavour-conserving couplings and gluon couplings will induce mixing with $\pi^0$, $\eta$ and $\eta'$.

\subsection{On the \texorpdfstring{$\boldsymbol{\kappa}$}{kappa} dependence}\label{app:kappa_dependence}

The mixing parameters $\theta_{\phi a}$ in Eq.~\eqref{eq:mixing_angles}  are unphysical, as they depend on the matrix $\boldsymbol{\kappa}$. As it has been noted recently~\cite{Ovchynnikov:2025gpx,Bai:2025fvl,Balkin:2025enj}, the na\"ive substitution of Eq.~\eqref{eq:phimatrix_alp} may lead to inconsistent results for certain processes once the formalism is extended to other mesons. The remedy is to use the $\boldsymbol{\kappa}$-invariant $\mathrm{U}(3)$ matrix given by~\cite{Ovchynnikov:2025gpx,Bai:2025fvl}
\begin{equation}
\begin{split}
    \boldsymbol{\Phi} &\to \boldsymbol{\tilde{\Phi}} = \boldsymbol{\Phi} +2\frac{F_0}{f_a}c_G\boldsymbol{\kappa} = \sum\hat{\phi}\boldsymbol{\hat{\phi}} + a\boldsymbol{\tilde{a}}\,,\\
    \boldsymbol{a} &\to\boldsymbol{\tilde{a}} = \boldsymbol{a} + 2\frac{F_0}{f_a}c_G \boldsymbol{\kappa}\,.
    \end{split}
\end{equation}

Indeed, let us write explicitly $\theta_{\phi a}$,
\begin{equation}
    \theta_{\phi a} = -\frac{F_0}{f_a}c_G \frac{4B_0 \langle \boldsymbol{\kappa} \boldsymbol{m}\boldsymbol{\hat{\phi}} \rangle + 2 m_a^2 \langle\boldsymbol{\kappa}\boldsymbol{\hat{\phi}}\rangle}{m_a^2 - m_\phi^2} -\frac{m_a^2}{m_a^2-m_\phi^2}\frac{F_0}{f_a}\langle\boldsymbol{c_A}\boldsymbol{\hat{\phi}}\rangle\,.
\end{equation}
Next, by applying the completion relation for the $\boldsymbol{\hat{\varphi}}$ matrices, we can write $\boldsymbol{\kappa} = \sum \langle\boldsymbol{\kappa}\boldsymbol{\hat{\varphi}}\rangle \boldsymbol{\hat{\varphi}}$.
\begin{equation}
    \theta_{\phi a} = -\frac{F_0}{f_a}c_G \frac{4B_0 \sum \langle\boldsymbol{\kappa}\boldsymbol{\hat{\varphi}}\rangle\langle  \boldsymbol{m}\boldsymbol{\hat{\phi}} \boldsymbol{\hat{\varphi}}\rangle + 2 m_a^2 \langle\boldsymbol{\kappa}\boldsymbol{\hat{\phi}}\rangle}{m_a^2 - m_\phi^2}  -\frac{m_a^2}{m_a^2-m_\phi^2}\frac{F_0}{f_a}\langle\boldsymbol{c_A}\boldsymbol{\hat{\phi}}\rangle\,.
\end{equation}
Since there is no mass mixing between the rotated mesons $\hat{\varphi}$, from Eqs.~\eqref{eq:mass_neutrmeson}-\eqref{eq:mixing_neutrmeson}
\begin{equation}
    2B_0\langle \boldsymbol{m}\boldsymbol{\hat{\phi}}\boldsymbol{\hat{\varphi}}\rangle = m_\phi^2 \delta_{\phi\varphi} - \frac{2}{3}m_{\eta_0}^2\langle\boldsymbol{\hat{\phi}}\rangle \langle\boldsymbol{\hat{\varphi}}\rangle\,,
\end{equation}
we obtain
\begin{equation}
\begin{split}
    \theta_{\phi a} &= -\frac{F_0}{f_a}c_G \frac{2 m_\phi^2\langle\boldsymbol{\kappa}\boldsymbol{\hat{\phi}}\rangle -\frac{4}{3}m_{\eta_0}^2\langle\boldsymbol{\hat{\phi}}\rangle \sum\langle\boldsymbol{\kappa}\boldsymbol{\hat{\varphi}}\rangle\langle \boldsymbol{\hat{\varphi}}\rangle + 2 m_a^2 \langle\boldsymbol{\kappa}\boldsymbol{\hat{\phi}}\rangle}{m_a^2 - m_\phi^2}  -\frac{m_a^2}{m_a^2-m_\phi^2}\frac{F_0}{f_a}\langle\boldsymbol{c_A}\boldsymbol{\hat{\phi}}\rangle\,\\
    &= 2\frac{F_0}{f_a}c_G \left(-\langle\boldsymbol{\kappa}\boldsymbol{\hat{\phi}}\rangle + \frac{2}{3}\frac{m_{\eta_0}^2}{m_a^2-m_\phi^2}\langle\boldsymbol{\hat{\phi}}\rangle\langle\boldsymbol{\kappa}\rangle  \right) -\frac{m_a^2}{m_a^2-m_\phi^2}\frac{F_0}{f_a}\langle\boldsymbol{c_A}\boldsymbol{\hat{\phi}}\rangle\,.
\end{split}
\end{equation}

Finally,
\begin{equation}
\boldsymbol{a} = \sum \theta_{\phi a}\boldsymbol{\hat{\phi}} = -2\frac{F_0}{f_a}c_G \boldsymbol{\kappa} +\frac{1}{3}\frac{F_0}{f_a}\sum \frac{4 m_{\eta_0}^2 c_G \langle\boldsymbol{\kappa}\rangle\langle\boldsymbol{\hat{\phi}}\rangle - 3 m_a^2 \langle\boldsymbol{c_A}\boldsymbol{\hat{\phi}}\rangle}{m_a^2 - m_\phi^2}\boldsymbol{\hat{\phi}}\,.
\end{equation}
Remembering that $\langle\boldsymbol{\kappa}\rangle = 1/2$ is invariant under reparameterisations of $\boldsymbol{\kappa}$, it is actually enough to remove the first term.

\subsection{ALP interactions with vector mesons}\label{sec:chiral_VMD}
\label{sec:VMD}
The $\mathrm{U}(3)$ matrix for the vector mesons is
\begin{equation}
    \boldsymbol{V} = \sum v \boldsymbol{v} = \frac{1}{2} \begin{pmatrix}
        \rho^0 + \omega^0 & \sqrt{2}\rho^+ & \sqrt{2}K^{*+} \\
        \sqrt{2}\rho^- & -\rho^0 + \omega^0 & \sqrt{2}K^{*0} \\
        \sqrt{2}K^{*-} & \sqrt{2}\bar{K}^{*0} & \sqrt{2}\phi^0
    \end{pmatrix}  \,,
\end{equation}
and the $\mathrm{U}(3)$ representation for each vector can be readily obtained from here.

The VMD Lagrangian~\cite{Fujiwara:1984mp} describing the interactions of vector mesons with pseudoscalar mesons and photons is:
\begin{equation}
    \mathcal{L}_\mathrm{VMD} =-\frac{3g^2}{4\pi^2F_0}\varepsilon^{\mu\nu\rho\sigma} \langle\partial_\mu \boldsymbol{V}_\nu \partial_\rho\boldsymbol{V}_\rho \boldsymbol{\Phi}\rangle+4g^2 F_0^2 \langle\boldsymbol{V}_\mu\boldsymbol{V}^\mu\rangle- 8egF_0^2 \langle\boldsymbol{V}_\mu \boldsymbol{Q}\rangle A^\mu\,,
\end{equation}
where $g \approx m_\rho/\sqrt{2}F_0 \approx\sqrt{12\pi}$.
The last term induces a mixing between the vector mesons and the photon. The result of the diagonalisation is 
\begin{equation}
    \rho_\mu^0 \to \rho_\mu^0 + \frac{e}{g} A_\mu\,,\qquad \omega_\mu^0 \to \omega_\mu^0 + \frac{e}{g} \frac{1}{3}A_\mu\,,\qquad \phi_\mu^0 \to \phi_\mu^0 - \frac{e}{g} \frac{\sqrt{2}}{3}A_\mu\,.
\end{equation}
Consequently, we can introduce a $\mathrm{U}(3)$ representation for the photon,
\begin{equation}
    \boldsymbol{A} = \frac{e}{g}\left(\boldsymbol{\rho^0} + \frac{1}{3}\boldsymbol{\omega^0} -\frac{\sqrt{2}}{3} \boldsymbol{\phi^0}\right)\,.
\end{equation}

By making the substitutions $\boldsymbol{\Phi} \to \boldsymbol{\Phi} + \boldsymbol{\tilde{a}} a$ and $\boldsymbol{V}_\mu \to \boldsymbol{V}_\mu + \boldsymbol{A} A_\mu$, the induced interaction of the ALP to photons is

\begin{equation}
    \mathcal{L}_\mathrm{VMD}^{a\gamma\gamma} =-\frac{3\alpha_\mathrm{em}}{2\pi F_0}\left[\langle\boldsymbol{\tilde{a}} \boldsymbol{\rho^0}\boldsymbol{\rho^0}\rangle + \frac{1}{9}\langle\boldsymbol{\tilde{a}} \boldsymbol{\omega^0}\boldsymbol{\omega^0}\rangle + \frac{2}{9} \langle\boldsymbol{\tilde{a}} \boldsymbol{\phi^0}\boldsymbol{\phi^0}\rangle +\frac{2}{3} \langle\boldsymbol{\tilde{a}} \boldsymbol{\rho^0}\boldsymbol{\omega^0}\rangle \right] a F^{\mu\nu} \tilde{F}_{\mu\nu}\,.
\end{equation}

The VMD contribution to the effective ALP coupling to photons is obtained by including the phenomenological suppression factor $\mathcal{F}(m_a)$ derived from $e^+e^-\to q \ov{q}$~\cite{Ilten:2018crw},
\begin{equation}\label{eq:VMDphotons}
    \mathcal{C}_\gamma^\mathrm{VMD} = -\mathcal{F}(m_a) \frac{6f_a}{F_0}\left[\langle\boldsymbol{\tilde{a}} \boldsymbol{\rho^0}\boldsymbol{\rho^0}\rangle + \frac{1}{9}\langle\boldsymbol{\tilde{a}} \boldsymbol{\omega^0}\boldsymbol{\omega^0}\rangle + \frac{2}{9} \langle\boldsymbol{\tilde{a}} \boldsymbol{\phi^0}\boldsymbol{\phi^0}\rangle +\frac{2}{3} \langle\boldsymbol{\tilde{a}} \boldsymbol{\rho^0}\boldsymbol{\omega^0}\rangle \right]\,.
\end{equation}

\subsection{ALP interactions with scalar and tensor mesons}\label{sec:alp_scalar_tensor}
The scalar mesons are organised in the nonet
\begin{equation}
    \boldsymbol{\mathcal{S}} = \begin{pmatrix}a_0^0 -s_s \sigma + c_s f_0 & \sqrt{2} a_0^+ & \sqrt{2}\kappa^+\\ \sqrt{2}a_0^- & -a_0^0 -s_s \sigma + c_s f_0 & \sqrt{2} \kappa^0\\ \sqrt{2}\kappa^- & \sqrt{2} \kappa^0 & c_s \sigma + s_s f_0
    \end{pmatrix}\,,
\end{equation}
where $s_s = \sin \theta_s$, $c_s = \cos\theta_s$ and $\theta_s = 21^\circ$ is the scalar mixing angle.
The phenomenological Lagrangian that describes the interactions of scalar and pseudoscalar mesons is given by~\cite{Black:1998wt,Fariborz:1999gr}
\begin{equation}
\begin{split}\label{eq:lagr_scalars}
    \mathcal{L}_{SPP} =&  \sqrt{2}A\langle\boldsymbol{\mathcal{S}}\partial_\mu \boldsymbol{\Phi}\partial^\mu\boldsymbol{\Phi}\rangle + \sqrt{2}(B-A)\langle\boldsymbol{\mathcal{S}}\rangle \langle\partial_\mu \boldsymbol{\Phi}\partial^\mu\boldsymbol{\Phi}\rangle\\
    &+\sqrt{2}(C-2A) \langle\boldsymbol{\mathcal{S}} \partial_\mu\boldsymbol{\Phi}\rangle \langle\partial^\mu\boldsymbol{\Phi}\rangle + 2\sqrt{2}(D+A)\langle\boldsymbol{\mathcal{S}}\rangle \langle\partial_\mu \boldsymbol{\Phi}\rangle\langle\partial^\mu \boldsymbol{\Phi}\rangle\,.
\end{split}
\end{equation}
The couplings $A,B,C,D$ are taken from the fit performed by~\cite{Ovchynnikov:2025gpx}.

In the case of tensor mesons, we focus exclusively on the $f_2$ meson, with representation $\boldsymbol{f_2} \approx \mathrm{diag}(1, 1,0)$. The relevant Lagrangian is~\cite{Guo:2011ir}
\begin{equation}\label{eq:lagr_f2}
    \mathcal{L}_{f_2} = \frac{2g_T}{F_0^2}  \langle \boldsymbol{f_2}\partial_\mu \boldsymbol{\Phi} \partial_\nu \boldsymbol{\Phi}\rangle f_2^{\mu\nu}\,,
\end{equation}
with $g_T = 39.4\,\mathrm{MeV}$.

With the substitution $\boldsymbol{\Phi} \to \boldsymbol{\Phi}+\boldsymbol{\tilde{a}} a$ in Eq.~\eqref{eq:lagr_scalars} and \eqref{eq:lagr_f2} one can obtain the interactions of scalar and tensor mesons with an ALP due to its mixing with pseudoscalar mesons.

\subsection{Weak interactions}\label{sec:chiral_weak}
The left-handed current $\bar{q}_L\gamma_\mu q_L$ is represented in the chiral theory by~\cite{Bauer:2021wjo,Cornella:2023kjq}
\begin{equation}
\begin{split}
L_\mu =& -\frac{i F_0^2}{2} \exp\!\!\left(-i \boldsymbol{\kappa}\frac{c_G a }{f_a}\right) [\boldsymbol{U}D_\mu\boldsymbol{U} ^\dagger]\exp\!\!\left(-i \boldsymbol{\kappa}\frac{c_G a }{f_a}\right)\,, \\
L_\mu^{ji}\supset& \frac{F_0}{2}\partial_\mu\boldsymbol{\Phi}^{ji} -\frac{i }{4} ([\boldsymbol{\Phi}, \partial_\mu\boldsymbol{\Phi}])^{ji}\ +i\frac{ F_0}{2} (\boldsymbol{\kappa}^{ii} -\boldsymbol{\kappa}^{jj})\frac{c_G a }{f_a} \partial_\mu\boldsymbol{\Phi}^{ji} \\
&+ \frac{F_0^2}{2}\frac{\partial_\mu a}{f_a} (\boldsymbol{\hat{c}_L}^{ji} - \boldsymbol{\hat{c}_R}^{ji}) - \frac{iF_0}{2}\frac{\partial_\mu a}{f_a}([\boldsymbol{\Phi},\boldsymbol{\hat{c}_R}])^{ji}\,.
\end{split} 
\end{equation}
The weak chiral Lagrangian includes the octet and 27-plet operators,~\cite{Cornella:2023kjq}
\begin{equation}
\begin{split}
\mathcal{L}_w^{(p^2)} &= - \frac{G_F}{\sqrt{2}} V_{ud}V_{us}^*\left[g_8 \mathcal{O}_8 + g_{27}^{1/2} \mathcal{O}_{27}^{1/2} + g_{27}^{3/2}\mathcal{O}_{27}^{3/2}\right] + \mathrm{h.c.}\,,\\
\mathcal{O}_8 &= [L_\mu L^\mu]^{sd} = L_\mu^{su}L^{\mu,ud} + L_\mu^{sd}L^{\mu,dd} +L_\mu^{ss}L^{\mu,sd}\,,\\
\mathcal{O}_{27}^{1/2} &= L_\mu^{sd} L^{\mu,uu} + L_\mu^{su} L^{\mu,ud}+2L_\mu^{sd}L^{\mu,dd} -3 L_\mu^{sd}L^{\mu,ss}\,,\\
\mathcal{O}_{27}^{3/2}&= L_\mu^{sd} L^{\mu,uu} + L_\mu^{su} L^{\mu,ud}-L_\mu^{sd}L^{\mu,dd}\,,
\end{split}
\end{equation}
with $g_8 = 3.61\pm 0.28$~\cite{Cirigliano:2011ny}, $g_{27}^{1/2} = 0.033 \pm 0.003$ and $g_{27}^{3/2} = 0.165\pm0.016$~\cite{Cornella:2023kjq}. In the limit of exact $S\mathrm{U}(3)$ symmetry, $g_{27}^{3/2} = 5 g_{27}^{1/2}$.

\subsection{ALP processes in chiral perturbation theory}
Following Refs.~\cite{Aloni:2018vki,Balkin:2025enj}, the validity regime of the $\chi$PT expressions is extended up to $\sim3\,\mathrm{GeV}$ by the use of data-driven functions $\mathcal{F}$ that interpolate to the perturbative QCD regime, and that depend only on the Lorentz structure and energy scale $\mu$ of the process. In the case of vertices involving the ALP, one pseudoscalar and two vector mesons, the interpolating function is obtained by fitting to $e^+e^- \to \rho\pi,\omega\pi, K^*K, \phi\eta$ data,
\begin{equation}
    \mathcal{F}_{PVV}(\mu) = \left\{\begin{matrix}
        1 & \mu<1.4\,\mathrm{GeV}\,,\\ \mathrm{interpolation} & 1.4\,\mathrm{GeV} <\mu<2\,\mathrm{GeV}\,\\(1.4\,\mathrm{GeV}/\mu)^4 & \mu>2\,\mathrm{GeV}\,.
    \end{matrix} \right.
\end{equation}
For the rest of the Lorentz structures, the corresponding $\mathcal{F}(\mu)$ functions are obtained by re-scaling $\mathcal{F}_{PVV}(\mu)$ to reproduce the correct $\mu$ scaling.

In our analysis we included the decays into three pseudoscalar mesons $a\to \pi^0\pi^+\pi^-$, $a\to3\pi^0$, $a\to\eta\pi^+\pi^-$, $a\to\eta\pi^0\pi^0$, $a\to\eta'\pi^+\pi^-$, $a\to\eta'\pi^0\pi^0$. In all channels, we include contributions from the contact terms resulting from expanding the chiral effective Lagrangian (both direct contributions and ALP-mixing effects), plus resonant contributions from scalar and tensor mesons.

Using the VMD Lagrangian, we can calculate the decay rates of the ALP into two vector mesons. In particular, the decay rate for $a\to \omega^0\omega^0$ is given by
\begin{equation}
    \Gamma(a\to \omega^0\omega^0) = \frac{9 g^4|\mathcal{F}(m_a)\langle\boldsymbol{\omega^0}\boldsymbol{\omega^0}\boldsymbol{\tilde{a}}\rangle|^2}{256\pi^5 F_0^2}m_a^3 \left(1-\frac{4m_\omega^2}{m_a^2}\right)^{3/2}\,.
\end{equation}
The rest of the vector mesons are wide resonances, and therefore it would be necessary to compute the four-body decays $a\to \rho\rho\to 4\pi$, $a\to \phi^0\phi^0\to 4K$ and $a\to K^* \overline{K}^*\to 2K 2\pi$. Since the calculations are more involved, and these decay channels are always subdominant, we do not consider them. The VMD Lagrangian also describes the decays of the ALP into one vector meson and one photon: we considered $a\to\gamma\rho^0(\to 2\pi)$.

Finally, following~\cite{Bauer:2021wjo,Cornella:2023zme}, we can calculate the ALP production in $K\to \pi a$ processes using the Lagrangian for weak interactions in the chiral perturbation theory.

\bibliographystyle{utphys}
\bibliography{biblio_pheno}

\end{document}